\documentclass[traditabstract]{aa}
\usepackage{epsfig}    
\usepackage{lscape}
\usepackage{natbib}
\bibpunct{(}{)}{;}{a}{}{,}    %% natbib format like A&A and ApJ
\usepackage{graphics}
\usepackage{graphicx,graphics}
\usepackage{color} 
\usepackage{times}
\usepackage{amsmath}
\usepackage{amssymb}
\begin{document}

\title {Searching for  Milky Way twins: Radial abundance distribution as a strict criterion} 

\author {
         L.~S.~Pilyugin\inst{\ref{ITPA},\ref{MAO}}            \and 
         G.~Tautvai\v{s}ien\.{e}\inst{\ref{ITPA}}             \and
         M.~A.~Lara-L\'{o}pez\inst{\ref{UCM}}
        }
\institute{
  Institute of Theoretical Physics and Astronomy, Vilnius University, Sauletekio av. 3, 10257, Vilnius, Lithuania \label{ITPA} \and
  Main Astronomical Observatory, National Academy of Sciences of Ukraine, 27 Akademika Zabolotnoho St, 03680, Kiev, Ukraine \label{MAO} \and
  Departamento de F\'{i}sica de la Tierra y Astrof\'{ı}sica, Instituto de F\'{ı}sica de Part\'{ı}culas y del Cosmos, IPARCOS. Universidad Complutense
  de Madrid (UCM), E-28040, Madrid, Spain \label{UCM}
  }

\abstract{
  We search for  Milky Way-like galaxies among a sample of approximately 500 galaxies. The characteristics we considered of the candidate galaxies are the following:
  stellar mass $M_{\star}$,  optical radius $R_{25}$, rotation velocity $V_{rot}$, central oxygen abundance (O/H)$_{0}$, and  abundance at the optical radius (O/H)$_{R_{25}}$.
  If the values of  $R_{25}$ and $M_{\star}$ of the galaxy were close to that of the Milky Way, then the galaxy  was referred to as a structural Milky Way analogue  (sMWA).
  The oxygen   abundance  at a given radius of a galaxy is defined by the evolution of that region (astration level, that is, the fraction of gas converted into stars, as well as gas
  exchange with the surroundings), and we then assumed that the similarity of  (O/H)$_{0}$ and (O/H)$_{R_{25}}$ in two galaxies suggests a similarity in their (chemical)
  evolution. If the values of  (O/H)$_{0}$ and (O/H)$_{R_{25}}$ in the galaxy were close to that of the Milky Way, then the galaxy was referred to as an evolutionary Milky Way
  analogue  (eMWA). If the galaxy was simultaneously an eMWA and sMWA, then the galaxy was considered a Milky Way twin. We find that the position of the Milky Way on the
  (O/H)$_{0}$ -- (O/H)$_{R_{25}}$ diagram shows a large deviation from the general trend in the sense that the (O/H)$_{R_{25}}$ in the Milky Way is appreciably lower than in other
  galaxies of similar (O/H)$_{0}$. This feature of the Milky Way evidences that its (chemical) evolution  is not typical. We identify four galaxies (NGC~3521, NGC~4651, NGC~2903,
  and MaNGA galaxy M-8341-09101) that are simultaneously sMWA and eMWA and can therefore be considered as  Milky Way twins.  In previous studies,    Milky Way-like galaxies
  were selected  using structural and morphological characteristics, that is, sMWAs were selected.  We find that the abundances at the centre and at the optical radius
  (evolutionary characteristics) provide a stricter criterion for selecting real Milky Way twins.
}


\titlerunning{Milky Way twins}
\authorrunning{Pilyugin et al.}
\maketitle

\section{Introduction}
%=====================

The position of the Milky Way in the context of the general galaxy population is an important subject of study, as it concerns identifying whether the Milky Way is really
a typical spiral galaxy and the way(s) in which it differs if it is not. In a number of investigations, there have been attempts to establish how typical the Milky Way is
amongst galaxies and how many Milky Way-like galaxies exist.  Milky Way-like galaxies are usually referred to as Milky Way analogues (MWAs). \citet{Boardman2020a} noted
that there is no single and commonly accepted definition of an MWA; rather, the definition can change depending on the goals of a particular study.  Different characteristics
of the Milky Way can be used when comparing it to other galaxies  \citep{deVaucouleurs1978,Hammer2007,Mutch2011,Licquia2015b,McGaugh2016,Licquia2016b,FraserMcKelvie2019,Boardman2020a,Fielder2021},
and galaxies may be identified as being Milky Way-like on the basis of having similar qualitative characteristics to the Milky Way or on the basis of their position relative
to the Milky Way in a given parameter space.

\citet{deVaucouleurs1978} identified four nearby galaxies (NGC~1073, NGC~4303, NGC~5921, and NGC~6744) as MWAs since their photometric parameters (morphological type $T$, 
isophotal diameter, effective diameter, absolute $B$ magnitude, and mean colour index ($B - V$)) are in remarkably close agreement with the corresponding values of the 
Milky Way. \citet{Mutch2011} selected MWAs using stellar mass and the structural parameter $f_{DeV}$ as the selection criteria, where the structural parameter $f_{DeV}$   
describes the fraction of light that corresponds to the de Vaucouleurs component of the light profile and is correlated with the morphological type of galaxy as well as 
with the bulge-to-total ratio ($B/T$). \citet{Licquia2015b} selected a large MWA sample from the Eighth Data Release \citep {Aihara2011} of the Sloan Digital Sky Survey III 
\citep[SDSS-III; ][]{York2000} on the basis of stellar mass and current star formation rate. \citet{Boardman2020a} presented a sample of 62 galaxies identified as MWAs on the
basis of their stellar masses and bulge-to-total ratios from the Mapping Nearby Galaxies at Apache Point Observatory (MaNGA) survey  \citep{Bundy2015}. The selection of MWAs
using only two characteristics of the Milky Way as the selection criteria is based on the Copernican assumption that the Milky Way is not extraordinary amongst galaxies, that is,
any property of the Milky Way is similar to that of the MWA galaxies selected \citep{Mutch2011,Licquia2015b,Boardman2020a}.

It is evident that when the process of selecting MWAs involves using a larger number of parameters as well as more stringent constraints on a given parameter, it yields
a sample of MWA galaxies with properties that are closer to the true properties of the Milky Way. However, the use of a greater number of selection criteria (or an overly
strict definition of “analogue”) yields few, if any, MWA galaxies \citep{Boardman2020b}. Indeed, \citet{FraserMcKelvie2019} found only 176 MWAs amongst over a million
galaxies in the Sloan Digital Sky Survey (SDSS) Data Release 7 \citep{Abazajian2009} when selecting based on stellar mass, bulge-to-total ratio ($B/T$), and morphology
(the presence of spiral arms and the presence of a bar).  \citet{Boardman2020a} did not find a single MWA in the  SDSS-IV MaNGA survey \citep{Bundy2015} when attempting
to select based on a combination of stellar mass, star formation rate,  bulge-to-total ratio ($B/T$), and disc scale length. Those results suggested that some parameter(s)
of the Milky Way galaxy may be unusual or that the combination of its properties may be rare.

\citet{Hammer2007} compared the Milky Way to disc galaxies within the same mass range, using its location in the Tully-Fisher (variant of $V_{rot}$ versus $k$-band luminosity) and
other diagrams. They found for all relationships that the Milky Way galaxy is systematically offset by 1$\sigma$, showing a significant deficiency in stellar mass, angular momentum,
disc radius, and Fe/H in the stars in its outskirts at a given $V_{rot}$, i.e. on the basis of its location in the ($L_{k}$, $V_{rot}$, and $R_{25}$) volume, the fraction of spirals
like the Milky Way was found to be around 7\%. In contrast,  \citet{McGaugh2016} found that the Milky Way appears to be a normal spiral galaxy that obeys scaling relations, such as
Tully-Fisher, and the size-mass relation. He noted that the galaxy NGC~3521 has a baryonic mass, rotation velocity, and scale disc length that are very similar to those of the
Milky Way; these quantities are identical within the uncertainties. \citet{McGaugh2016} concluded that NGC~3521 is a near twin to the Milky Way. \citet{Licquia2016b} considered
the three variants of the Tully-Fisher relation (rotation velocity versus $i$-band luminosity, stellar mass, and baryonic mass) and found that our Galaxy’s properties are in excellent
agreement with those Tully-Fisher relations. They also examined the three-dimensional relation  (rotation velocity-luminosity-radius) using the disc scale length as a measure of
size for spiral galaxies. \citet{Licquia2016b} found that the Milky Way lies farther from the relation in comparison to 90\% of other spiral galaxies, yielding evidence that it is
unusually compact for its rotation velocity and luminosity.  The expected disc scale length for the Milky Way from the relation is approximately 5 kpc, nearly twice as large as the
observed value. Thus, there is a hint that some parameter(s) of the Milky Way (e.g. size) may not be  perfectly typical and that the combination of the parameters may be rather rare.

The characteristics of a galaxy can be conditionally divided into two types. The parameters of the first type (e.g. morphology, luminosity, stellar mass, rotation velocity) describe
the structure and global characteristics of a galaxy  at the present-day epoch and can be called "structural parameters." The parameter of the second type is related to the oxygen
abundance of a galaxy. The oxygen abundance at a given radius of a galaxy is defined by the evolution of this region of a galaxy (fraction of gas converted into stars, i.e. astration
level, and matter exchange with the surroundings). Then the oxygen abundance can be considered as an indicator of a galaxy's evolution and can be called   an "evolutionary parameter."
A galaxy located close to the Milky Way in the field(s) of the first type of parameters is referred to in this work as a structural Milky Way analogue (sMWA). A galaxy located close to
the Milky way in the field of the second type of parameters is referred to as an evolutionary Milky Way analogue (eMWA). If a galaxy is simultaneously an sMWA and an eMWA, then
such a galaxy is  considered a twin of the Milky Way.

In fact, sMWAs were selected and examined in the papers cited above. The oxygen abundance at the optical radius of the Milky Way is appreciably lower in comparison to other galaxies
with similar central oxygen abundance (see below). This feature of the Milky Way evidences that its (chemical) evolution  is not typical. Therefore, it is highly useful to study the
evolutionary analogues of the Milky Way. The goal of the current study is to search for and examine the galaxies that are simultaneously  sMWAs and  eMWAs. We considered three structural
parameters: optical radius, stellar mass, and rotation velocity. Since rotation curve data are not available for some galaxies or the rotation velocity has been measured with a large
uncertainty (e.g. in face-on galaxies with low-inclination angles) the position in the $M_{\star}$ -- $R_{25}$ -- $V_{rot}$ has not been determined. The stellar mass -- optical radius
diagram was therefore used to select the sMWAs. The central abundance versus abundance at the optical radius diagram served in the search for eMWAs.   We compared the Milky Way
with a sample of spiral galaxies for which radial abundance distributions, optical radii, stellar masses, and  rotation curves were derived by us or compiled from the literature. Our
sample of the comparison galaxies involves the nearby galaxies and the galaxies from the MaNGA survey.  The obtained candidates of Milky Way twins are examined in more detail.

This paper is organised in the following way: the Milky Way characteristics are described in Sect.~2. In Sec.~3, a sample of (comparison) galaxies is reported.
The selection of the Milky Way-like galaxies is given in Sect.~4. The discussion is in Sect.~5, and Sect.~6 contains a brief summary.

\section{Characteristics of the Milky Way Galaxy}
%=====================

\subsection{Milky Way optical radius R$_{25}$  }
%=====================

The maximum size of the Galactic stellar disc of the Milky Way is not yet known. Some studies have suggested an abrupt drop-off of the stellar density of the disc at
galactocentric distances $R$ $\ga$ 15 kpc. \citet{Minniti2011}  considered the clump giants of the disc as standard candles, calibrated from Hipparcos parallaxes.
They concluded that there is an edge in the stellar disc of the Milky Way at $R$ = 13.9$\pm$0.5~kpc along various lines of sight across the Galaxy.  Based on the Two
Micron All Sky Survey (2MASS) data, \citet{Amores2017}  found a disc truncation at approximately 16.1$\pm$1.3~kpc. If age dependence for disc parameters is considered,
then the disc truncation is 19.4$\pm$1.4 kpc or 18.7$\pm$1.6~kpc. The density drops by 90\% within 1~kpc, and therefore the cutoff is sharp. In practice, this means
that no disc stars, or very few of them, should be found beyond this limit.

In contrast, \citet{LopezCorredoira2018}  revealed the presence of disc stars at $R > 26$~kpc. \citet{Chrobakova2020}  calculated the stellar density using star counts
obtained from Gaia Data Release 2 up to a galactocentric distance of $R$ = 20 kpc. They found that the stellar density maps can be fitted by an exponential disc in the
radial direction $h_{d}$ = 2.07$\pm$0.07 kpc, with a weak dependence on the azimuth, up to 20 kpc without any cutoff. The flare and warp are clearly visible. They found
the stellar density in the solar neighbourhood $\rho$  = 0.064 stars/pc$^{3}$ where the solar neighbourhood is defined as the area where 7.5~kpc $< R <$ 8.5~kpc and
$-0.05 < z < 0.05$~kpc.  \citet{Chrobakova2020} noted that this does not mean that radial truncations are not possible in spiral galaxies, as there are other galaxies
in which they have been observed many years ago \citep{vanderKruit1981}, though the Milky Way is not one of them.  Is should be noted that an exponential distribution was
also observed for the gas density of the Milky Way without any truncation up to a distance of 40~kpc from the centre \citep{Kalberla2008}.
 
The optical radius can be estimated from the radial distributions of the surface stellar mass density in the disc described by Eq.~\ref{equation:r-star} by adopting ($M/L$)$_{B}$ =
1.4 \citep{Flynn2006}. The radial distribution of the surface stellar mass density in the disc is described by the expression   
\begin{equation}
\Sigma_{\star} = \Sigma_{\star,R_{0}} \exp \left (-\frac{R - R_{0}}{h_{d}} \right )  ,
\label{equation:r-star}
\end{equation}
where  $\Sigma_{\star,R_{0}}$ is the surface stellar mass density at the solar galactocentric distance \citep[$R_{0}$ = 8.178$\pm$0.013 kpc, ][]{Gravity2019} and $h_{d}$ is the disc
scale length.  It should be noted that a two-component disc (thin and thick discs of different scale lengths) is usually considered \citep[e.g.][]{McMillan2017}. The single exponential
disc is adopted here.

\citet{Flynn2006} found a local stellar disc surface density of 35.5 $M_{\sun}$/pc$^{2}$, while \citet{McKee2015} found the value of 33.4 $M_{\sun}$/pc$^{2}$. The solar neighbourhood
is located in the interarm region, and  the “counted” local stellar disc surface density should be corrected for the spiral arm enhancement in order to find the azimuthal average of
the surface stellar mass density at the solar galactocentric distance, that is, a 10\% enhancement should be added to the “counted”   local stellar disc surface density
\citep{Flynn2006,Kubryk2015}. This results in a stellar disc surface density at the solar galactocentric distance of 39~$M_{\sun}$/pc$^{2}$ \citep{Flynn2006}  and 37~$M_{\sun}$/pc$^{2}$
\citep{McKee2015}. We adopted $\Sigma_{\star,R_{0}}$ = 38~$M_{\sun}$/pc$^{2}$. \citet{BlandHawthorn2016} analysed 130 papers on disc parameters, with scale lengths ranging from 1.8 to 6.0~kpc.
Their analysis of the main papers (15 in all) on this topic led to $h_{d}$ = 2.6 $\pm$ 0.5~kpc.

Regarding the conversion of the solar units $L_{\sun}$/pc$^{2}$ to  mag$_{B}$/arcsec$^{2} $, we note that the $\mu_{B}$ = 25.0~mag$_{B}$/arcsec$^{2}$ corresponds to 6.44 $L_{\sun}$/pc$^{2}$
with the adopted $M_{B,\sun}$ = 5.45 (the relation is $\mu_{B}$(mag/arcsec$^{2}$) = --2.5log$L_{B}$ ($L_{\sun}$/pc$^{2}$) +27.022). The value of the surface brightness of
$\mu_{B}$ = 25.0 mag$_{B}$/arcsec$^{2}$ corresponds to the stellar surface mass density of $\sim$9.0 $M_{\sun}$/pc$^{2}$, assuming ($M/L$)$_{B}$ = 1.4 \citep{Flynn2006}. The value of the
optical radius $R_{25}$ is around 12 kpc for the $h_{disc}$ = 2.6 kpc and the stellar disc surface mass density at the solar galactocentric distance of 38~$M_{\sun}$/pc$^{2}$. 
This value of the optical radius is close to the early estimation of the optical radius ($R_{25}$ = 11.5 kpc) obtained by \citet{deVaucouleurs1978}.
We adopted $R_{25}$ = 12.0 kpc for the Milky Way.

\subsection{Stellar mass  $M_{\star}$  and black hole mass M$_{BH}$ of the Milky Way}
%=====================

The stellar mass of the Milky Way is the sum of the stellar masses of disc and bulge. Two comments on the determination of the stellar mass of the Milky Way should be presented.
First, the mass of a single component exponential disc of a given scale length can be estimated through the simple relation \citep{McMillan2017}: 
\begin{equation}
M_{\star,d} = 2 \pi \Sigma_{\star,0} h_{d}^{2}, 
\label{equation:mdisk}
\end{equation}
where  $\Sigma_{\star,0}$ is the central surface mass density of the disc and $h_{d}$ is the scale length of the surface mass density distribution. However, a two-component disc (thin and
thick disc) is usually considered. Therefore, the above expression should be applied to each component of the disc. Second, it is often used to derive not the mass of the pure
bulge, but the sum of the mass of the bulge and disc within the bulge region.

%++++++++++++++++++ Table 1   values of the stellar mass from the literature 
%\setcounter{table}{1}
\begin{table}
\caption[]{\label{table:star}
    Estimations of the stellar mass of the Milky Way from different publications.
}
\begin{center}
\begin{tabular}{lccc} \hline \hline
%\multicolumn{3}{|c|}{scatter in 12+logO/H)$_{R_{x}}$}      \\           
                      &
disc                  &
bulge                 &
total                 \\
                       &
(10$^{10}$M$_{\sun}$)   &
(10$^{10}$M$_{\sun}$)   &
(10$^{10}$M$_{\sun}$)   \\
    \hline         
\citet{Flynn2006}          &                    &                  &  5.18$\pm$0.32      \\ 
\citet{Bovy2013}           &  4.6$\pm$0.3       &   0.6            &  5.2                \\ 
\citet{Licquia2015}        &  5.17$\pm$1.11     &   0.91$\pm$0.07  &  6.08$\pm$1.14      \\ 
\citet{Licquia2016}        &  4.8$^{+1.5}_{-1.1}$  &   0.91$\pm$0.07  &  5.7$^{+1.5}_{-1.1}$   \\ 
\citet{McMillan2017}       &                    &   0.91$\pm$0.09  &  5.43$\pm$0.57       \\ 
\citet{Cautun2020}         &  4.10$\pm$0.40     &   0.94$\pm$0.10  &  5.04$\pm$0.50       \\ 
                    \hline
\end{tabular}\\
\end{center}
\end{table}

Estimations of the stellar mass of the Milky Way have been reported in many papers \citep[][among others]{Flynn2006, Bovy2013, Licquia2015, Licquia2016, BlandHawthorn2016, McMillan2017, Cautun2020}. 
Estimations from some papers are listed in Table.~\ref{table:star}. We adopted the following value of the stellar mass of the Milky Way: $M_{\star}$ = 5.2$\times$10$^{10}$ $M_{\sun}$ or
log($M_{\star}/M_{\sun}$) = 10.716.
The mass of the black hole at the centre of the Milky Way is equal to $M_{BH}$ = 4.15$\times$10$^{6}$ $M_{\sun}$ or log($M_{BH}/M_{\sun}$) = 6.618 \citep{Gravity2019}.

\subsection{Rotation velocity V$_{rot}$}
%=====================

%===============    Fig  No  1            Rotation curve 
\begin{figure}
\resizebox{1.00\hsize}{!}{\includegraphics[angle=000]{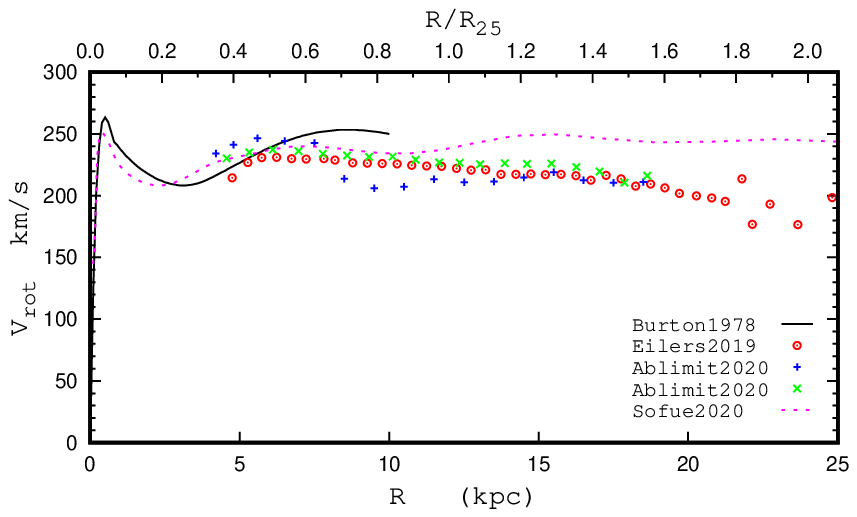}}
\caption{
 Rotation curves of the Milky Way from  \citet{Burton1978}, \citet{Eilers2019}, \citet{Ablimit2020} (two variants), and  \citet{Sofue2020}. 
}
\label{figure:rc}
\end{figure}

The rotation curve of the Milky Way has been obtained by many studies \citep[e.g.][]{Burton1978,Eilers2019,Ablimit2020,Sofue2020}. Fig.~\ref{figure:rc} shows the rotation curves from
one early investigation \citep{Burton1978} and from three recent papers \citep{Eilers2019,Ablimit2020,Sofue2020}. \citet{Burton1978} have determined the Galactic rotation curve in
the galactocentric distance range up to 10 kpc using  molecular gas velocities combined with atomic gas velocities. \citet{Eilers2019} have measured the circular velocity
curve of the Milky Way with the highest precision to date across galactocentric distances of 5 $\la R \la$ 25 kpc. They used precise parallaxes, spectral data, and photometric
information from the Apache Point Observatory Galactic Evolution Experiment (APOGEE), the Wide-field Infrared Survey Explorer (WISE), 2MASS data, and Gaia for
23,000 luminous red giant stars. \citet{Ablimit2020}  analysed about 3,500 classical Cepheids identified from different surveys. They used two kinematical methods to measure the
Galactic rotation curve in the galactocentric distance range from $\sim$4 to $\sim$19 kpc. \citet{Sofue2020}  constructed an unified rotation curve of the Milky Way from the Galactic
centre to the galactocentric distance of about 100 kpc by averaging the published rotation curves that are based on the molecular gas velocities and on the velocities of
approximately 16,000 red clump giants in the outer disc combined with velocities of around 5,700 halo K giants.

The obtained values of the maximum rotation velocity are between $\sim$230~km\,s$^{-1}$ and $\sim$245~km\,s$^{-1}$. The value of the  rotation velocity of the Milky Way that we adopted  is
$V_{rot}$ = 235~km\,s$^{-1}$, or log$V_{rot}$ = 2.371.

\subsection{Oxygen abundance}
%=====================

%===============    Fig  No 2            OH - R  Milky Way 
\begin{figure}
\resizebox{1.00\hsize}{!}{\includegraphics[angle=000]{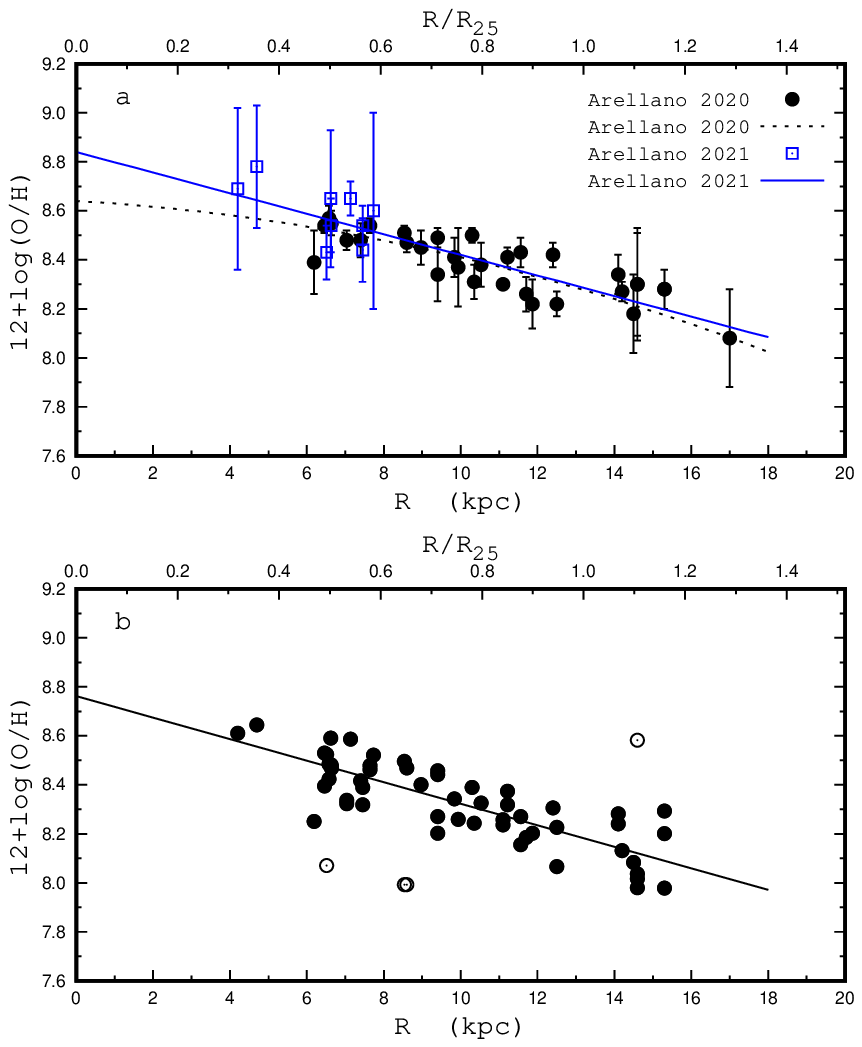}}
\caption{
  Radial oxygen abundance distribution in the Milky Way.
  {\sl Panel {\bf a}:} Radial abundance gradients obtained by \citet{Arellano2020} and \citet{Arellano2021}. The black circles designate the  H\,{\sc ii} regions from
  \citet{Arellano2020}. The dashed black curve is the quadratic fit to those data. The blue squares mark the  H\,{\sc ii} regions from \citet{Arellano2021}. The solid
  blue line shows the linear fit to all the data. 
  {\sl Panel {\bf b}:} Radial abundance distribution in the Milky Way obtained in this work (see text). The black circles denote the  H\,{\sc ii} regions used in the determination
  of the radial abundance gradient (solid line).   The open circles designate the  H\,{\sc ii} regions excluded from the determination of the abundance gradient. 
}
\label{figure:r-ohte}
\end{figure}

The oxygen abundances of 42 H\,{\sc ii} regions of the Milky Way determined through the direct $T_{e}$ method have been recently reported by \citet{Arellano2020,Arellano2021}. 
Panel (a) in Fig.~\ref{figure:r-ohte} shows the radial distribution of the oxygen abundances in the Milky Way obtained by the cited authors. The black circles denote the H\,{\sc ii}
regions from \citet{Arellano2020}. The dashed black curve is the quadratic fit to those data
\begin{equation}
12+\log(\rm O/H) = 8.64 -0.009 \times R -0.0014 \times R^{2}   .
\label{equation:OH2020}
\end{equation}
The blue squares in panel (a) of Fig.~\ref{figure:r-ohte} designate the  H\,{\sc ii} regions from \citet{Arellano2021}. The solid blue line is the linear fit to all the data
\begin{equation}
12+\log(\rm O/H) = 8.84(\pm 0.09) -0.042(\pm 0.009) \times R  .
\label{equation:OH2021}
\end{equation}

It is believed that the $T_{e}$ method, which is based on the measurements of temperature sensitive line ratios, should give accurate oxygen abundances. In practice, however,
$T_{e}$-based oxygen abundances of the same H\,{\sc ii} region derived in different works can differ for two reasons: First, there may be errors in the line intensity
measurements. Second, the $T_{e}$-based oxygen abundances depend on the relations used to convert the values of the line fluxes to the electron temperatures and to the ion abundances.
The determined abundances also  depend on the adopted relationship between the electron temperature in low-ionisation zones and electron temperatures in the high-ionisation part
of the nebula. Differences between the oxygen abundances in a given H\,{\sc ii} region produced by different relations used in the $T_{e}$ method can be appreciable. For example,
\citet{Esteban2017}  found an oxygen abundance of 12 + log(O/H)$_{T_{e}}$ = 8.14$\pm$0.05 in the Galactic H\,{\sc ii} region Sh 2-83, while \citet{Arellano2020}  found
12 + log(O/H)$_{T_{e}}$ = 8.28$\pm$0.08 in the same region using the same spectroscopic measurements. \citet{Berg2015}  detected auroral lines in 45  H\,{\sc ii} regions in the nearby
galaxy NGC~628. They determined the $T_{e}$-based abundances in those H\,{\sc ii} regions and estimated the radial abundance gradient. They found the central (intersect) oxygen abundance
12 + log(O/H)$_{0}$ = 8.83$\pm$0.07 in  NGC~628. In their recent paper \citep{Berg2020}, they recalculated the ionic and total $T_{e}$-based abundances in the same   H\,{\sc ii} regions
and determined the central (intersect) oxygen abundance 12 + log(O/H)$_{0}$ =  8.71$\pm$0.06 in  NGC~628.

In this paper, the oxygen abundance of the Milky Way is compared to the oxygen abundances in galaxies determined through the $R$ calibration from \citet{Pilyugin2016}. The $T_{e}$-based
oxygen abundances in H\,{\sc ii} regions used as the calibrating data points in the construction of the $R$ calibration were derived using the $T_{e}$-method equations reported
in \citet{Pilyugin2012}. Therefore the $T_{e}$-based oxygen abundance in the Milky Way and the $R$ calibration-based abundances in other galaxies correspond to this unique
abundance scale \citep{Pilyugin2016,Pilyugin2022}  if only the $T_{e}$-based oxygen abundances in H\,{\sc ii} regions in the Milky Way are determined  using the $T_{e}$-method equations
from \citet{Pilyugin2012}. We then redetermined the oxygen abundances in H\,{\sc ii} regions of the Milky Way.  The line intensity measurements  in the spectra of the H\,{\sc ii}
regions were taken from the same sources as in \citet{Arellano2020} and \citet{Arellano2021}. If the measurements of two auroral lines  ([O\,{\sc iii}]$\lambda$4363 and
[N\,{\sc ii}]$\lambda$5755) were available for the H\,{\sc ii} region, then the two values of the oxygen abundance were determined. The first value of the abundance is based on the
electron temperature in high-ionisation parts of the nebula estimated using the auroral line [O\,{\sc iii}]$\lambda$4363 and the electron temperature in the low-ionisation parts
of the nebula obtained from the relationship between the electron temperature in low-ionisation zones and electron temperatures in high-ionisation parts of the nebula.
The second value of the abundance is based on the electron temperature in low-ionisation zones estimated from the measured auroral line [N\,{\sc ii}]$\lambda$5755 and the electron
temperature in high-ionisation zones determined from the relationship between electron temperatures.

The obtained oxygen abundances are presented in the lower panel of Fig.~\ref{figure:r-ohte}. Because there is no visual evidence for a break in the O/H gradient, we characterised
the O/H gradient in the Milky Way with a single linear regression. The points with deviations in excess of 0.3~dex from the regression (open circles  in the lower panel of
Fig.~\ref{figure:r-ohte}) were rejected when determining  the final relation. The obtained relation,  
\begin{equation}
12+\log(\rm O/H) = 8.76(\pm 0.04) -0.044(\pm 0.004) \times R  ,
\label{equation:OHhere}
\end{equation}
is shown by the solid line  in the lower panel of Fig.~\ref{figure:r-ohte}. The scatter in O/H around this relation is 0.092 dex for the 54 data points. The  derived O/H -- $R$ relation
(Eq.~\ref{equation:OHhere}) resulted in the value of the oxygen abundance at the optical radius ($R_{25}$ = 12 kpc) of 12 + log(O/H)$_{R_{25}}$ = 8.24. 

We emphasise that we do not pretend that the values of the $T_{e}$-based abundances for the Milky Way recomputed here are more accurate than the original abundance
values derived by \citet{Arellano2020} and \citet{Arellano2021}. Our aim is that the abundances of all the galaxies used here correspond to the same metallicity scale.
This is mandatory for the current study. 

We also note that the derived value of the central (intersect) oxygen abundance in the Milky Way may be slightly overestimated. The measured radial distributions of the surface
mass density of atomic  H\,{\sc i} and molecular H$_{2}$ hydrogen in the Milky Way suggests that the gas can be entirely exhausted (fully converted into stars) within a radius of
approximately 1 kpc \citep{McMillan2017}. If this is the case, then the gas-phase oxygen abundance reaches the maximum (attainable) value at the radius of around 1 kpc and remains
constant within this radius.

Thus, we adopted the following values for the Milky Way characteristics. 
The stellar mass is $M_{\star}$ = 5.2$\times$10$^{10}$ M$_{\sun}$ or log$M_{star}$ = 10.716. %\\
The rotation velocity is $V_{rot}$ = 235 km\,s$^{-1}$ or log$V_{rot}$ = 2.371.  %\\
The optical radius is $R_{25}$ = 12 kpc or log$R_{25}$ = 1.079. %\\ 
The mass of the black hole is M$_{BH}$ = 4.15$\times$10$^{6}$ M$_{\sun}$ or logM$_{BH}$ = 6.618.  %\\ 
The central oxygen abundance is 12 + log(O/H)$_{0}$ = 8.76.   %\\
The oxygen abundance at the optical radius is 12 + log(O/H)$_{R_{25}}$ = 8.24.

\section{A sample of comparison galaxies}
%=========================================

%++++++++++++++++++ Table     Characteristics of nearby galaxies 
%\setcounter{table}{1}
\begin{table*}
\caption[]{\label{table:nearby}
    Characteristics of nearby galaxies.
}
\begin{center}
\begin{tabular}{lrccrcrrccc} \hline \hline
%\multicolumn{3}{|c|}{scatter in 12+logO/H)$_{R_{x}}$}      \\           
galaxy                  &
$R_{25}$                 &
$i$                     &
PA                      &
$d$                       &
$V_{rot}$                &
log$M_{\star}$           &
$R_{25}$                &
12+log(O/H)$_{0}$       &
12+log(O/H)$_{R_{25}}$   &
log$M_{BH}$            \\
                       &
arcmin                 &
degree                 &
degree                 &
Mpc                    &
km\,s$^{-1}$                   &
$M_{\sun}$              &
kpc                    &
                       &
                       &
$M_{\sun}$              \\
\hline
%             R25min    i      PA        d        Vrot      logM     R25kpc     OHo           OH25         logMbh                                                                                    
NGC~0224   &  95.27  &  77  &   38   &   0.82  &  253    &  11.00  &  22.72  &  8.715      &  8.337       &  8.15    \\ 
NGC~0253   &  13.77  &  76  &  235   &   3.70  &  211    &  10.54  &  14.82  &  8.493      &  8.414       &  7.00    \\ 
NGC~0300   &  10.94  &  42  &  106   &   1.94  &   85    &   9.32  &   6.17  &  8.444      &  8.019       &  --      \\ 
NGC~0598   &  35.40  &  54  &  201   &   0.94  &  125    &   9.69  &   9.68  &  8.489      &  8.037       &  --      \\ 
NGC~0628   &   5.24  &   6  &   25   &   9.91  &   --    &  10.36  &  15.09  &  8.667      &  8.257       &  --      \\ 
NGC~0753   &   1.26  &  50  &  128   &  72.40  &  210    &  10.91  &  26.45  &  8.586      &  8.522       &  --      \\ 
NGC~0925   &   5.24  &  66  &  287   &   9.29  &  115    &   9.75  &  14.15  &  8.440      &  8.080       &  --      \\ 
NGC~1058   &   1.51  &  15  &  145   &  10.60  &   --    &   9.40  &   4.66  &  8.649      &  8.384       &  --      \\ 
NGC~1068   &   3.54  &  35  &   73   &  13.97  &   --    &  10.91  &  14.38  &  8.696      &  8.594       &  6.75    \\ 
NGC~1097   &   4.67  &  46  &  134   &  13.58  &  253    &  10.76  &  18.43  &  8.661      &  8.523       &  8.38    \\ 
NGC~1232   &   3.71  &  30  &  270   &  21.50  &  220    &  10.67  &  23.18  &  8.741      &  8.254       &  --      \\ 
NGC~1313   &   6.12  &  48  &    0   &   4.32  &  112    &   9.46  &   7.69  &  8.163      &  8.016       &  --      \\ 
NGC~1365   &   5.61  &  46  &  222   &  19.60  &  300    &  10.78  &  31.98  &  8.619      &  8.411       &  6.60    \\ 
NGC~1512   &   4.46  &  42  &  262   &  18.83  &  179    &  10.72  &  24.41  &  8.741      &  8.362       &  7.78    \\ 
NGC~1598   &   0.72  &  55  &  123   &  55.80  &  110    &  10.21  &  11.73  &  8.665      &  8.390       &  --      \\ 
NGC~1672   &   3.30  &  43  &  134   &  19.40  &   --    &  10.73  &  18.64  &  8.639      &  8.510       &  7.08    \\ 
NGC~2403   &  10.94  &  63  &  124   &   3.19  &  134    &   9.65  &  10.15  &  8.465      &  7.934       &  --      \\ 
NGC~2442   &   2.75  &  29  &   27   &  21.50  &   --    &  10.84  &  17.18  &  8.606      &  8.556       &  7.28    \\ 
NGC~2805   &   3.16  &  36  &  123   &  28.70  &  106    &   9.98  &  26.34  &  8.477      &  8.106       &  --      \\ 
NGC~2835   &   3.30  &  41  &    1   &  12.22  &   --    &  10.00  &  11.74  &  8.498      &  8.193       &  6.72    \\ 
NGC~2903   &   5.87  &  65  &  204   &   8.90  &  215    &  10.52  &  15.21  &  8.707      &  8.369       &  7.06    \\ 
NGC~2997   &   4.46  &  33  &  108   &  11.30  &  185    &  10.49  &  14.66  &  8.725      &  8.362       &  5.84    \\ 
NGC~3031   &  10.69  &  59  &  330   &   3.63  &  215    &  10.69  &  11.29  &  8.638      &  8.489       &  7.81    \\ 
NGC~3184   &   3.71  &  16  &  179   &  11.62  &  210    &  10.41  &  12.53  &  8.721      &  8.395       &  --      \\ 
NGC~3351   &   3.71  &  41  &  192   &   9.96  &  196    &  10.37  &  10.74  &  8.696      &  8.619       &  6.52    \\ 
NGC~3359   &   3.62  &  53  &  350   &  22.60  &  145    &  10.18  &  23.81  &  8.401      &  7.966       &  --      \\ 
NGC~3521   &   4.16  &  73  &  340   &  10.70  &  227    &  10.70  &  12.94  &  8.78$^{a}$  &  8.29$^{a}$  &  6.85    \\ 
NGC~3621   &   4.88  &  65  &  345   &   7.06  &  140    &  10.06  &  10.03  &  8.704      &  8.124       &  6.00    \\ 
NGC~4254   &   2.68  &  34  &   68   &  13.10  &  183    &  10.42  &  10.23  &  8.662      &  8.408       &  --      \\ 
NGC~4258   &   9.31  &  72  &  331   &   7.58  &  200    &  10.71  &  20.53  &  8.600      &  8.443       &  7.58    \\ 
NGC~4303   &   3.23  &  27  &  318   &  16.99  &  150    &  10.51  &  15.95  &  8.658      &  8.261       &  6.58    \\ 
NGC~4321   &   3.71  &  27  &  153   &  15.21  &  270    &  10.75  &  16.40  &  8.616      &  8.537       &  6.67    \\ 
NGC~4395   &   6.59  &  46  &  324   &   4.51  &   80    &   9.42  &   8.65  &  8.063      &  8.078       &  5.64    \\ 
NGC~4501   &   3.46  &  64  &  141   &  16.80  &  280    &  11.00  &  16.90  &  8.774      &  8.445       &  7.13    \\ 
NGC~4625   &   1.09  &  31  &  303   &  11.75  &   60    &   9.08  &   3.74  &  8.624      &  8.489       &  --      \\ 
NGC~4651   &   1.99  &  53  &   77   &  19.00  &  215    &  10.42  &  11.00  &  8.652      &  8.275       &  --      \\ 
NGC~5055   &   5.87  &  59  &  102   &   8.87  &  192    &  10.73  &  15.16  &  8.693      &  8.414       &  8.92    \\ 
NGC~5068   &   3.62  &  35  &  342   &   5.20  &  --     &   9.41  &   5.48  &  8.474      &  8.316       &  --      \\ 
NGC~5194   &   5.61  &  22  &  173   &   8.58  &  219    &  10.66  &  14.00  &  8.700      &  8.557       &  --      \\ 
NGC~5236   &   6.44  &  24  &  225   &   4.89  &  190    &  10.53  &   9.16  &  8.688      &  8.567       &  --      \\ 
NGC~5248   &   3.08  &  47  &  109   &  14.87  &  196    &  10.41  &  13.34  &  8.517      &  8.550       &  6.30    \\ 
NGC~5457   &  14.42  &  18  &   37   &   6.85  &   --    &  10.58  &  28.73  &  8.688      &  7.895       &  6.41    \\ 
NGC~6384   &   3.08  &  55  &   31   &  25.90  &  230    &  10.76  &  23.23  &  8.783      &  8.410       &  --      \\ 
NGC~6744   &   9.98  &  50  &   16   &   9.39  &  200    &  10.77  &  27.25  &  8.840      &  8.200       &  6.89    \\ 
NGC~6946   &   5.74  &  33  &  243   &   7.34  &  186    &  10.45  &  12.26  &  8.648      &  8.390       &  --      \\ 
NGC~7331   &   5.24  &  76  &  168   &  14.70  &  244    &  11.00  &  22.39  &  8.590      &  8.545       &  8.02    \\ 
NGC~7518   &   0.71  &  47  &  294   &  47.56  &  --     &  10.17  &   9.77  &  8.675      &  8.572       &  --      \\ 
NGC~7529   &   0.43  &  29  &  157   &  63.20  &  --     &   9.87  &   7.82  &  8.637      &  8.374       &  --      \\ 
NGC~7591   &   0.98  &  68  &  148   &  67.30  &  199    &  10.57  &  19.08  &  8.638      &  8.553       &  --      \\ 
NGC~7793   &   5.24  &  50  &  290   &   3.62  &   95    &   9.40  &   5.51  &  8.477      &  8.134       &  --      \\ 
IC~0342    &  10.69  &  31  &   37   &   3.45  &  170    &  10.37  &  10.73  &  8.713      &  8.287       &  --      \\ 
IC~5201    &   4.26  &  67  &  206   &   9.20  &   98    &   9.88  &  11.39  &  8.349      &  7.819       &  --      \\ 
IC~5309    &   0.67  &  63  &   20   &  55.70  &  152    &  10.21  &  10.93  &  8.614      &  8.565       &  --      \\ 
%             R25min    i      PA        d        Vrot      logM     R25kpc     OHo           OH25         logMbh                                                                                    
                    \hline
\end{tabular}\\
\end{center}
Notes. The notation "a" indicates that the abundance distribution was traced using P calibration-based abundances.
\end{table*}

The characteristics of the Milky Way are compared with   a sample of  galaxies.  Our sample of the comparison involves two subsamples:
a sample of nearby galaxies and a sample of galaxies from the MaNGA survey. Those subsamples are described in this section.

\subsection{Nearby galaxies}
%=====================

The spectral measurements of H\,{\sc ii} regions in nearby galaxies are compiled in \citet{Pilyugin2004,Pilyugin2014}. This data compilation served as the base for the construction of the
comparison sample of nearby galaxies. To have the oxygen abundances in a unique metallicity scale, only the galaxies where the radial abundance  gradient was determined through
the $R$ calibration from \citet{Pilyugin2016} were considered, with one exception (see below). Therefore, only the spectra with measured emission
lines [O\,{\sc ii}]$\lambda\lambda$3727,3729 from the compilation in \citet{Pilyugin2014} were taken into consideration. New spectral measurements of H\,{\sc ii}
regions in the galaxies were added (if available).

The galaxy NGC~3521 is the one exception, as the abundances in ten  H\,{\sc ii} regions were estimated through the $P$ calibration and were added to the abundances in three  H\,{\sc ii}
regions determined through the $R$ calibration (Section 4.4.1). The relative accuracy of the $P$ calibration-based abundances is around 0.1 dex, that is, the oxygen abundances estimated
through the $P$ calibration agree with the $T_{e}$-based abundances within $\sim$0.1 dex \citep{Pilyugin2005}. However, the $P$ calibration encounters the following difficulty: It is well
known that the relation between the oxygen abundance and the strong oxygen line intensities is double valued, with two distinct parts, traditionally known as the upper and lower branches
of the $R_{23}$ -- O/H diagram \citep[e.g.][]{Pagel1979,Pilyugin2000,Pilyugin2001}.  Two distinct relations between the oxygen abundance and the strong oxygen line intensities have been established:
one for the upper branch (the high-metallicity calibration) and one for the lower branch (the low-metallicity calibration). There is a transition zone between the upper and lower
branches (from 12 + log(O/H) $\sim$ 8.3 to 12 + log(O/H) $\sim$ 8.0) where the $P$ calibration cannot be applied \citep[e.g.][]{Pilyugin2005}. Thus, one has to know a priori on which of the two
branches the  H\,{\sc ii} region lies. In the case of NGC~3521, we overcame this difficulty in the following way. It is known that  discs of spiral galaxies show radial oxygen abundance
gradients, in the sense that the oxygen abundance is higher in the central part of the disc and decreases with galactocentric distance.  Abundances determined through the $R$ calibration in three  H\,{\sc ii} regions in the
inner part of NGC~3521  are 12 + log(O/H) $\sim$ 8.6.  We thus started from the  H\,{\sc ii} regions in the inner part of  NGC~3521 and moved outward
until the radius where the oxygen abundance decreases to 12 + log (O/H) $\sim$ 8.3 (the start of the transition zone between the upper and lower branches). An unjustified use of the upper
branch calibration in the determination of the oxygen abundance in low-metallicity  H\,{\sc ii} regions would result in wrong estimations.  Therefore, we used the $R$ calibration  everywhere,
which is workable over the whole metallicity scale, while we used the $P$ calibration in one galaxy only.

The distances to galaxies, the angular optical radii, the angles of the orientation of the galaxy in space, the stellar masses, the rotation curves, and the masses of
the central black holes were taken from the literature. The rotation curves of a number of nearby galaxies (e.g. with small inclination angles) were either not available or
estimated with large uncertainties. The estimations of the masses of the central black holes were available for a small fraction of galaxies only.

The sample of nearby galaxies are reported in Table.~\ref{table:nearby}; the list includes 53 galaxies.
Table~\ref{table:nearby} lists the general characteristics of each galaxy. 
  The first column provides the name for each galaxy.
  The optical (isophotal) radius $R_{25}$ in arcmin of each galaxy is reported in the second column. 
  The inclination and the position angle of the major axis are listed in columns three and four.
  The distance is reported in column five.
  The rotation velocity is given in column six.
  The stellar mass is reported in column seven. 
  The optical radius in kiloparsecs, estimated from the data in columns two and five, is given in column eight. 
  The oxygen abundances at the centre 12+log(O/H)$_{0}$ and at the optical radius  12+log(O/H)$_{R_{25}}$ are listed in columns nine and ten.
  The mass of the black hole at the centre of the galaxy is reported in column 11. 
  Notes on the individual galaxies and the references to the sources of the data are given in Appendix A. The radial abundance distributions and rotation curves of several candidate
    Milky Way twins are presented in Section 4.4.

\subsection{MaNGA galaxies}
%=====================

%===============    Fig  No  3           M-8466-12702 - maps
\begin{figure*}
\resizebox{1.00\hsize}{!}{\includegraphics[angle=000]{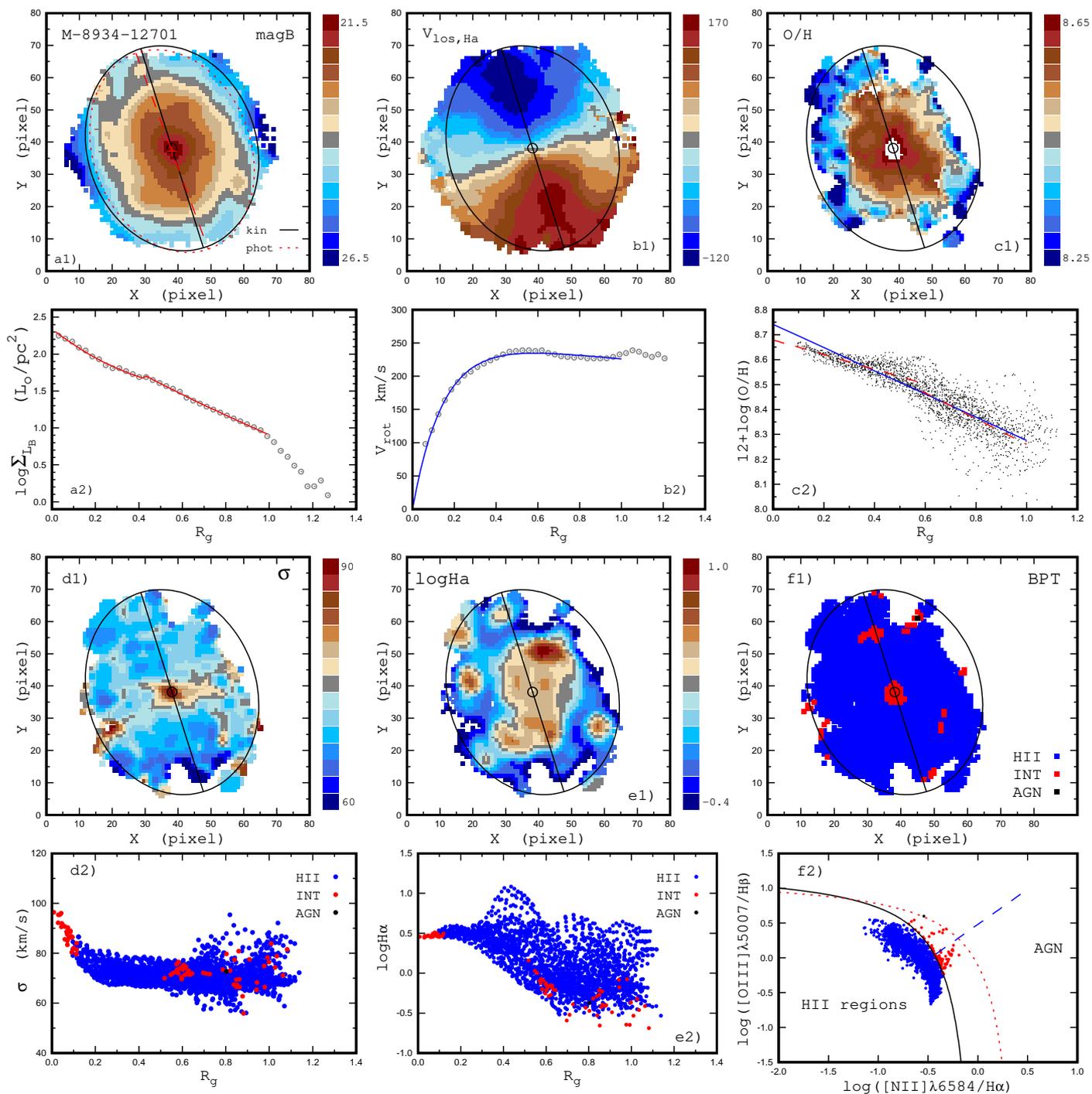}}
\caption{
Properties of the MaNGA galaxy M-8934-12701.
{\em Panel} {\bf a1:} Surface brightness distribution across the image of the galaxy in sky coordinates (pixels). The value of the surface brightness is colour-coded.
The circle shows the kinematic centre of the galaxy, the dark solid line indicates the position of the major kinematic axis of the galaxy, and the solid ellipse is the
optical radius for kinematic angles. The red cross is the photometric centre, the red dashed line indicates the position of the major photometric axis, and the red dashed
ellipse is the optical radius for photometric angles.
{\em Panel} {\bf a2:} Photometric profile (points) and its fit (line) within the optical radius. 
{\em Panel} {\bf b1:} Line-of-sight velocity field in sky coordinates.
{\em Panel} {\bf b2:} Rotation velocity curve (points) and its fit (line) within the optical radius. 
{\em Panel} {\bf c1:} Oxygen abundance map.
{\em Panel} {\bf c2:} Radial abundance distribution. The grey points denote the abundances for the individual spaxels, the solid line is the simple linear fit to those data,
while the dashed line is the broken linear fit to the same data. 
{\em Panel} {\bf d1:} Map of the gas velocity dispersion $\sigma$. 
{\em Panel} {\bf d2:} Radial distribution of the gas velocity dispersion $\sigma$ in the disc. The coloured circles denote the gas velocity dispersion for the individual spaxels of
H\,{\sc ii} region-like (blue symbols), intermediate (red symbols), and AGN-like (black  symbols) spectra classification. 
{\em Panel} {\bf e1:} Map of the measured flux in the H$\alpha$ emission line per spaxel in units of 10$^{-17}$ erg/s/cm$^{2}$/spaxel. 
{\em Panel} {\bf e2:} Radial distribution of the measured H$\alpha$ flux. The coloured circles denote the flux for the individual spaxels of
H -region-like (blue symbols), intermediate (red symbols), and AGN-like (black  symbols) spectra classification. 
{\em Panel} {\bf f1:} Map of the BPT spectra classification. The BPT types of radiation for the individual spaxels are colour-coded.
{\em Panel} {\bf f2:} BPT diagram for the individual spaxels colour-coded by H\,{\sc ii} region-like (blue symbols), intermediate (red symbols), and AGN-like (black symbols) spectra.
Solid and short-dashed curves mark the demarcation line between AGNs and H\,{\sc ii} regions defined by \citet{Kauffmann2003}
and \citet{Kewley2001}, respectively. The long-dashed line is the dividing line between Seyfert galaxies and LINERs defined by \citet{CidFernandes2010}.
}
\label{figure:m-8934-12701}
\end{figure*}

%===============    Fig  No 4            M-8466-12702 - maps
\begin{figure*}
\resizebox{1.00\hsize}{!}{\includegraphics[angle=000]{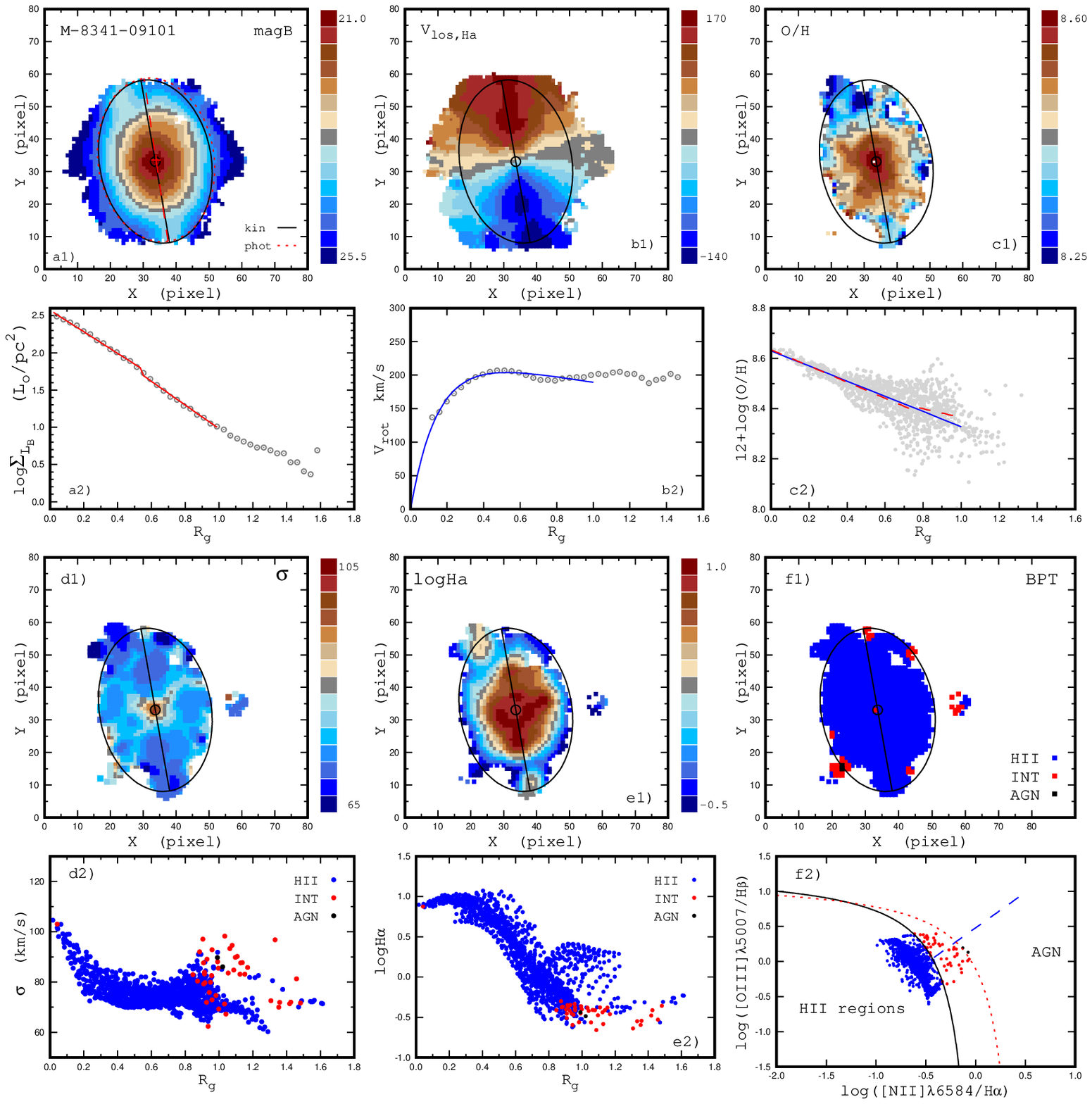}}
\caption{
Properties of the MaNGA galaxy M-8341-09101. The notations are the same as in Fig.~\ref{figure:m-8934-12701}. 
}
\label{figure:m-8341-09101}
\end{figure*}

Using the publicly available spectroscopy obtained by the MaNGA survey (\citet{Bundy2015}, \citet{Albareti2017}, Data Release 15), the rotation curves, surface brightness
profiles, radial distributions of the oxygen abundance, and the gas velocity dispersion were determined for a large sample of late-type galaxies
\citep{Pilyugin2018,Pilyugin2019,Pilyugin2020,Pilyugin2021}. As an example, Fig.~\ref{figure:m-8934-12701} and Fig.~\ref{figure:m-8341-09101} show the obtained maps and radial
distributions of different characteristics of the MaNGA galaxies M-8934-12701
and M-8341-09101. The characteristics of the MaNGA galaxies were determined using our own emission line measurements. The emission line parameters for those galaxies
are also available from the MaNGA Data Analysis Pipeline (DAP) measurements. The geometrical parameters of the galaxies (coordinates of the rotating centre, the position angle of
the major kinematic angle, and the inclination angle) were obtained for a number of galaxies using both our measurements and the DAP measurements  (we used the datacubes from the publicly available
Data Release 16) \citep{Pilyugin2021}. We also carried out a quantitative comparison between spaxel properties (oxygen abundance and gas velocity dispersion) based on our
measurements and the DAP measurements. The  geometrical parameters and spaxel properties of the galaxies based on our measurements agree with those based on the DAP measurements  \citep{Pilyugin2021}.
Thus, our data on the MaNGA galaxies are robust.

The distances to the galaxies were taken from  the NASA/IPAC Extragalactic Database ({\sc ned})\footnote{The NASA/IPAC Extragalactic Database is operated by the
Jet Propulsion Laboratory, California Institute of Technology, under contract with the National Aeronautics and Space Administration.  {\tt http://ned.ipac.caltech.edu/}}.
The {\sc ned} distances use flow corrections for Virgo, the Great Attractor, and the Shapley Supercluster infall (adopting a cosmological model with $H_{0} = 73$ km/s/Mpc,
$\Omega_{m} = 0.27$, and $\Omega_{\Lambda} = 0.73$).
We note that we considered a rather nearby sample of galaxies, and therefore the choice of  parameters for the cosmological model is not very crucial.   
We chose the spectroscopic stellar masses $M_{\star}$  of the SDSS and BOSS \citep[i.e. the Baryon Oscillation
Spectroscopic Survey in SDSS-III; see][]{Dawson2013}. The spectroscopic masses were taken from the table {\sc stellarMassPCAWiscBC03} and were determined using the Wisconsin
method \citep{Chen2012} with the stellar population synthesis models from \citet{Bruzual2003}.
 
The oxygen abundances were determined through the $R$ calibration \citep{Pilyugin2016} using the spaxel spectra.
The measured line fluxes  $F_{\lambda}^{obs}$ were corrected for the interstellar reddening $F_{\lambda}$ = $F_{\lambda}^{obs} \times 10^{C_{{\rm H}{\beta}} f_{\lambda}}$, where  $C_{{\rm H}{\beta}}$ is
the logarithmic extinction at H$\beta$. The value of   $C_{{\rm H}{\beta}}$ was estimated through a comparison between the measured and the theoretical $F_{{\rm}H\alpha}/F_{{\rm H}\beta}$ ratios,
\begin{equation}
 C_{{\rm H}\beta} = (\log(F_{{\rm}H\alpha}/F_{{\rm H}\beta})^{teor} - \log(F_{{\rm H}\alpha}/F_{{\rm H}\beta})^{obs})/(f_{{\rm H}\alpha} - f_{{\rm H}\beta}), 
\label{equation:chb}
\end{equation}
using the reddening law (function $f_{\lambda}$) of \citet{Cardelli1989} with $R_{V}$ = 3.1.  The theoretical value for the $(F_{{\rm}H\alpha}/F_{{\rm H}\beta})^{teor}$ line ratio (= 2.87)
was taken from \citet{Osterbrock2006}, assuming the case $B$ recombination. If the measured value of the ratio $F_{{\rm}H\alpha}/F_{{\rm H}\beta}$ was lower than the theoretical one, then the
reddening was assumed to be zero.

We classified the excitation of the spaxel spectrum using its position on the  standard diagnostic Baldwin--Phillips--Terlevich (BPT) diagram  [N\,{\sc ii}]$\lambda$6584/H$\alpha$
versus the [O\,{\sc iii}]$\lambda$5007/H$\beta$, as suggested by \citet{Baldwin1981}. As in our previous studies \citep{Zinchenko2019,Pilyugin2020,Pilyugin2021}, the spectra located
to the left (below) the demarcation line of \citet{Kauffmann2003} are referred to as the SF-like or H\,{\sc ii} region-like spectra; those located to the right (above) the demarcation
line of \citet{Kewley2001} are referred to as the AGN-like spectra; and the spectra located between both demarcation lines are classified as intermediate (INT) spectra.
The oxygen abundances were determined only in spaxels with H\,{\sc ii} region-like spectra.

The final sample of  MaNGA galaxies was selected by visual inspection, considering the obtained surface brightness profile, rotation curve, and abundance distribution for each galaxy.
The spaxels with measured emission lines and surface brightness needed to be well distributed across the galactic disc, that is, covering more than $\sim$2/3$R_{25}$. This condition
allowed us to estimate values of the rotation velocity, the surface brightness, and the oxygen abundance both at the centre and at the optical radius since the extrapolation is
relatively small (if any). Unfortunately, this condition resulted in the loss of galaxies where ongoing star formation does not occur over a significant fraction of galaxy. 
We note that only the spaxel spectra where the used lines are measured with a $S/N$ $>$ 3 were considered. Therefore, the spaxels with reliable measured spectra can cover less than
$\sim$2/3$R_{25}$, even if the total spaxel spectra beyond 2/3$R_{25}$ are available. The deprojected fractional galactocentric distances (normalised to the optical radius $R_{25}$)
of each spaxel were estimated using the kinematic angles for galaxies with  inclination angles of $i$ $\ga$ 30 degrees and photometric angles for face-on galaxies with inclination
angles of $i$ $\la$ 30 degrees. 

For the current study,  galaxies from the MaNGA  sample were selected when the radial oxygen abundance distribution was satisfactory approximated  by a single linear relation. The
approximations of the radial abundance distribution in each galaxy for both the simple linear fit and  the broken linear fit were estimated (panels (c2) in Fig.~\ref{figure:m-8934-12701}
and Fig.~\ref{figure:m-8341-09101}). We used the differences between the intersect values of the oxygen abundances given by the single linear relation and the abundances given by the
broken linear relation at the galaxy centre $\Delta$(O/H)$_{0}$ and at the optical radius $\Delta$(O/H)$_{R_{25}}$ as the selection criteria. The absolute values of $\Delta$(O/H)$_{0}$
and $\Delta$(O/H)$_{R_{25}}$ are usually within 0.05 dex for the galaxies \citep{Zinchenko2016,Pilyugin2017}. The galaxies where the sum of the absolute values of $\Delta$(O/H)$_{0}$ and
$\Delta$(O/H)$_{R_{25}}$ exceed 0.1 dex were rejected. 

Thus, our final sample of  comparison galaxies includes 504 galaxies, of which 451 are MaNGA galaxies  and 53 are nearby galaxies. Unfortunately, the rotation curves are available
only for 423 galaxies  (e.g. rotation curves are not derived for low-inclination MaNGA galaxies), and the stellar mass estimations are available only for 495 out of 504 galaxies.

\section{Milky Way among galaxies}

%=====================

\subsection{Comparison of the Milky Way properties with our reference sample }

%===============    Fig  No  5           OH-R-HII 
\begin{figure*}
\resizebox{1.00\hsize}{!}{\includegraphics[angle=000]{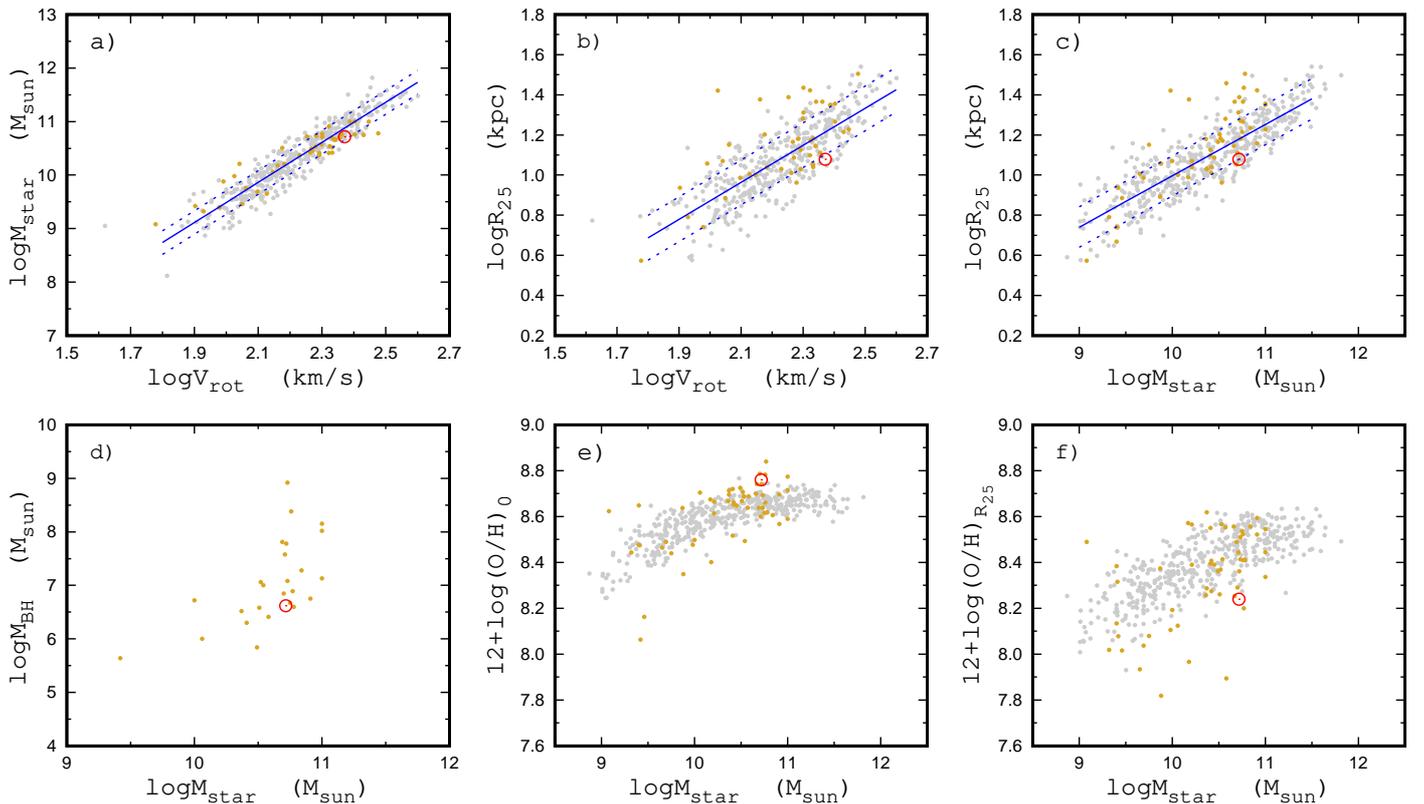}}
\caption{
  Comparison of the characteristics of the Milky Way and galaxies from our sample.
  {\sl Panel {\bf a}}:  Stellar masses $M_{\star}$ as a function of rotation velocity $V_{rot}$  (Tully-Fisher diagram).   The grey points in each panel denote
  individual MaNGA galaxies, while the goldenrod points correspond to nearby galaxies in our sample. The solid line is the linear fit to all the data, while the dotted lines
  are $\pm\sigma$ deviations.
  The red circle marks the position of the Milky Way.
  {\sl Panel {\bf b}}:  Optical radius $R_{25}$ as a function of rotation velocity $V_{rot}$. 
  {\sl Panel {\bf c}}:  Optical radius $R_{25}$ as a function of stellar mass $M_{\star}$. 
  {\sl Panel {\bf d}}:  Black hole mass $M_{BH}$ as a function of stellar mass $M_{\star}$. 
  {\sl Panel {\bf e}}:  Central oxygen abundance (O/H)$_{0}$ as a function of stellar mass $M_{\star}$.
  {\sl Panel {\bf f}}:  Oxygen abundance at the optical radius (O/H)$_{R_{25}}$ as a function of stellar mass $M_{\star}$.
}
\label{figure:sample}
\end{figure*}

Figure~\ref{figure:sample} shows the comparison between the characteristics of the Milky Way and the galaxies in our sample. In panel (a), we plot the stellar mass of the galaxy $M_{\star}$
as a function of its rotation velocity $V_{rot}$  (Tully-Fisher diagram). Since the gas mass in spiral galaxies with stellar masses similar to the Milky Way is lower than the stellar mass
by around an order of magnitude \citep{Parkash2018},  the stellar mass can also be considered in the first approximation as the representative of the baryonic mass of the galaxy.
The grey points  in the figure denote the data for individual MaNGA galaxies, while the goldenrod points correspond to nearby galaxies from our sample. The linear fit to all the  data is
given  by the expression  
\begin{equation}
\log M_{\star} = 2.003(\pm 0.152) + 3.742(\pm 0.068) \times \log V_{rot}  
\label{equation:m-v}
\end{equation}
 and is shown with the solid line in panel (a) of Fig.~\ref{figure:sample}. The mean value of the scatter in the $M_{\star}$ around the relation is $\sigma$ = 0.220 dex for the 414 points.
The red circle marks the position of the Milky Way. The deviation of the Milky Way stellar mass from the relation is --0.170 dex, or --0.77$\sigma$. 

Panel (b) of Fig.~\ref{figure:sample} shows the optical radius of the galaxy $R_{25}$ in kiloparsecs  as a function of its rotation velocity $V_{rot}$. The linear fit to all the data, 
\begin{equation}
\log R_{25} = -0.973(\pm 0.075) + 0.922(\pm 0.034) \times \log V_{rot}  
\label{equation:r-v},
\end{equation}
is shown with the solid line in panel (b) of Fig.~\ref{figure:sample}. The mean value of the scatter in the $R_{25}$ values around the relation is $\sigma$ = 0.112~dex for the 423 points.
The red circle marks the position of the Milky Way. The deviation of the Milky Way optical radius from the relation is $-0.134$~dex, or $-1.21\sigma$. 

In panel (c) of Fig.~\ref{figure:sample}, we plot the optical radius of the galaxy $R_{25}$ in kiloparsecs versus the stellar mass $M_{\star}$. The obtained $R_{25}$ -- $M_{\star}$ relation is given by
\begin{equation}
\log R_{25} = -1.561(\pm 0.074) + 0.256(\pm 0.007) \times \log M_{\star}  
\label{equation:r-m}
\end{equation}
and is shown with the solid line in panel (c) of Fig.~\ref{figure:sample}. The mean value of the scatter in the $R_{25}$ values around the relation is $\sigma$ = 0.100~dex for the 495 points.
The red circle marks the position of the Milky Way. The deviation of the Milky Way optical radius from the relation is $-0.101$~dex, or $-1.00\sigma$. 

Panel (d) of Fig.~\ref{figure:sample} shows the black hole mass $M_{BH}$ as a function of stellar mass $M_{\star}$. Unfortunately, the estimations for the central black hole mass
are available only for a small fraction of the nearby galaxies.

Panel (e) of Fig.~\ref{figure:sample} shows the  central oxygen abundance (O/H)$_{0}$ in the galaxy as a function of stellar mass $M_{\star}$. The points designate the data for individual galaxies
(grey -- MaNGA and goldenrod -- nearby galaxies) from our sample. The red circle marks the position of the Milky Way. Panel (f) of Fig.~\ref{figure:sample} shows the oxygen abundance
at the optical radius (O/H)$_{R_{25}}$  as a function of stellar mass $M_{\star}$.

%===============    Fig  No  6           Gist grad OH
\begin{figure}
\resizebox{1.00\hsize}{!}{\includegraphics[angle=000]{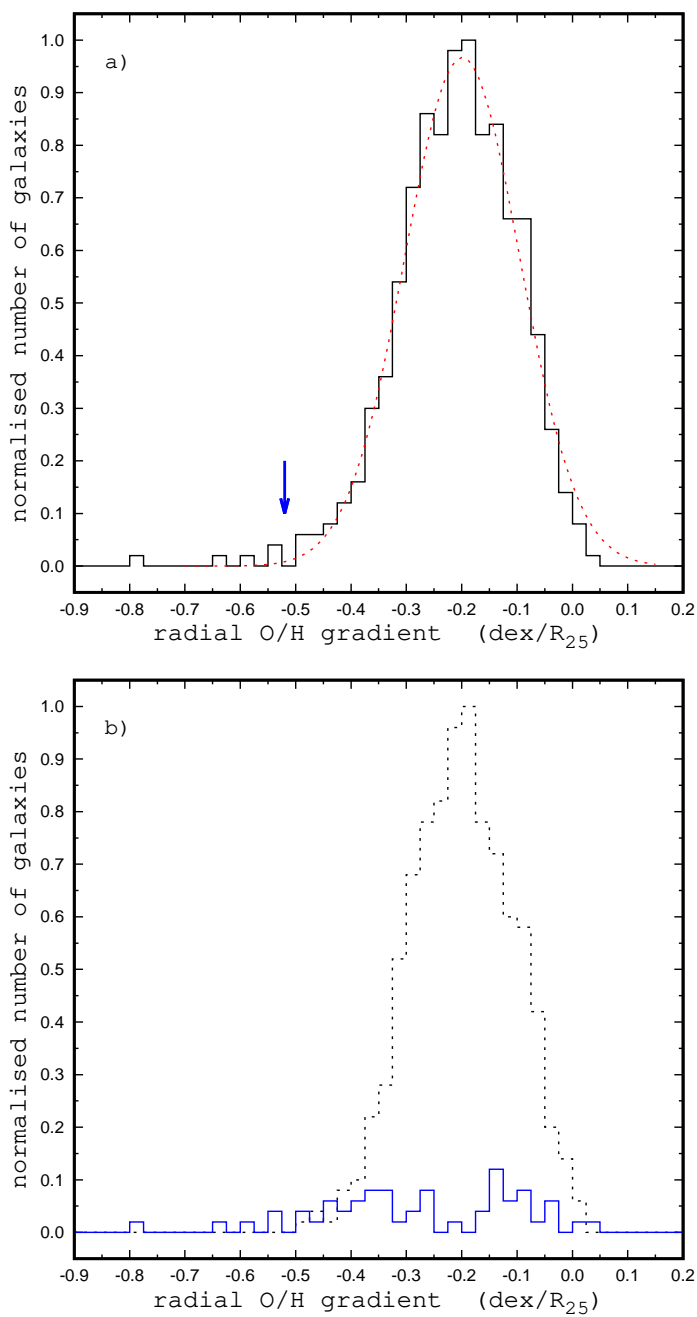}}
\caption{ Histograms of the radial oxygen abundance gradients for our sample of galaxies.  
    {\sl Panel {\bf a}:} Normalised histogram of the radial oxygen abundance gradients for our sample of galaxies (MaNGA + nearby).
    The solid line denotes the histogram of the obtained gradients, while the dashed line
    corresponds to the Gaussian fit to those data. The arrow marks the value of the radial abundance gradient of the Milky Way.  
    {\sl Panel {\bf b}:} Normalised histograms of the radial oxygen abundance gradients for subsamples of MaNGA (dotted line) and nearby (solid line) galaxies.
    The histograms were normalised to the same value  as the histogram in panel (a). 
  }
\label{figure:gist-grad}
\end{figure}

Inspection of Fig.~\ref{figure:sample} showed that the most prominent deviation of the position of the Milky Way from the general trends of the galaxies in our sample takes place
in the (O/H)$_{0}$ -- $M_{\star}$ and  (O/H)$_{R_{25}}$ -- $M_{\star}$ diagrams. The difference between oxygen abundances at the optical radius and at the centre
log(O/H)$_{R_{25}}$ -- log(O/H)$_{0}$ is, by definition, the radial abundance gradient in the galaxy expressed in units of dex/$R_{25}$. Panel (a) in
Fig.~\ref{figure:gist-grad} shows the normalised histogram of the radial oxygen abundance gradients for our sample of galaxies. The solid line denotes the histogram of the
measured gradients in bins of 0.025 dex/$R_{25}$, and the dashed line is the Gaussian fit to those data. The arrow marks the value of the radial abundance gradient
of the Milky Way. The mean value of the gradients in our sample of galaxies is --0.20 dex/$R_{25}$, and the scatter is $-0.10$~dex/$R_{25}$. \citet{Sanchez2014} examined the
radial abundance gradients in the sample of CALIFA galaxies and found that all galaxies without clear evidence of an interaction present a common gradient in the oxygen abundance
with a characteristic slope of $-0.16$~dex/$R_{25}$  and a dispersion of  0.12 dex/$R_{25}$. Thus, the distribution of the gradients for our sample of galaxies is similar to
the distribution from \citet{Sanchez2014}. Panel (a) in
Fig.~\ref{figure:gist-grad} shows  that the gradient in the Milky Way is in the far wing (or in the tail)  of the distribution of gradients.  
The deviation of the gradient in the Milky Way ($-0.5$~dex/$R_{25}$) from the mean value of the gradients for our sample of galaxies ($-0.2$~dex/$R_{25}$) is $-0.3$~dex/$R_{25}$,
which is --3$\sigma$. The ratio of the oxygen abundance at the centre to the oxygen abundance at the optical radius is (O/H)$_{0}$/(O/H)$_{R_{25}}$ $\sim$3 for the Milky Way,
which is twice as large as the mean value of ratios for our sample of galaxies. It should be noted that the  (O/H)$_{0}$/(O/H)$_{R_{25}}$ ratio in the well studied nearby
galaxy NGC~5457 (M~101) is twice as high as that in the Milky Way.
Panel (b) in Fig.~\ref{figure:gist-grad} shows the normalised histograms of the radial oxygen abundance gradients for subsamples of MaNGA (dotted line) and nearby (solid line) galaxies.
The histograms have been normalised to the same value  as the histogram in panel (a).

\subsection{Searching for Milky Way twin candidates using strict criteria}

%==================================================================================

The optical radius, stellar mass, and rotation curve specify the structure of a galaxy. These characteristics of a galaxy are related to each other. In general, any two of these
characteristics can be used to search for galaxies with a similar structure.   Rotation velocities are not available for a fraction of the galaxies in our sample. Therefore, a stellar
mass versus optical radius diagram was used as the basic diagram to compare the structure of the Milky Way and other spiral galaxies. Galaxies located close to the Milky Way on the
$M_{\star}$ -- $R_{25}$ diagram are referred to as structural Milky Way analogues (sMWAs).

The oxygen abundance at a given radius of a galaxy is defined by the evolutionary stage (the astration level) of that region and by the matter exchange with the surroundings. One
can expect that if the oxygen abundances at the centre and at the optical radius (evolutionary stages at the centre and at the optical radius) are similar in two galaxies, then those
galaxies have evolved in a more or less similar way. Then the central abundance - abundance at the optical radius diagram can be used to compare the evolution of the Milky Way and
other spiral galaxies. Galaxies located close to the Milky Way on the (O/H)$_{0}$ -- (O/H)$_{R_{25}}$ diagram are referred to as evolutionary Milky Way analogues (eMWAs).
If a galaxy is simultaneously an  sMWA and  an eMWA, then such a galaxy  can be  considered a Milky Way twin candidate. 

We defined the distance between the positions of a galaxy and the Milky Way in the $R_{25}$ --$M_{\star}$ diagram 
\begin{equation}
D_{MR} = [(\log R_{25} - \log R_{25,MW})^{2} + (\log M_{\star} - \log M_{\star,MW})^{2}]^{\frac{1}{2}}.  
\label{equation:dmr}
\end{equation}
The value of  D$_{MR}$ is some kind of  measure (index) of the  structural difference between a galaxy and the Milky Way.
We also defined the distance between the position of each galaxy and the Milky Way in the (O/H)$_{R_{25}}$ -- (O/H)$_{0}$ diagram 
\begin{equation}
D_{\rm OH} = [(Z_{0}-Z_{0,MW})^{2} + (Z_{R_{25}} - Z_{R_{25},MW})^{2}]^{\frac{1}{2}},  
\label{equation:doh}
\end{equation}
where the notation Z = 12 + log(O/H) is used for the sake of brevity.
The value of the D$_{\rm OH}$ is some kind of measure (index) of the evolutionary difference between a galaxy and the Milky Way.

%===============    Fig  No  7           eMWA  --  sMWA 
\begin{figure}
\resizebox{0.97\hsize}{!}{\includegraphics[angle=000]{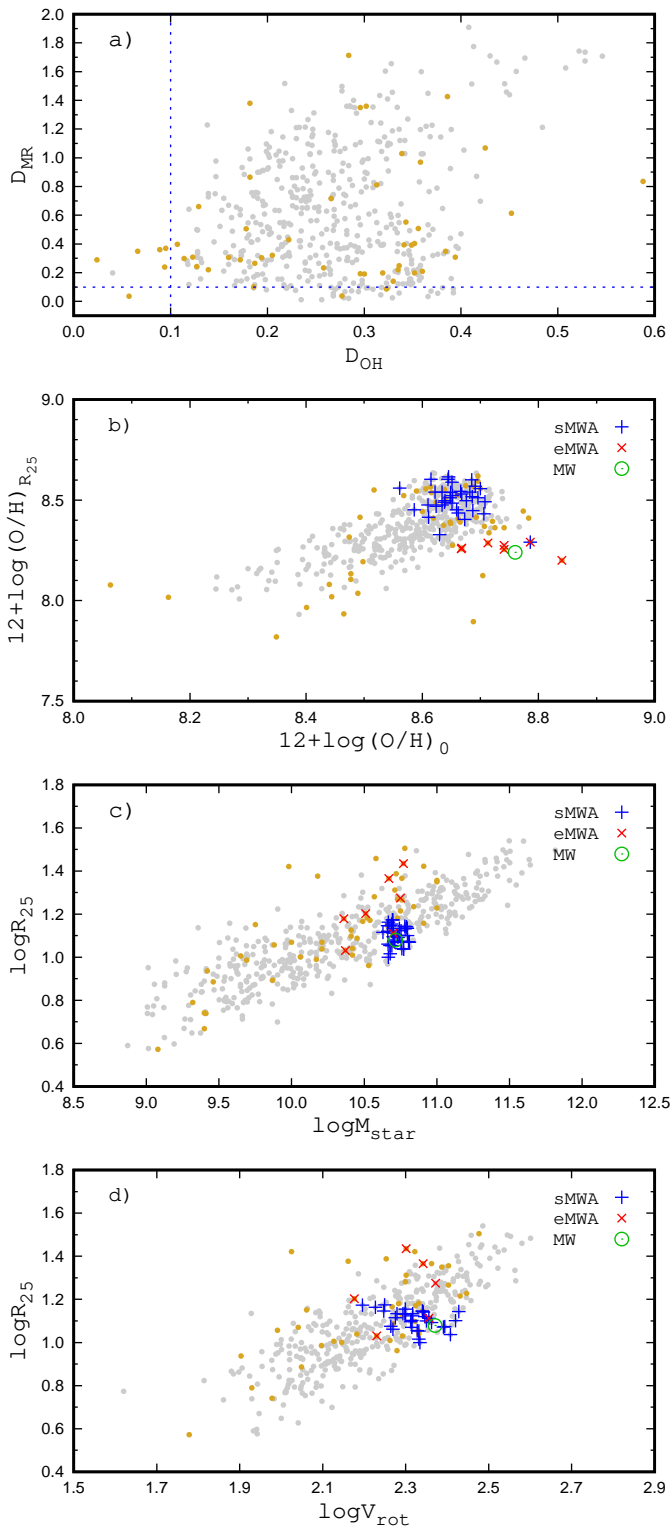}}
\caption{
   Milky Way twin candidates according to the strict criteria. 
  {\sl Panel {\bf a}:}  Index of the  structural differences between galaxies and the Milky Way D$_{MR}$ (Eq.~\ref{equation:dmr}) as a function of the index of the evolutionary
  difference D$_{\rm OH}$ (Eq.~\ref{equation:doh}). The points are individual MaNGA (grey) and nearby (goldenrod) galaxies. The dashed lines indicate D$_{MR}$ = 0.1 and D$_{\rm OH}$ = 0.1.   
  {\sl Panel {\bf b}:} Central oxygen abundance versus abundance at the optical radius. The galaxies with  D$_{\rm OH}$ $\le$ 0.1
  (eMWAs) are marked with red crosses.  The galaxies with  D$_{MR}$ $ <$ 0.1   (sMWAs) are indicated with  blue plus signs.
  The green circle denotes the Milky Way.
  {\sl Panel {\bf c}:} Optical radius versus stellar mass diagram. The notations are the same as in panel (b).
  {\sl Panel {\bf d}:} Optical radius versus rotation velocity diagram. The notations are the same as in panel (b).
}
\label{figure:emwa-smwa}
\end{figure}

We plot the index of the  structural difference between the galaxy and the Milky Way D$_{MR}$ as a function of the index of the evolutionary   difference D$_{\rm OH}$  for our sample of
galaxies in panel (a) of Fig.~\ref{figure:emwa-smwa}.  In this section, we select the Milky Way-like galaxies using a strict criteria for similarity. We adopted galaxies with
D$_{\rm OH}$ $\le$ 0.1 as  eMWAs and  galaxies with D$_{MR}$ $\le$ 0.1 as sMWAs. 
Panel (b) of Fig.~\ref{figure:emwa-smwa} shows the (O/H)$_{0}$ -- (O/H)$_{R_{25}}$ diagram. Panel (c) shows the $R_{25}$ -- $M_{\star}$ diagram, and panel (d) shows the $R_{25}$ -- $V_{rot}$
diagram.

Inspection of Fig.~\ref{figure:emwa-smwa} shows that the eMWAs are rather rare, while the sMWAs are more numerous. This implies that the Milky Way shows a rather atypical (chemical) evolution.
A prominent feature of the Milky Way is its high metallicity (oxygen abundance) at the centre of the disc, which is close to the maximum attainable oxygen abundances in galaxies,
while the oxygen abundance at the optical radius is significantly lower in comparison to other galaxies of similar central oxygen abundances.
We found that seven galaxies from our sample (six nearby galaxies, NGC~628, NGC~1232, NGC~3521, NGC~4303, NGC~6744, and IC~342, plus the MaNGA galaxy M-8934-12701) satisfy  the condition
D$_{\rm OH}$ $\le$ 0.1 and are hence considered eMWAs. The characteristics of nearby galaxies are listed in Table.~\ref{table:nearby}. The MaNGA galaxy M-8934-12701 is at a distance
of 240.8 Mpc. The characteristics of  M-8934-12701 are the following: the stellar mass is log$M_{\star}$ = 10.75; the rotation velocity is $V_{rot}$ = 235 km\,s$^{-1}$; the optical radius is
$R_{25}$ = 18.85 kpc;  the central oxygen abundance is 12 + log(O/H)$_{0}$ = 8.74; and the oxygen abundance at the optical radius is 12 + log(O/H)$_{R_{25}}$ = 8.28. 
Fig.~\ref{figure:m-8934-12701} shows the maps and radial distributions of the properties in the MaNGA galaxy M-8934-12701. 

At the same time, we found that 35 galaxies from our sample satisfy  the condition D$_{RM}$ $\le$ 0.1. Close examination of Fig.~\ref{figure:emwa-smwa} showed that six out of seven
eMWAs are not sMWAs. The values of the stellar mass and rotation velocity in three eMWAs are relatively close to that
in the Milky Way, but their optical radii are significantly larger in comparison to the optical radius of the Milky Way ($R_{25}$ = 23.18 kpc in NGC~1232, $R_{25}$ = 27.25 kpc in MGC~6744, and
$R_{25}$ = 18.85 in the MaNGA galaxy M-8934-12701 versus  $R_{25}$ = 12.0 in the Milky Way). Three other eMWAs have noticeably  lower values of stellar mass (log$M_{\star}$ =10.36 in NGC~628,
log$M_{\star}$ =10.51 in NGC~4303, and log$M_{\star}$ =10.37 in IC~342 versus log$M_{\star}$ =10.72 in the Milky Way) and rotation velocity ($V_{rot}$ = 150 km\,s$^{-1}$ in NGC~4303 and
$V_{rot}$ = 170 km\,s$^{-1}$ in IC~342 versus $V_{rot}$ = 235 km\,s$^{-1}$ in the Milky Way). We note that in previous studies,  Milky Way-like galaxies were selected using the
structural and morphological characteristics of  galaxies, that is, sMWAs were selected. We found that the abundances at the centre and at the optical radius (evolutionary characteristics)
provide a more strict criterion for selecting real Milky Way twins.

We found that only the galaxy NGC~3521 is simultaneously both an sMWA and an eMWA. All the characteristics considered (stellar mass, optical radius, rotation velocity, central
oxygen abundance, and abundance at the optical radius) of  NGC~3521 are close to that of the Milky Way. The masses of black holes in those galaxies are also close to each other.
Thus, the galaxy NGC~3521 can be a real Milky Way twin candidate. A detailed comparison of the galaxy NGC~3521 with the Milky Way is given in Section 4.4.

\subsection{Searching for Milky Way twin candidates using weak criteria}

%==================================================================================

%===============    Fig  No 8            eMWA with a weak criteria 
\begin{figure}
\resizebox{1.00\hsize}{!}{\includegraphics[angle=000]{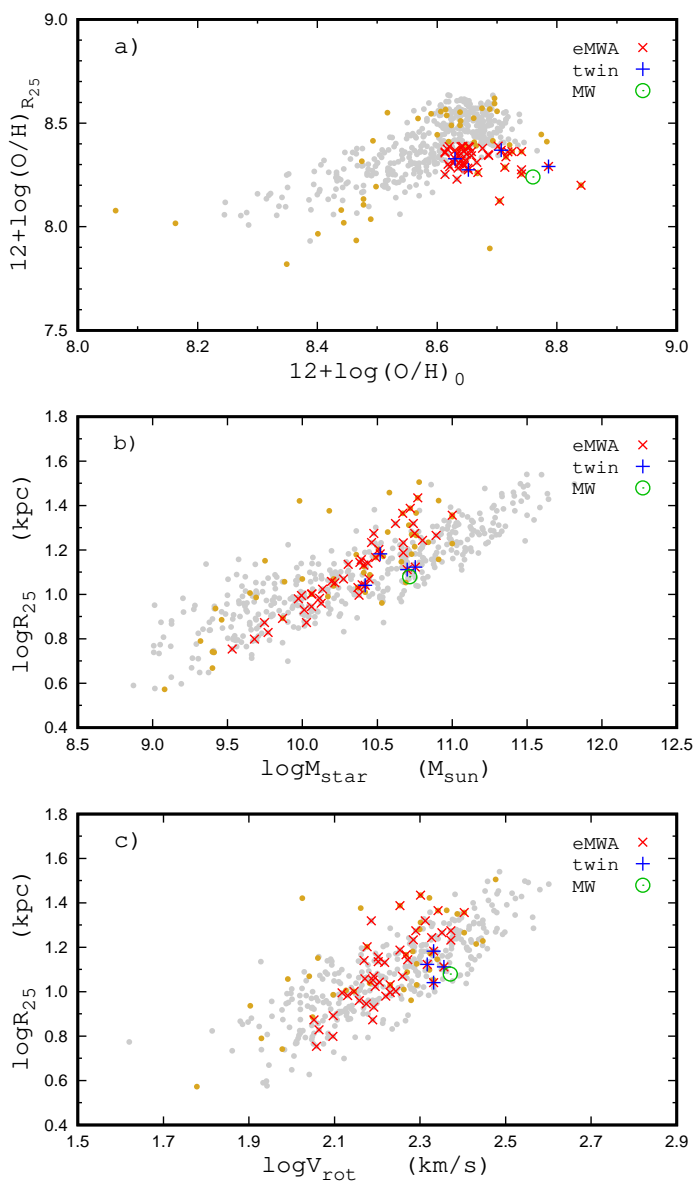}}
\caption{
    Milky Way twin candidates according to the weak criteria. 
  {\sl Panel {\bf a}:} Central oxygen abundance versus abundance at the optical radius diagram. The grey points are individual MaNGA galaxies, and the goldenrod pints denote the
  nearby galaxies. The galaxies with  $\Delta$(O/H)$_{0} \le$ 0.1~dex 
  and $\Delta$(O/H)$_{R_{25}} \le$ 0.1 dex (eMWAs) are marked with the red crosses.  The eMWAs with   $\Delta R_{25} \le$ 0.114~dex, 
  $\Delta M_{\star} \le$ 0.3~dex, and $\Delta V_{rot} \le$ 0.079 dex  (Milky Way twin candidates) are marked with the blue   plus signs. The green circle denotes the Milky Way.
  {\sl Panel {\bf b}:} Optical radius versus stellar mass diagram. The notations are the same as in panel (a).
  {\sl Panel {\bf c}:} Optical radius versus rotation velocity diagram. The notations are the same as in panel (a).
}
\label{figure:emwa-extend}
\end{figure}

\citet{Boardman2020a} noted that the selection of Milky Way analogues using a larger number of simultaneous selection parameters, as well as more stringent constraints on a given
parameter, yields a sample of Milky Way-like galaxies with properties that are closer to the true properties of the Milky Way. At the same time, the use of a larger number of the
selection criteria (or an overly strict definition of the “analogue”) yields few to no Milky Way-like galaxies \citep{Boardman2020b}. The fact that we found only one Milky Way
candidate twin in the  previous section can be the result of an overly strict definition of a Milky Way twin. In this section we search for  Milky Way twins using a more weak
(less strict) criteria.

The difference between the determined values of the parameter for two galaxies does not necessarily coincide with the real  difference of the parameter for those galaxies, but it
can also involve the uncertainty in the parameter determinations. Therefore, in the search for Milky Way twins, the uncertainty in the parameter determinations should be taken into
account in the choice of the allowed  difference. Unfortunately, the characteristics of many nearby galaxies can involve significant uncertainties, as noted in many papers. For instance,
\citet{McQuinn2017} noted ``surprisingly, many of the best-studied spiral galaxies in the Local Volume have distance uncertainties that are much larger than can be achieved with
modern observation techniques.'' The uncertainty in the distance to the galaxies results in  uncertainties in the optical radius in kiloparsecs, as well as in the luminosity
and consequently in the stellar mass. The uncertainty in the luminosity to stellar mass conversion also makes a significant (maybe dominant) contribution to the  uncertainty in the
estimation of the stellar mass of the galaxy.

We noted previously that for this current study, we selected our sample of  MaNGA galaxies where radial oxygen abundance distribution could be approximated satisfactorily by a
single linear relation. \citet{Belfiore2017}  found that the galaxy inclination and the point spread function can generate a flattening in the radial abundance gradient in the MaNGA
galaxies, since flux from different galactocentric radii is summed up when the galaxy is projected in the plane of the sky. The metallicity depletion at the galaxy centre depends on
the PSF and galaxy inclination, and it can be as large as $\sim$0.04~dex. 

It is difficult to specify the rotation of a galaxy with a single parameter because  there is a wide variety of  shapes in the rotation curves of galaxies. The rotation velocity
of the flat part of the rotation curve V$_{flat}$ is often used as the characteristic value for the galaxy rotation. However, there is not a commonly accepted way to  determine V$_{flat}$.
Even for galaxies with well measured rotation curves, the estimations of  V$_{flat}$ in  different publications can differ by around 20~km$s^{-1}$. For example, there are several measurements
for the rotation curve of NGC~5055  \citep{Thornley1997, BlaisOuellette2004, deBlok2008}. The estimations of the V$_{flat}$ for NGC~5055 are V$_{flat}$ = 192~km\,s$^{-1}$ \citep{Leroy2008},
V$_{flat}$ = 197~km\,s$^{-1}$ \citep{Frank2016}, and V$_{flat}$ = 179 km\,s$^{-1}$ \citep{Lelli2016}.

In this section, we search for  Milky Way twins using the following (weak) selection criteria: 
1) The absolute value of the difference between the oxygen abundance at the centre of the galaxy and that for the Milky Way should be lower than 0.15~dex,
   $\Delta$(O/H)$_{0}$ (= abs(log(O/H)$_{0}$ -- log(O/H)$_{0,MW}$))    $\le$ 0.15.  % \\ 
2) The absolute value of the difference between the oxygen abundance at the optical radius  of the galaxy and that for the Milky Way should be lower than 0.15 dex,
    $\Delta$(O/H)$_{R_{25}}$ (= abs(log(O/H)$_{R_{25}}$ --  log(O/H)$_{R_{25},MW}$))  $\le$ 0.15. % \\ 
3) The optical radii of the galaxy and the Milky Way should agree within $\sim$30\%, $\Delta R_{25}$ (= abs(log$R_{25}$ -- log$R_{25,MW}$)) $\le$ 0.114. % \\
4) The difference between the stellar mass of the galaxy and that of the Milky Way should be  less than a factor of approximately two, $\Delta M_{\star}$
(= abs(log$M_{\star}$ -- log$M_{\star,MW}$)) $\le$ 0.3. % \\
5) The rotation velocities of the galaxy and that of the Milky Way agree within $\sim$20\%,  $\Delta V_{rot}$ (= abs(log$V_{rot}$ -- log$V_{rot,MW}$)) $\le$ 0.079. %\\
The allowed differences of the parameters (criteria) that we adopted in the search for  Milky Way twins are somewhat arbitrary. 

All five parameters are available for 414 galaxies from our sample (the rotation curve is not determined for low-inclination galaxies). We found that 46 galaxies satisfy
criteria 1 and 2. Those galaxies (eMWAs) are shown in Fig.~\ref{figure:emwa-extend} with  red crosses. At the same time, only four out of 46 eMWAs (NGC~3521, NGC~4651, NGC~2903, and
MaNGA galaxy M-8341-09101) satisfy   criteria 3 through 5. Those galaxies (Milky Way twin candidates) are shown in Fig.~\ref{figure:emwa-extend} with  blue plus signs.

A comparison between  Fig.~\ref{figure:emwa-smwa} and Fig.~\ref{figure:emwa-extend} showed that the general behaviour of a sample of eMWAs selected with the weak criteria is similar
to that of the sample of eMWAs selected with the strict criteria. Only a small fraction of the weak eMWAs (4 out of 46) are  sMWAs. For the majority of  eMWAs, either the optical
radii is significantly larger in comparison to the optical radius of the Milky Way or  the stellar mass (and/or rotation velocity) is significantly lower than  that of the Milky Way.

\subsection{Abundance distributions and rotation curves of the Milky Way twin candidates}
%=====================

In this section, we compare the radial abundance distributions and rotation curves of the Milky Way twin candidates (NGC~3521, NGC~4651, NGC~2903, and MaNGA galaxy M-8341-09101)
with that of the Milky Way. \citet{Zhou2023}  found that the MaNGA galaxy M-8983-03703 is a Milky Way-like galaxy in the context of their star formation and chemical
evolution histories. Hence, this galaxy is also considered here. 

\subsubsection{NGC~3521}
% ========================

%===============    Fig  No 9            NGC 3521  vs  MW 
\begin{figure}
\resizebox{1.00\hsize}{!}{\includegraphics[angle=000]{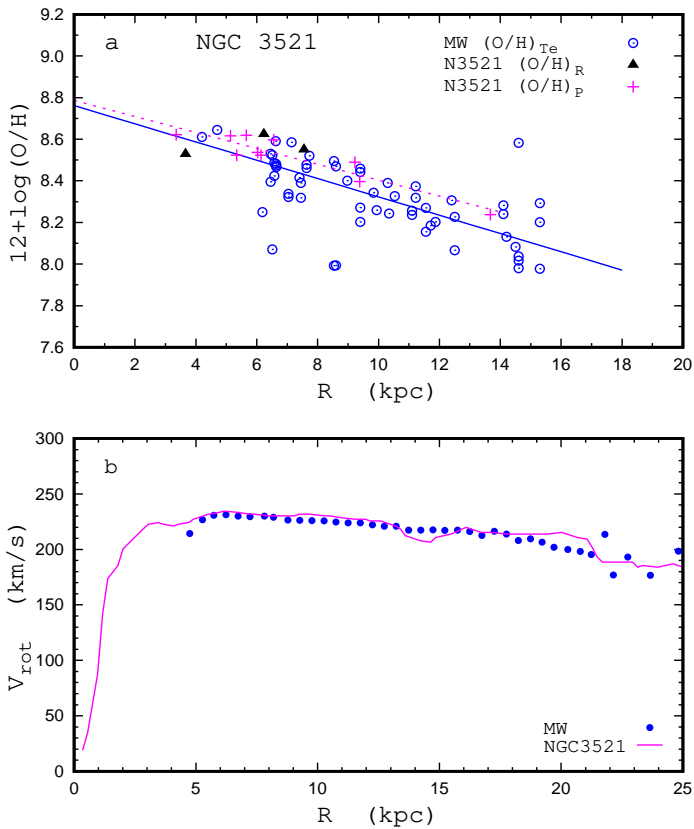}}
\caption{Comparison between the properties of  NGC~3521 and the Milky Way.
  {\sl Panel {\bf a}:} Comparison between the radial oxygen abundance distributions. The blue circles denote the $T_{e}$-based abundances in individual H\,{\sc ii} regions in the Milky Way
  (comes from panel (b) of Fig.~\ref{figure:r-ohte}), and the solid line is the best fit to those data.  The dark triangles mark the $R$ calibration-based abundances in H\,{\sc ii} regions
  in NGC~3521. The crimson plus signs are the $P$ calibration-based abundances in H\,{\sc ii} regions in NGC~3521, and the dotted line is the linear fit to those data points. 
  {\sl Panel {\bf b}:} Comparison between rotation curves. The points denote the rotation curve of the Milky Way from  \citet{Eilers2019}.   The line shows the rotation curve of the
  NGC~3521 from \citet{deBlok2008}.
}
\label{figure:ngc3521-r-oh}
\end{figure}

The galaxy NGC~3521 is an SABb galaxy  (morphological type code $T$ = 4.0$\pm$0.2).  It is included in The  H\,{\sc i} Nearby Galaxy Survey (THINGS)  list \citep{Walter2008}. We adopted
the following parameters of NGC~3521 used or obtained by the THINGS survey: distance $d$ = 10.7 Mpc;\ optical radius of 4.16 arcmin; physical optical radius
$R_{25}$ = 12.94 kpc \citep{Walter2008}; inclination angle $i$ = 73$\degr$;  position angle of the major axis PA = 340$\degr$ \citep{deBlok2008}; stellar mass
$M_{\star}$ = 5.01 $\times$ 10$^{10}$ M$_{\sun}$, or log($M_{\star}/M_{\sun}$) = 10.70 \citep{Leroy2008}; and  rotation velocity $V_{rot}$ = 227~km\,s$^{-1}$ \citep{deBlok2008,Leroy2008}.   
The black hole  mass in  NGC~3521 is log($M_{BH}/M_{\sun}$) = 6.85$\pm$0.58 \citep{Davis2014}.

The measurements of the emission lines, which are necessary for the determination of the abundance through the $R$ calibration, are available for only three H\,{\sc ii} regions in
the NGC~3521 \citep{Bresolin1999}. However,  NGC~3521 was taken into consideration for two reasons: First, \citet{McGaugh2016} noted that NGC~3521
is the closest structural analogue to the Milky Way, having a similar luminosity, scale length, and rotation curve. Second, the measurements of the emission lines $R_{2}$ and $R_{3}$,
which allow for estimation of
the abundance through the $P$ calibration, are available for ten H\,{\sc ii} regions in  NGC~3521 \citep{Zaritsky1994}.
The oxygen abundances determined through the $R$ calibration in three  H\,{\sc ii} regions from \citet{Bresolin1999} are denoted with the triangles in panel (a) of Fig.~\ref{figure:ngc3521-r-oh}. 
In order to estimate the oxygen abundances using the oxygen lines $R_{2}$ and $R_{3}$ measured by  \citet{Zaritsky1994} in ten H\,{\sc ii} regions in NGC~3521, we obtained the
calibration relation (O/H)$_{P}$ = $f$($R_{2},R_{3}$) for the upper branch \citep{Pilyugin2001,Pilyugin2005} using the calibration data points from \citet{Pilyugin2016} and supplemented by the
recent measurements of H\,{\sc ii} regions in NGC~5457 from \citet{Croxall2016} and \citet{Esteban2020}
\begin{eqnarray}
       \begin{array}{lll}
        12+\log(O/H)_P   & =  &  8.6059 + 0.4143\,P                \\
                         & +  & (0.2662 - 1.2014\,P)\,X            \\
                         & -  & (0.6786 - 0.8121\,P)\,X^2   \\
     \end{array}
\label{equation:ohph},
\end{eqnarray}
where $P$ = $R_{3}/R_{23}$ is the excitation parameter, $R_{23}$ = $R_{3}$ + $R_{3}$, and $X$ = log(1 + $R_{23}$). The oxygen abundances determined through the $P$ calibration in H\,{\sc ii}
regions from \citet{Zaritsky1994} are shown with plus signs in panel (a) of Fig.~\ref{figure:ngc3521-r-oh}. The radial distribution of those oxygen abundances was approximated by the
relation
\begin{equation}
12+\log(\rm O/H) = 8.786(\pm 0.041) - 0.494(\pm 0.069) \times R_{g}   
\label{equation:n3521-oh-compil}
\end{equation}
with a scatter of 0.042 dex for the 10 data points.

Thus, the values of the optical radii of  NGC~3521 ($R_{25}$ = 12.9~kpc) and the Milky Way  ($R_{25}$ = 12.0~kpc) are close to each other. The value of the stellar mass of the
NGC~3521 (log($M_{\star}/M_{\sun}$) = 10.7) is close to the value of the stellar mass of the Milky Way  (log($M_{\star}/M_{\sun}$) = 10.716). The mass of the black hole at the centre of NGC~3521
(log($M_{BH}/M_{\sun}$) = 6.85) is close to that in the Milky Way (log($M_{BH}/M_{\sun}$) = 6.62). The radial distributions of the oxygen abundances (panel (a) of
Fig.~\ref{figure:ngc3521-r-oh}) and rotation curves (panel (b) of Fig.~\ref{figure:ngc3521-r-oh}) in those galaxies agree satisfactorily with each other. However, in order to make a
solid conclusion as to whether the galaxy NGC~3521 is a twin of the Milky Way, the radial distribution of the $R$ calibration-based abundances in  NGC~3521 should be established,
that is, the measurements of the spectra including emission lines necessary for abundance determinations through the $R$ calibration should be carried out. The distance to
NGC~3521 and its optical radius should also be more precise.

\subsubsection{NGC~4651}
%===========================

%===============    Fig  No  10        NGC 4651  vs  MW 
\begin{figure}
\resizebox{1.00\hsize}{!}{\includegraphics[angle=000]{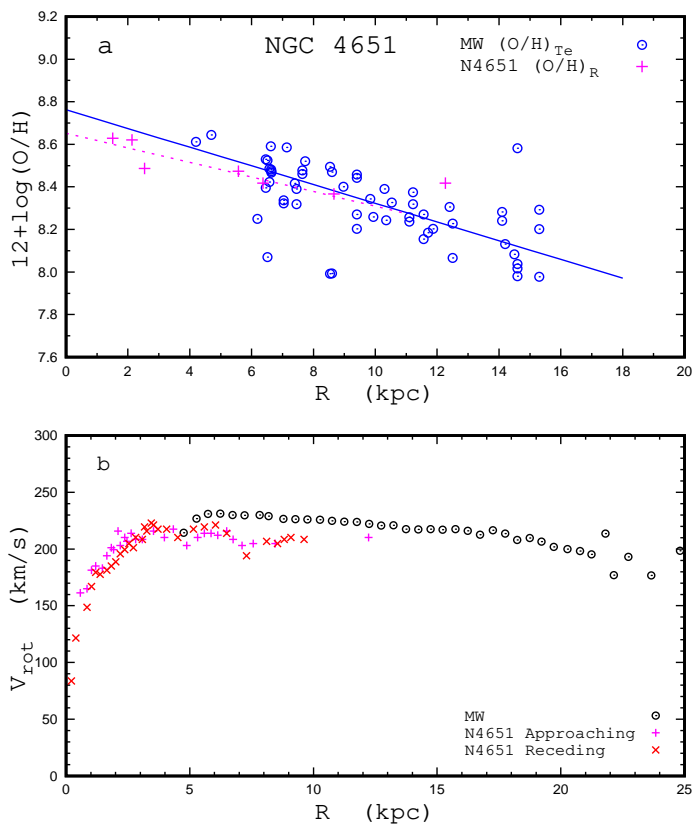}}
\caption{Comparison of properties between NGC~4651 and the Milky Way.
  {\sl Panel {\bf a}:} Comparison between radial oxygen abundance distributions. The blue circles denote the T$_{e}$-based abundances in individual H\,{\sc ii} regions in the Milky Way
  (comes from panel (b) of Fig.~\ref{figure:r-ohte}), and the solid line is the best fit to those data.   The crimson plus signs are the $R$ calibration-based abundances in H\,{\sc ii} regions
  in NGC~4651, and the dotted line is the linear fit to those data points. 
  {\sl Panel {\bf b}:} Comparison between rotation curves.   The points denote the rotation curve of the Milky Way from  \citet{Eilers2019}.   The rotation curve of  NGC~4651
  from \citet{Epinat2008} is shown with plus signs (approaching side) and crosses (receding side).
}
\label{figure:ngc4651-r-oh}
\end{figure}

The galaxy NGC~4651 is an Sc galaxy  (morphological type code $T$ = 5.1$\pm$0.6). The inclination angle of  NGC~4651 is $i$ = 53$\degr$, and the position angle of the major axis is PA = 77$\degr$
\citep{Epinat2008}. The optical radius is 1.99 arcmin \citep{RC3}. At the distance  of $d$ = 19.0 Mpc \citep{Foster2014}, the physical optical radius is $R_{25}$ = 11.00 kpc.
The stellar mass  is $M_{\star}$ = 2.61 $\times$ 10$^{10}$ $M_{\sun}$, or log($M_{\star}/M_{\sun}$) = 10.42 \citep{Leroy2019} (or $M_{\star}$ = 1.7 $\times$ 10$^{10}$ $M_{\sun}$ or
log($M_{\star}/M_{\sun}$) = 10.23 \citep{Foster2014}).  The rotation velocity of  NGC~4651 is 215~km\,s$^{-1}$ \citep{Epinat2008}.   

The radial distribution of the oxygen abundances estimated through the $R$ calibration in H\,{\sc ii} regions from \citet{Skillman1996} was approximated by the relation
\begin{equation}
12+\log(\rm O/H) = 8.652(\pm 0.039) - 0.377(\pm 0.083) \times R_{g}   
\label{equation:n4651-oh-compil}
\end{equation}
with a scatter of 0.039~dex for the six data points   (panel (a) in Fig.~\ref{figure:ngc4651-r-oh}). 

The values of the optical radii of  NGC~4651 ($R_{25}$ = 11.0 kpc) and the Milky Way  ($R_{25}$ = 12.0~kpc) are close to each other. The radial distributions of the oxygen abundances
(panel (a) of Fig.~\ref{figure:ngc4651-r-oh}) and the rotation curves (panel (b) of Fig.~\ref{figure:ngc4651-r-oh}) in those galaxies agree satisfactorily to each other.
While the rotation velocity of  NGC~4651 is rather close to that of the Milky Way, the value of the stellar mass of  NGC~4651 (log($M_{\star}/M_{\sun}$) = 10.42) is appreciably lower
(by a factor of approximately two, or even more) than the stellar mass of the Milky Way (log($M_{\star}/M_{\sun}$) = 10.716). If the estimations of the stellar mass of NGC~4651 are correct, then 
NGC~4651 can be considered a low(-stellar) mass evolutionary analogue of the Milky Way.

\subsubsection{NGC~2903}
%===========================

%===============    Fig  No  11         NGC 2903  vs  MW 
\begin{figure}
\resizebox{1.00\hsize}{!}{\includegraphics[angle=000]{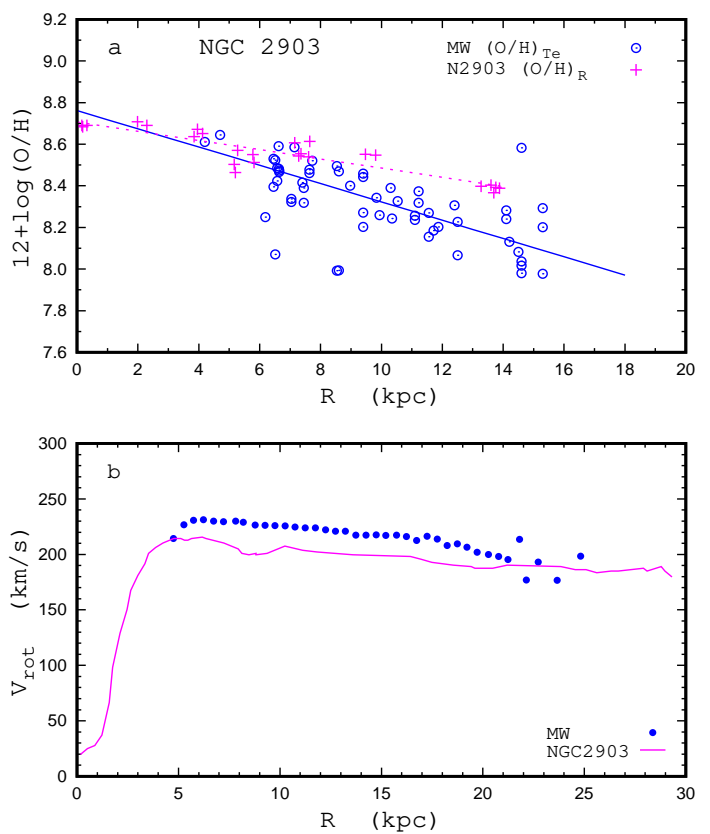}}
\caption{Comparison of properties between NGC~2903 and the Milky Way.
  {\sl Panel {\bf a}:} Comparison between radial oxygen abundance distributions.   The blue circles denote the T$_{e}$-based abundances in individual H\,{\sc ii} regions in the Milky Way
  (comes from panel (b) of Fig.~\ref{figure:r-ohte}), and the solid line is the best fit to those data.   The crimson plus signs are the $R$ calibration-based abundances in H\,{\sc ii} regions
  of NGC~2903, and the dotted line is the linear fit to those data points. 
  {\sl Panel {\bf b}:} Comparison between rotation curves.  The points denote the rotation curve of the Milky Way from  \citet{Eilers2019}.   The rotation curve of  NGC~2903
  from \citet{deBlok2008} is shown with the line.
}
\label{figure:ngc2903-r-oh}
\end{figure}

The galaxy NGC~2903 is an Sbc spiral galaxy (morphological type code $T$ = 4.0$\pm$0.1). Its inclination angle is $i$ = 65$\degr$, and the position angle of the major axis is PA = 204$\degr$
\citep{deBlok2008}. The optical radius of  NGC~2903 is $R_{25}$ = 5.87 arcmin \citep{Walter2008}. The distance to  NGC~2903 is $d$ = 8.9~Mpc \citep{Drozdovsky2000}, and it  results
in a physical optical radius of $R_{25}$ = 15.21~kpc. The stellar mass of  NGC~2903 based on the mean value from \citet{Jarrett2019} and \citet{Leroy2021} and rescaled to the adopted distance
is  $M_{\star}$ = 3.33 $\times$ 10$^{10}$ $M_{\sun}$, or log($M_{\star}/M_{\sun}$) = 10.52. The mass of the black hole in  NGC~2903 is log($M_{BH}/M_{\sun}$) = 7.06$^{+0.28}_{-7.06}$
\citep{vandenbosch2016}. The rotation velocity of  NGC~2903 is 215 km\,s$^{-1}$ \citep{deBlok2008}. 

The abundance gradient traced by the $R$ calibration-based abundances in the H\,{\sc ii} regions from the compilation in \citet{Pilyugin2014} is 
\begin{equation}
12+\log(\rm O/H) = 8.707(\pm 0.018) - 0.338(\pm 0.035) \times R_{g}   
\label{equation:n2903-oh-compil}
\end{equation}
with a scatter of 0.047 dex for the 25 data points.

The values of the optical radii of  NGC~2903 ($R_{25}$ = 15.21 kpc) and the Milky Way  ($R_{25}$ = 12.0~kpc) are close to each other, and the rotation curves (panel (b) of
Fig.~\ref{figure:ngc2903-r-oh}) in those galaxies agree satisfactorily to each other. However, the value of the stellar mass of  NGC~2903 (log($M_{\star}/M_{\sun}$) = 10.52)
is appreciably lower than the stellar mass of the Milky Way (log($M_{\star}/M_{\sun}$) = 10.716). There is also an appreciable difference between the radial distributions of
the oxygen abundances (panel (a) of Fig.~\ref{figure:ngc2903-r-oh}) in the galaxies. Thus,  NGC~2903 is a Milky Way twin only to some extent.

\subsubsection{MaNGA galaxy M-8341-09101}
%===========================

%===============    Fig  No  12           M-8341-09101  vs  MW 
\begin{figure}
\resizebox{1.00\hsize}{!}{\includegraphics[angle=000]{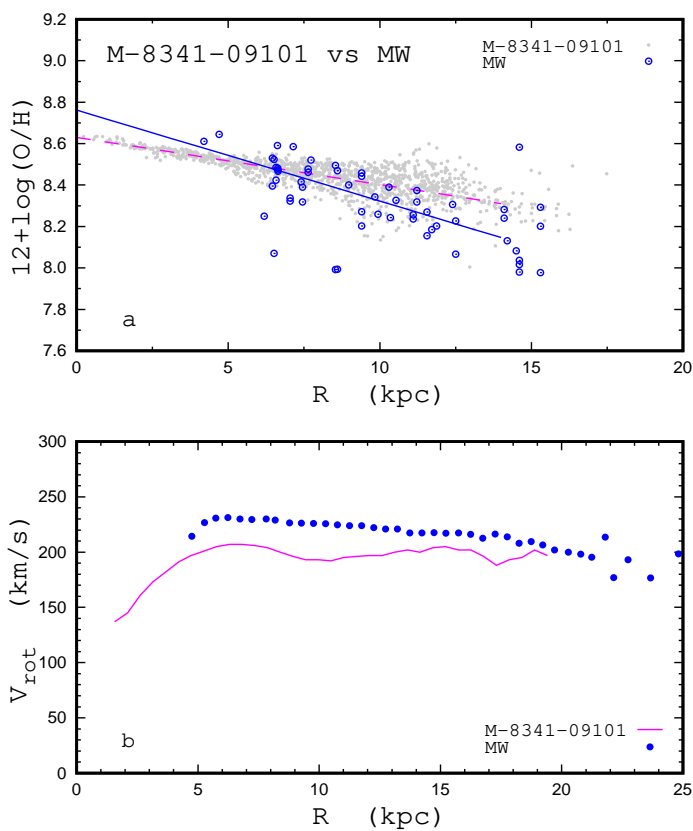}}
\caption{Comparison between the properties of the MaNGA galaxy M-8341-09101 and the Milky Way.
  {\sl Panel {\bf a}:} Comparison between the radial oxygen abundance distributions. The blue circles denote the T$_{e}$-based abundances in individual H\,{\sc ii} regions in the MW
  (comes from panel (b) of Fig.~\ref{figure:r-ohte}) and the solid line is the best fit to those data.   The grey points are the $R$ calibration-based abundances in  the spaxels of M-8341-09101
  and the dashed line is the linear fit to those data points. 
  {\sl Panel {\bf b}:} Comparison between rotation curves. The points denote the rotation curve of the Milky Way from  \citet{Eilers2019}. The rotation curve of  M-8341-09101
  is shown by the line.
}
\label{figure:M-8341-09101-r-oh}
\end{figure}

The distance to the MaNGA galaxy M-8341-09101 is 216.3~Mpc. The central oxygen abundance in  M-8341-09101 is 12 + log(O/H)$_{0}$ = 8.63, and the oxygen abundance at the optical radius
is 12 + log(O/H)$_{R_{25}}$ = 8.33. The comparison of the radial distribution of the oxygen abundances in the galaxy M-8341-09101 and in the Milky Way is shown
in panel (a) of Fig.~\ref{figure:M-8341-09101-r-oh}. One can see that the difference between the radial distribution of the oxygen abundances in the two galaxies is significant. 

The optical radius of  M-8341-09101 ($R_{25}$ = 13.3~kpc) is close to the radius of the Milky Way ($R_{25}$ = 12.0~kpc). The stellar mass of  M-8341-09101 (log($M_{\star}/M_{\sun}$) =
10.75) is close to that of the Milky Way  (log($M_{\star}/M_{\sun}$) = 10.716). The comparison between the rotation curves in  M-8341-09101 and  the Milky Way is shown in panel (b) of
Fig.~\ref{figure:M-8341-09101-r-oh}. Comparison between  Fig.~\ref{figure:M-8341-09101-r-oh} and Fig.~\ref{figure:ngc2903-r-oh} suggested that the properties (rotation curve and the
radial abundance distribution) of the galaxy M-8341-09101 and  NGC~2903 are close to each other, that is, the galaxy M-8341-09101 is more similar to NGC~2903 than to the Milky Way.

\subsubsection{MaNGA galaxy M-8983-03703}
%===========================

%===============    Fig  No  13           OH-R    M-8983-03703
\begin{figure}
\resizebox{1.00\hsize}{!}{\includegraphics[angle=000]{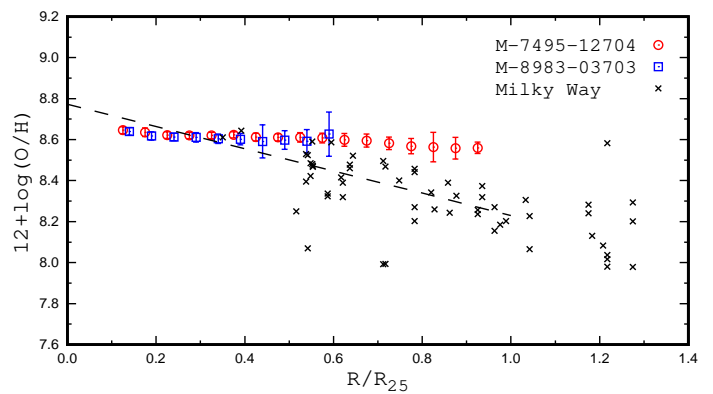}}
\caption{Comparison  between the radial oxygen abundance distributions in the MaNGA galaxy M-8983-03703 and the Milky Way.
  The blue squares are the median values of  O/H in bins of 0.05 in the fractional radius $R/R_{25}$ of M-8983-03703, and the bars show the scatter in  O/H around the median
  values of the bins. The red circles are abundances estimated from other MaNGA measurements (M-7495-12704) of this  galaxy. The median values of the oxygen abundances for different
  measurements were obtained for the same bin, but the positions of the symbols (circles and squares) were shifted slightly along the x-axis for the sake of clarity.
  The black crosses denote the $T_{e}$-based abundances of individual H\,{\sc ii} regions in the Milky Way, while the solid line is the best fit to those data within the optical radius.  
}
\label{figure:M-8983-03703-r-oh}
\end{figure}

\citet{Zhou2023}  found that the MaNGA galaxy M-8983-03703 is a Milky Way-like galaxy in the context of their star formation and chemical
evolution histories. Fig.~\ref{figure:M-8983-03703-r-oh} shows the comparison  between the radial oxygen abundance distributions in the MaNGA galaxy M-8983-03703 and the Milky Way.
The field-of-view of  M-8983-03703 measurement covers the central part of the galaxy only. There is another MaNGA measurement (M-7495-12704) of this  galaxy that covers
a much larger fraction of the galaxy. The radial oxygen abundance distribution obtained from  M-7495-12704 is also shown in  Fig.~\ref{figure:M-8983-03703-r-oh}.
One can see that the difference between the radial distributions of the oxygen abundances in  M-8983-03703 (and  M-7495-12704) and in the Milky Way is very large. Hence  M-8983-03703
cannot be considered as Milky Way twin candidate.

\section{Discussion}
%=====================

We found that the position of the Milky Way in the central abundance versus the abundance at optical radius diagram, Fig.~\ref{figure:emwa-smwa}, shows a significant shift from 
the general trend. A prominent feature of the Milky Way is a high metallicity (oxygen abundance) at the centre of the disc, which is close to the maximum attainable oxygen abundance
in galaxies, while the oxygen abundance at the optical radius is significantly lower in comparison to  galaxies of similar central oxygen abundances, that is, the radial abundance
gradient (in dex/$R_{25}$) in the Milky Way is significantly steeper than in other galaxies of similar (high) central metallicity. This evidences in favour of a rather atypical
(chemical) evolution of the Milky Way. In this section, we discuss a possible reason for the atypical radial abundance gradient in the Milky Way. 

The chemical evolution of a given region of a galaxy is defined by the star formation history and mass exchange with its surroundings. In the standard $\Lambda$ cold dark matter
($\Lambda$CDM) scenario,  successive mergings are considered one of the main mechanisms for assembling mass in galaxies \citep{WhiteRees1978,Blumenthal1984}. Galaxy-galaxy interactions
and mergers can have a significant effect on galaxy evolution, altering different characteristics of a galaxy. In particular, the gas inflows are predicted to lead to a redistribution
of metals, reducing the central gas-phase metallicity and producing an overall flatter abundance gradient \citep{Rupke2010a,Sillero2017,Bustamante2018}. These predictions are in
agreement with observational data. Measurements of central metallicities of interacting galaxies and galaxy pairs confirm that they have lower central metallicities than isolated
galaxies of similar stellar mass \citep[e.g.][]{Rupke2008,Ellison2008,Garduno2021}. However, \citet{BarreraBallesteros2015} found central oxygen abundances in a sample of interacting
galaxies similar to those in a control sample. It has been found \citep{Rupke2010b,Kewley2010,Rosa2014,Croxall2015,TorresFlores2020} that merging and interacting systems exhibit
shallow oxygen abundance gradients compared to isolated spiral galaxies.  \citet{Boardman2022,Boardman2023} concluded that galaxies that have experienced smooth gas accretion histories
produce negative metallicity gradients over time and that the increased merging activity disrupts this process, leading to flatter metallicity gradients. One could suggest that the
lack of  mergers and interactions is a necessary condition  for a  galaxy to have a high abundance at the centre and a steep radial abundance gradient. In such a case, the high central
metallicity and steep gradient in the Milky Way would be evidence that the Milky Way has evolved without mergers and interactions.

Empirical investigations of the merger history of the Milky Way have been carried out in the past decades. The Sagittarius \citep{Ibata1994} and Cetus mergers \citep{Newberg2009}
were discovered. Large stellar surveys, such as Gaia \citep{GaiaCollaboration2016,GaiaCollaboration2018,GaiaCollaboration2021} and the Apache Point Observatory Galactic Evolution
Experiment, which is one of the programmes in the SDSS \citep{Majewski2017,Abdurrouf2022}, have provided homogeneous astrometric, photometric, and spectroscopic data sampling for
a large amount of stars in the Milky Way. This has allowed for the detection and characterisation of substructures in the Milky Way that are remnants of the progenitor galaxies
that merged with the Milky Way and contributed to its stellar population. Hence, it has become increasingly possible to investigate the merging history of the Milky Way in great
detail. \citet{Belokurov2018} and  \citet{Helmi2018} argue that the inner halo is dominated by debris from a major accretion event that occurred between 8 and 11 Gyr ago. 
A captured galaxy was slightly more massive than the Small Magellanic Cloud. This object is referred to as the Gaia-Enceladus-Sausage. Several other merger galaxies have been
discovered: Thamnos \citep{Koppelman2019}, Sequoia \citep{Myeong2019}, Kraken \citep{Kruijssen2019,Kruijssen2020}, and Pontus \citep{Malhan2022}. The Gaia-Enceladus-Sausage is
the most recent major merger \citep{Borre2022,Dropulic2023}. Since this event, the Milky Way has evolved without significant mergers for the last $\sim$10 Gyr and, consequently,
one can expect that the redistribution of metals does not take place in the Milky Way disc. Thus, the Milky Way meets the  necessary conditions  (i.e. lack of the mergers and
interactions) in order to have a high abundance at its centre and a steep radial abundance gradient. The presence of close satellites (Magellanic Clouds) may appear to be in conflict
with this picture. However, \citet{vandenBergh2006}  argues that the Large and Small Magellanic Clouds may be interlopers from a remote part of the Local Group rather than true satellites
of the Milky Way (i.e. the Large Magellanic Cloud is on its first approach to the Milky Way).

Close examination of Fig.~\ref{figure:emwa-smwa} shows that among galaxies with central oxygen abundances close to  the Milky Way, the radial abundance gradients
(in terms of dex/$R_{25}$, difference between the oxygen abundance at the optical radius and at the centre) for two galaxies in our sample exceed the radial abundance gradient
of the Milky Way. Those galaxies are  NGC~6744, with 12 + log(O/H)$_{0}$ = 8.84 and 12 + log(O/H)$_{R_{25}}$ = 8.20 and consequently grad(O/H) = --0.64 dex/$R_{25}$,  and 
 NGC~5457 (M~101), with 12 + log(O/H)$_{0}$ = 8.71 and 12 + log(O/H)$_{R_{25}}$ = 7.87 and consequently grad(O/H) = $-0.84$~dex/$R_{25}$. These galaxies are huge compared to
the Milky Way, as the optical radius of  NGC~6744 is $R_{25}$ = 27.25 kpc, while the optical radius of  NGC~5457 is $R_{25}$ = 28.73 kpc. The stellar masses of  NGC~5457
(3.81$\times$10$^{10}$ $M_{\sun}$) and NGC~6744  (5.92$\times$10$^{10}$ $M_{\sun}$) are slightly lower or similar to that of the Milky Way  (5.2$\times$10$^{10}$ $M_{\sun}$).   
It should be emphasised that it is not necessary for a large galaxy to show a steep radial abundance gradient. Indeed, the galaxy NGC~753 of radius $R_{25}$ = 26.45 kpc shows
a radial gradient of $-0.12$~dex/$R_{25}$,  and the galaxy NGC~1365 of radius $R_{25}$ = 31.98 kpc shows a radial abundance gradient of $-0.21$~dex/$R_{25}$.   
The stellar masses of  NGC~753 (8.13$\times$10$^{10}$ $M_{\sun}$) and NGC~1365  (6.0$\times$10$^{10}$ $M_{\sun}$) are slightly higher  or similar to that of the Milky Way.   

One can expect that the shallow gradients in the majority of large galaxies can be attributed to the fact that the abundances within those galaxies are redistributed due to the
mergers or interactions.  \citet{Robotham2014} used a subset of the Galaxy And Mass Assembly II redshift sample to compare the effect of in situ star formation versus mass
accretion through mergers and found that galaxies of low masses are likely to obtain most of their mass through star formation, whilst massive galaxies are likely to obtain
most of their mass build-up through the accretion of smaller galaxies.
\citet{Conselice2022}  measured the pair fraction and merger fractions for galaxy mergers of different mass ratios and quantified the merger rate for massive galaxies (with
stellar masses higher than 10$^{11}$M$_{\sun}$). They found that over the last $\sim$10 Gyr, major mergers with mass ratios greater than 1:4 occurred 0.85$^{+0.19}_{-0.20}$
times on average, while minor mergers down to ratios of 1:10 occurred on average  1.43$^{+0.5}_{-0.3}$ times per galaxy.

Thus, one can suggest that the steep radial abundance gradient in the Milky Way can be attributed to the fact that the Milky Way has evolved without significant mergers and
interactions for the last $\sim$10~Gyr. However, it is not clear whether the evolution without mergers and interactions is a sufficient condition for a steep abundance gradient
in a galaxy.

\section{Conclusions}
%=====================

We searched for  Milky Way-like galaxies, comparing the following characteristics: stellar mass $M_{\star}$, optical radius $R_{25}$, rotation velocity $V_{rot}$, central oxygen
abundance (O/H)$_{0}$, and abundance at the optical radius (O/H)$_{R_{25}}$. Our sample of  comparison galaxies contained 504 galaxies: 53 nearby galaxies and 451 MaNGA galaxies.

If the values of the optical radius and the stellar mass of the galaxy were close to that of the Milky Way, then the galaxy  was referred to as a structural Milky Way analogue,
sMWA. The oxygen  abundance at a given radius of a galaxy is defined by the evolution of this region (fraction of gas converted into stars, i.e. astration level, and gas
exchange with the surroundings). One can expect that the similarity of the central oxygen abundance and the abundance at the optical radius in two galaxies suggests a  similar
(chemical) evolution. If the values of  (O/H)$_{0}$ and (O/H)$_{R_{25}}$ in a galaxy were close to that of the Milky Way, then the galaxy was referred to as an evolutionary Milky Way
analogue, eMWA. If the galaxy was simultaneously an eMWA and an sMWA, then it could be considered a Milky Way twin.

We find that the position of the Milky Way on the (O/H)$_{0}$ -- (O/H)$_{R_{25}}$ diagram shows a large  deviation from the general trend outlined by galaxies of our comparison sample
in the sense that the (O/H)$_{R_{25}}$ of the Milky Way is appreciably lower than in other galaxies of similar (O/H)$_{0}$. This feature in the Milky  Way evidences that its (chemical)
evolution  is not typical. 

The majority of the eMWAs are not sMWAs. Either the optical radii of the eMWAs are significantly larger in comparison to the  Milky Way or  the eMWAs have  appreciably lower values of
stellar mass and  rotation velocity than  the Milky Way.

We found four galaxies (NGC~3521, NGC~4651, NGC~2903, and MaNGA galaxy M-8341-09101) that are simultaneously an sMWA and  an eMWA, that is, they can be considered Milky Way twins.
The galaxy NGC~3521 is possibly the most similar to the Milky Way.  The characteristics we considered (stellar mass, optical radius, rotation velocity, oxygen abundance) of  NGC~3521
are similar to those of the Milky Way. The masses of black holes in those galaxies are also similar to each other.

In previous studies,    Milky Way-like galaxies were selected   using structural and morphological characteristics of galaxies, that is,  sMWAs were selected.  We find that
the abundances at the centre and at the optical radius  (evolutionary characteristics) provide a stricter criterion for selecting real Milky Way twins.

%==========================
\section*{Acknowledgements}
We are grateful to the referee, Dr. N.~Boardman,  for his constructive comments. \\
LSP acknowledges support from the Research Council of Lithuania (LMTLT), grant no. P-LU-PAR-23-28. \\
MALL acknowledges support from the Spanish grant PID2021-123417OB-I00, and the Ramón y Cajal program funded by the Spanish Government (RYC2020-029354-I) \\
This research has made use of the NASA/IPAC Extragalactic Database (NED), which
is funded by the National Aeronautics and Space Administration and operated by
the California Institute of Technology.  \\
We acknowledge the usage of the HyperLeda database (http://leda.univ-lyon1.fr). \\
Funding for SDSS-III has been provided by the Alfred P. Sloan Foundation,
the Participating Institutions, the National Science Foundation,
and the U.S. Department of Energy Office of Science.
The SDSS-III web site is http://www.sdss3.org/. \\
Funding for the Sloan Digital Sky Survey IV has been provided by the
Alfred P. Sloan Foundation, the U.S. Department of Energy Office of Science,
and the Participating Institutions. SDSS-IV acknowledges
support and resources from the Center for High-Performance Computing at
the University of Utah. The SDSS web site is www.sdss.org. \\
SDSS-IV is managed by the Astrophysical Research Consortium for the 
Participating Institutions of the SDSS Collaboration including the 
Brazilian Participation Group, the Carnegie Institution for Science, 
Carnegie Mellon University, the Chilean Participation Group,
the French Participation Group, Harvard-Smithsonian Center for Astrophysics, 
Instituto de Astrof\'isica de Canarias, The Johns Hopkins University, 
Kavli Institute for the Physics and Mathematics of the Universe (IPMU) / 
University of Tokyo, Lawrence Berkeley National Laboratory, 
Leibniz Institut f\"ur Astrophysik Potsdam (AIP),  
Max-Planck-Institut f\"ur Astronomie (MPIA Heidelberg), 
Max-Planck-Institut f\"ur Astrophysik (MPA Garching), 
Max-Planck-Institut f\"ur Extraterrestrische Physik (MPE), 
National Astronomical Observatories of China, New Mexico State University, 
New York University, University of Notre Dame, 
Observat\'ario Nacional / MCTI, The Ohio State University, 
Pennsylvania State University, Shanghai Astronomical Observatory, 
United Kingdom Participation Group,
Universidad Nacional Aut\'onoma de M\'exico, University of Arizona, 
University of Colorado Boulder, University of Oxford, University of Portsmouth, 
University of Utah, University of Virginia, University of Washington, University of Wisconsin, 
Vanderbilt University, and Yale University.

\appendix
%=========

\section{Characteristics of nearby galaxies}
%=====================

In this section we report the general characteristics (the inclination angle of a galaxy, the position angle of the major axis, distance, angular and physical optical radii, 
 stellar mass, rotation velocity, and the black hole  mass) and the source of the data for the sample of nearby galaxies. We estimated the abundances in  H\,{\sc ii}
regions through the $R$ calibration from \citet{Pilyugin2016} and determined the O/H - $R_{g}$  relation for each galaxy where $R_{g}$ is a fractional radius normalised
to the optical radius $R_{25}$. The mean value of the uncertainty in the (O/H)$_{R}$ abundance obtained through the $R$ calibration is $\sim$0.05 dex \citep{Pilyugin2016}.
We assumed a deviation of the abundance from the  O/H-$R_{g}$ relation larger than 0.15 dex could be attributed to the large uncertainty in the line measurements or
could be an indicator of a peculiarity in the abundance of the H\,{\sc ii} region. Therefore, the points with deviations larger than 0.15 dex in (O/H) were rejected and
not used in the determination of the final  O/H-$R_{g}$ relation. The nearby galaxies discussed in the Section 4.4 (NGC~2903, NGC~3521, and NGC~4651)
are not considered in this section. The compiled and derived characteristics of nearby galaxies are listed in Table.~\ref{table:nearby}. 

\subsection{NGC~224 (M~31)}
%=====================

Galaxy NGC~224 (M~31, Andromeda galaxy) is an Sb galaxy  (morphological type code $T$ = 3.0$\pm$0.4). The inclination angle of  NGC~224 is $i$ = 77$\degr$, and the position angle
of the major axis is PA = 38$\degr$ \citep{Corbelli2010}. The optical radius is 95.27 arcmin \citep{RC3}. At the distance  of $d$ = 0.82 Mpc \citep{Anand2021},
the physical optical radius of  the NGC~224 is $R_{25}$ = 22.72 kpc. The stellar mass is $M_{\star}$ = 1.0 $\times$ 10$^{11}$ $M_{\sun}$, or log($M_{\star}/M_{\sun}$) = 11.00 \citep{Jarrett2019}.  
The black hole  mass  is log($M_{BH}/M_{\sun}$) = 8.15$^{+0.22}_{-0.11}$ \citep{Davis2019}. The rotation velocity of  NGC~224 is 253 km\,s$^{-1}$ \citep{Corbelli2010}. 
The radial distribution of the oxygen abundances based on the H\,{\sc ii} regions from the compilation in \citet{Pilyugin2014} was approximated by the relation
\begin{equation}
12+\log(\rm O/H) = 8.715(\pm 0.016) - 0.377(\pm 0.028) \times R_{g}   
\label{equation:n0224-oh-compil}
\end{equation}
with a scatter of 0.060 dex for 204 data points. 

\subsection{NGC~253}
%=====================

Galaxy NGC~253 is an SABc galaxy  (morphological type code $T$ = 5.1$\pm$0.4). The inclination angle of  NGC~253 is $i$ = 76$\degr$, and the position angle of the major axis
PA = 235$\degr$ \citep{Lucero2015}. The optical radius is 13.77 arcmin \citep{RC3}. At a distance  of $d$ = 3.70 Mpc \citep{Anand2021}, the physical optical
radius of  NGC~253 is $R_{25}$ = 14.82 kpc. The stellar mass (mean value) is $M_{\star}$ = 3.50 $\times$ 10$^{10}$ $M_{\sun}$, or log($M_{\star}/M_{\sun}$) = 10.54 \citep{Jarrett2019,Leroy2021}.  
The black hole  mass  is log($M_{BH}/M_{\sun}$) = 7.00$\pm$0.19 \citep{Davis2019}. The rotation velocity of  NGC~253 is 211 km\,s$^{-1}$ \citep{Hlavacek2011a,Lucero2015}.   
The radial distribution of the oxygen abundances traced by the H\,{\sc ii} regions from \citet{Webster1983} was approximated by the relation
\begin{equation}
12+\log(\rm O/H) = 8.493(\pm 0.144) - 0.079(\pm 0.182 \times R_{g}   
\label{equation:n0253-oh-compil}
\end{equation}
with a scatter of 0.061 dex for eight data points. 

\subsection{NGC~300}
%=====================

Galaxy NGC~300 is an Scd galaxy  (morphological type code $T$ = 6.9$\pm$0.4). The inclination angle of  NGC~300 is $i$ = 42$\degr$, and the position angle of the major axis
PA = 106$\degr$ \citep{Carignan1985}. The optical radius is 10.94 arcmin \citep{RC3}. At a distance  of $d$ = 1.94 Mpc \citep{Bono2010}, the physical optical radius of NGC~300
is $R_{25}$ = 6.17 kpc. The stellar mass (the mean value for estimations from \citet{Jarrett2019} and \citet{Leroy2021} rescaled to the adopted distance) is
$M_{\star}$ = 2.10 $\times$ 10$^{9}$ $M_{\sun}$, or log($M_{\star}/M_{\sun}$) = 9.32. The rotation velocity of  NGC~300 is 85 km\,s$^{-1}$ \citep{Hlavacek2011b,Ponomareva2016}  
The radial distribution of the oxygen abundances estimated using the H\,{\sc ii} regions from the compilation in \citet{Pilyugin2014} and supplemented by the measurements from
\citet{Toribio2016} was approximated by the relation
\begin{equation}
12+\log(\rm O/H) = 8.444(\pm 0.017) - 0.424(\pm 0.034 \times R_{g}   
\label{equation:n0300-oh-compil}
\end{equation}
with a scatter of 0.062 dex for 46 data points. 

\subsection{NGC~598 (M~33)}
%=====================

Galaxy NGC~598 (M~33) is an Sc galaxy  (morphological type code $T$ = 5.9$\pm$0.4). The inclination angle of  NGC~598 is $i$ = 54$\degr$, and the position angle of the major axis is
PA = 201$\degr$ \citep{Kam2017}. The optical radius is 35.40 arcmin \citep{RC3}. At a distance  of $d$ = 0.94 Mpc \citep{Anand2021}, the physical optical radius of  NGC~598
is $R_{25}$ = 9.68 kpc. The stellar mass is $M_{\star}$ = 4.87 $\times$ 10$^{9}$ $M_{\sun}$, or log($M_{\star}/M_{\sun}$) = 9.69 \citep{Jarrett2019}. The rotation velocity of the NGC~598
is 125 km\,s$^{-1}$ \citep{Kam2017}. The radial abundance distribution traced by the H\,{\sc ii} regions from the compilation in \citet{Pilyugin2014} was approximated by the relation
\begin{equation}
12+\log(\rm O/H) = 8.489(\pm 0.020) - 0.452(\pm 0.053) \times R_{g}   
\label{equation:n0598-oh-compil}
\end{equation}
with a scatter of 0.056 dex for 35 data points. 

\subsection{NGC~628 (M~74)}
%=====================

The nearby galaxy NGC~628 (M~74, the Phantom Galaxy) is an isolated Sc spiral galaxy (morphological type code $T$ = 5.2$\pm$0.5).  It is a face-on galaxy.
Its inclination angle is $i$ = 6$\degr$, and the position angle of the major axis PA = 25$\degr$ \citep{Kamphuis1992}. The optical radius of  NGC~628 is $R_{25}$ = 5.24
arcmin \citep{RC3}.  The  H\,{\sc i} disc extends out to more than three times the optical radius \citep{Kamphuis1992}. There are recent distance estimations for  NGC~628
obtained through the tip of the red giant branch method based on Hubble Space Telescope measurements. \citet{Jang2014} found the distance to NGC~628 to be 10.19 $\pm$ 0.14
(random) $\pm$ 0.56 (systematic) Mpc. \citet{McQuinn2017} measured the distance to NGC~628 to be 9.77 $\pm$ 0.17 (statistical uncertainty) $\pm$ 0.32 (systematic uncertainty) Mpc.
\citet{Sabbi2018} determined the distances for the central pointing ($d$ = 8.6 $\pm$ 0.9 Mpc) and for the outer field ($d$ = 8.8 $\pm$ 0.7 Mpc). In this work, we adopted the distance
to the NGC~628 used in our previous paper: $d$ = 9.91 Mpc \citep{Pilyugin2014}, which is close to the values obtained by \citet{Jang2014} and \citet{McQuinn2017}.  The optical
radius of  NGC~628 is $R_{25}$ = 15.09 kpc with the adopted distance. The mean value of the stellar mass from \citet{Leroy2008} and \citet{Leroy2021} rescaled to the adopted
distance is $M_{\star}$ = 2.3 $\times$ 10$^{10}$ $M_{\sun}$, or log($M_{\star}/M_{\sun}$) = 10.36. The radial abundance distribution for the H\,{\sc ii} regions from
\citet{McCall1985, Ferguson1998, vanZee1998, Bresolin1999, Berg2015} was approximated by the relation
\begin{equation}
12+\log(\rm O/H) = 8.667(\pm 0.016) - 0.409(\pm 0.028) \times R_{g}   
\label{equation:n0628-oh-compil}
\end{equation}
with a scatter of 0.047 dex for 78 data points. 

\subsection{NGC~753}
%=====================

Galaxy NGC~753 is an SABc galaxy  (morphological type code $T$ = 4.9$\pm$1.0). The inclination angle of  NGC~753 is $i$ = 50$\degr$, the position angle of the major axis is
PA = 128$\degr$ \citep{Amram1994}. The optical radius is 1.26 arcmin \citep{RC3}. At a distance of $d$ = 72.4 Mpc \citep{Leroy2019}, the physical optical radius  of NGC~753  is
$R_{25}$ = 26.45 kpc. The value of the rotation velocity is $\sim$210 km\,s$^{-1}$ \citep{Amram1994}. The stellar mass is $M_{\star}$ = 8.13 $\times$ 10$^{10}$ $M_{\sun}$, or
log($M_{\star}/M_{\sun}$) = 10.91 \citep{Leroy2019}. The abundance gradient traced by the H\,{\sc ii} regions from  \citet{Henry1996} is 
\begin{equation}
12+\log(\rm O/H) = 8.586(\pm 0.065) - 0.064(\pm 0.095 \times R_{g}   
\label{equation:n0753-oh-compil}
\end{equation}
with a scatter of 0.055 dex for eight data points. 

\subsection{NGC~925}
%=====================

Galaxy NGC~925 is an Scd galaxy  (morphological type code $T$ = 7.0$\pm$0.3). The inclination angle of  NGC~925 is $i$ = 66$\degr$, and the position angle of the major axis is
PA = 287$\degr$ \citep{deBlok2008}. The optical radius is 5.24 arcmin \citep{RC3}. At a distance   of $d$ = 9.29 Mpc \citep{Saha2006}, the physical optical radius of NGC~925
is $R_{25}$ = 14.15 kpc. The stellar mass is $M_{\star}$ = 5.62 $\times$ 10$^{9}$ $M_{\sun}$, or log($M_{\star}/M_{\sun}$) = 9.75 \citep{Leroy2019}. The rotation velocity of  NGC~925 is
115 km\,s$^{-1}$ \citep{Ponomareva2016}. The radial abundance distribution based on the H\,{\sc ii} regions from \citet{vanZee1998} was approximated by the relation
\begin{equation}
12+\log(\rm O/H) = 8.440(\pm 0.017) - 0.360(\pm 0.027) \times R_{g}   
\label{equation:n0925-oh-compil}
\end{equation}
with a scatter of 0.047 dex for 34 data points. 

\subsection{NGC~1058}
%=====================

Galaxy NGC~1058 is an Sc galaxy  (morphological type code $T$ = 5.1$\pm$0.9). The inclination angle of NGC~1058 is $i$ = 15$\degr$, and the position angle of the major axis is
PA = 145$\degr$ \citep{GarciaGomez2004}. The optical radius is 1.51 arcmin \citep{RC3}. At a distance    of $d$ = 10.6 Mpc \citep{Schmidt1994}, the physical optical radius of NGC~1058
is $R_{25}$ = 4.66 kpc. The stellar mass (rescaled to adopted distance) is $M_{\star}$ = 2.51 $\times$ 10$^{9}$ $M_{\sun}$, or log($M_{\star}/M_{\sun}$) = 9.40 \citep{Leroy2019}.  
The radial abundance distribution traced by the H\,{\sc ii} regions from \citet{Ferguson1998,Bresolin2019} was approximated by the relation
\begin{equation}
12+\log(\rm O/H) = 8.649(\pm 0.023) - 0.265(\pm 0.043) \times R_{g}   
\label{equation:n1058-oh-compil}
\end{equation}
with a scatter of 0.046 dex for 22 data points. 

\subsection{NGC~1068 (M~77)}
%=====================

Galaxy NGC~1068 (M~77) is an Sb galaxy  (morphological type code $T$ = 3.0$\pm$0.3). The inclination angle of  NGC~1068 is $i$ = 35$\degr$, and the position angle of the major axis PA = 73$\degr$
\citep{Leroy2021}. The optical radius is 3.54 arcmin \citep{RC3}. At a distance    of $d$ = 13.97 Mpc \citep{Anand2021}, the physical optical radius of NGC~1068 is $R_{25}$ =
14.38 kpc. The stellar mass is $M_{\star}$ = 8.13 $\times$ 10$^{10}$ $M_{\sun}$, or log($M_{\star}/M_{\sun}$) = 10.91 \citep{Leroy2021}. The mass of the black hole in the NGC~1068 is
log($M_{BH}/M_{\sun}$) = 6.75$\pm$0.08 \citep{Davis2019}. The radial distribution of the oxygen abundances in the H\,{\sc ii} regions from the compilation in \citet{Pilyugin2014} was
approximated by the relation
\begin{equation}
12+\log(\rm O/H) = 8.696(\pm 0.018) - 0.102(\pm 0.062) \times R_{g}   
\label{equation:n1068-oh-compil}
\end{equation}
with a scatter of 0.035 dex for 13 data points. 

\subsection{NGC~1097}
%=====================

Galaxy NGC~1097 is an SBb galaxy  (morphological type code $T$ = 3.3$\pm$0.5). The inclination angle of  NGC~1097 is $i$ = 46$\degr$, and the position angle of the major axis is PA = 134$\degr$
\citep{Ondrechen1989b}. The optical radius is 4.67 arcmin \citep{RC3}. At a distance    of $d$ = 13.58 Mpc \citep{Anand2021}, the physical optical radius of NGC~1097 is $R_{25}$ =
18.43 kpc. The stellar mass is $M_{\star}$ = 5.75 $\times$ 10$^{10}$ $M_{\sun}$, or log($M_{\star}/M_{\sun}$) = 10.76 \citep{Leroy2021}.  The black hole  mass  is log($M_{BH}/M_{\sun}$) = 8.38$\pm$0.04
\citep{Davis2019}.  The rotation velocity of  NGC~1097 is 253 km\,s$^{-1}$ \citep{Ondrechen1989b}. The abundance gradient estimated using the  H\,{\sc ii} regions from the compilation in
\citet{Pilyugin2014} was approximated by the relation
\begin{equation}
12+\log(\rm O/H) = 8.661(\pm 0.009) - 0.138(\pm 0.040) \times R_{g}   
\label{equation:n1097-oh-compil}
\end{equation}
with a scatter of 0.022 dex for 15 data points. 

\subsection{NGC~1232}
%=====================

Galaxy NGC~1232 is an SABc galaxy  (morphological type code $T$ = 5.0$\pm$0.4). The inclination angle of  NGC~1232 is $i$ = 30$\degr$, and the position angle of the major axis is PA = 270$\degr$
\citep{vanZee1999}. The optical radius is 3.71 arcmin \citep{RC3}. At a distance    of $d$ = 21.5 Mpc \citep{vanZee1999}, the physical optical radius of NGC~1232 is $R_{25}$ = 23.18 kpc.
The stellar mass rescaled to the adopted distance is $M_{\star}$ = 4.70 $\times$ 10$^{10}$ $M_{\sun}$, or log($M_{\star}/M_{\sun}$) = 10.67 \citep{Leroy2019}. The rotation velocity of the NGC~1232 is
220 km\,s$^{-1}$ \citep{vanZee1999}. The radial abundance distribution traced by the H\,{\sc ii} regions from the compilation in \citet{Pilyugin2014} was approximated by the relation
\begin{equation}
12+\log(\rm O/H) = 8.741(\pm 0.028) - 0.487(\pm 0.048) \times R_{g}   
\label{equation:n1232-oh-compil}
\end{equation}
with a scatter of 0.050 dex for 29 data points. 

\subsection{NGC~1313}
%=====================

Galaxy NGC~1313 is an SBcd galaxy  (morphological type code $T$ = 7.0$\pm$0.4). The inclination angle of   NGC~1313 is $i$ = 48$\degr$, and the position angle of the major axis PA = 0$\degr$
\citep{Ryder1995b}. The optical radius is 6.12 arcmin or 367.2 arcsec \citep{Ho2011}. At a distance    of $d$ = 4.32 Mpc \citep{Anand2021}, the physical optical radius of NGC~1313
is $R_{25}$ = 7.69 kpc. The stellar mass (mean value) is $M_{\star}$ = 2.91 $\times$ 10$^{9}$ $M_{\sun}$, or log($M_{\star}/M_{\sun}$) = 9.46 \citep{Jarrett2019,Leroy2019}.  The rotation velocity
of the NGC~1313 is 112 km\,s$^{-1}$ \citep{Ryder1995b}. The abundance gradient based on the H\,{\sc ii} regions from the compilation in \citet{Pilyugin2014} was approximated by the relation
\begin{equation}
12+\log(\rm O/H) = 8.163(\pm 0.025) - 0.148(\pm 0.049) \times R_{g}   
\label{equation:n1313-oh-compil}
\end{equation}
with a scatter of 0.071 dex for 37 data points. 

\subsection{NGC~1365}
%=====================

Galaxy NGC 1365 is an Sb galaxy  (morphological type code $T$ = 3.2$\pm$0.7). Its inclination angle is $i$ = 46$\degr$, and the position angle of the major axis is PA = 222$\degr$
\citep{Ondrechen1989}. The optical radius of   NGC~1365 is $R_{25}$ = 5.61 arcmin \citep{RC3}. The distance to   NGC~1365 is $d$ = 19.6 Mpc \citep{Saha2006}. The physical
optical radius at the adopted distance is $R_{25}$ = 31.98 kpc. The stellar mass rescaled to the adopted distance is $M_{\star}$ = 6.00 $\times$ 10$^{10}$ $M_{\sun}$, or
log($M_{\star}/M_{\sun}$) = 10.78 \citet{Jarrett2019}. The mass of the black hole in the NGC~1365 is log($M_{BH}/M_{\sun}$ = 6.60$\pm 0.30$ \citep{Davis2019}.
We adopted $V_{rot}$ = 300 km\,s$^{-1}$ as the rotation velocity of NGC~1365, which is the mean value of  $V_{rot}$ $\sim$ 280 km\,s$^{-1}$ determined by \citet{Ondrechen1989} and
$V_{rot}$ $\sim$ 320 km\,s$^{-1}$ obtained by \citet{Zanmar2008}. The abundance gradient traced by the H\,{\sc ii} regions from the compilation in \citet{Pilyugin2014} was approximated
by the relation
\begin{equation}
12+\log(\rm O/H) = 8.619(\pm 0.015) - 0.208(\pm 0.027) \times R_{g}   
\label{equation:n1365-oh-compil}
\end{equation}
with a scatter of 0.051 dex for 79 data points. 

\subsection{NGC~1512}
%=====================

Galaxy NGC~1512 is an Sa galaxy  (morphological type code $T$ = 1.2$\pm$0.5). The inclination angle of   NGC~1512 is $i$ = 42$\degr$, and the position angle of the major axis is PA = 262$\degr$
\citep{Lang2020}. The optical radius is 4.46 arcmin \citep{RC3}. At a distance    of $d$ = 18.83 Mpc \citep{Anand2021}, the physical optical radius of NGC~1512 is $R_{25}$ = 24.41 kpc.
The stellar mass  is $M_{\star}$ = 5.25 $\times$ 10$^{10}$ $M_{\sun}$, or log($M_{\star}/M_{\sun}$) = 10.72 \citep{Leroy2021}. The black hole  mass  is log($M_{BH}/M_{\sun}$) = 7.78$\pm$0.19
\citep{Davis2014}.  The rotation velocity of   NGC~1512 is 179 km\,s$^{-1}$ \citep{Lang2020}. The radial distribution of the oxygen abundances for the H\,{\sc ii} regions from
\citet{Bresolin2012,LopezSanchez2015} is approximated by the relation
\begin{equation}
12+\log(\rm O/H) = 8.741(\pm 0.019) - 0.379(\pm 0.032) \times R_{g}   
\label{equation:n1512-oh-compil}
\end{equation}
with a scatter of 0.050 dex for 59 data points. 

\subsection{NGC~1598}
%=====================

Galaxy NGC~1598 is an SBc galaxy  (morphological type code $T$ = 4.8$\pm$0.6). The inclination angle of   NGC~1598 is $i$ = 55$\degr$, the position angle of the major axis is PA = 123$\degr$,
and the optical radius is 0.72 arcmin \citep{RC3}. At a distance    of $d$ = 55.80 Mpc \citep{Springob2009}, the physical optical radius of NGC~1598 is $R_{25}$ = 11.73 kpc.
The stellar mass is $M_{\star}$ = 1.62 $\times$ 10$^{10}$ $M_{\sun}$, or log($M_{\star}/M_{\sun}$) = 10.21 \citep{Lapi2018}. The rotation velocity of   NGC~1598 is 110 km\,s$^{-1}$ \citep{Lapi2018}.   
The radial abundance distribution based on the H\,{\sc ii} regions from \citet{Storchi1996} was approximated by the relation
\begin{equation}
12+\log(\rm O/H) = 8.665(\pm 0.023) - 0.275(\pm 0.044 \times R_{g}   
\label{equation:n1598-oh-compil}
\end{equation}
with a scatter of 0.030 dex for nine data points. 

\subsection{NGC~1672}
%=====================

Galaxy NGC~1672 is an Sb galaxy (morphological type code $T$ = 3.3$\pm$0.6). The inclination angle of   NGC~1672 is $i$ = 43$\degr$, and the position angle of the major axis is PA = 134$\degr$
\citep{Lang2020}. The optical radius is 3.30 arcmin \citep{RC3}. At a distance    of $d$ = 19.40 Mpc \citep{Anand2021}, the physical optical radius of NGC~1672 is $R_{25}$ = 18.64 kpc.
The stellar mass is $M_{\star}$ = 5.37 $\times$ 10$^{10}$ $M_{\sun}$, or log($M_{\star}/M_{\sun}$) = 10.73 \citep{Leroy2021}. The mass of the black hole in   NGC~1672 is log($M_{BH}/M_{\sun}$)
= 7.08$\pm 0.90$ \citep{Davis2014}. The radial distribution of the oxygen abundance estimated using the H\,{\sc ii} regions from \citet{Storchi1996}  was approximated by the relation
\begin{equation}
12+\log(\rm O/H) = 8.639(\pm 0.012) - 0.130(\pm 0.037) \times R_{g}   
\label{equation:n1672-oh-compil}
\end{equation}
with a scatter of 0.028 dex for 15 data points. 

\subsection{NGC~2403}
%=====================

Galaxy NGC~2403 is an SABc galaxy  (morphological type code $T$ = 6.0$\pm$0.3). The inclination angle of   NGC~2403 is $i$ = 63$\degr$, and the position angle of the major axis is PA = 124$\degr$
\citep{deBlok2008}. The optical radius is 10.94 arcmin \citep{RC3}. At a distance    of $d$ = 3.19 Mpc \citep{Anand2021}, the physical optical radius of NGC~2403 is $R_{25}$ = 10.15 kpc.
The stellar mass is $M_{\star}$ = 4.50 $\times$ 10$^{9}$ $M_{\sun}$ or log($M_{\star}/M_{\sun}$) = 9.65 \citep{Leroy2008,Jarrett2019}. The rotation velocity of   NGC~2403 is 134 km\,s$^{-1}$
\citep{deBlok2008,Leroy2008}. The abundance gradient traced by the H\,{\sc ii} regions from the compilation in \citet{Pilyugin2014} is 
\begin{equation}
12+\log(\rm O/H) = 8.465(\pm 0.018) - 0.531(\pm 0.034) \times R_{g},   
\label{equation:n2403-oh-compil}
\end{equation}
with a scatter of 0.068 dex for 47 data points. 

\subsection{NGC~2442}
%=====================

Galaxy NGC~2442 is an Sbc galaxy  (morphological type code $T$ = 3.7$\pm$0.6). The inclination angle of   NGC~2442 is $i$ = 29$\degr$, and the position angle of the major axis is PA = 27$\degr$
\citep{Pilyugin2014}. The optical radius is 2.75 arcmin \citep{RC3}. At a distance    of $d$ = 21.5 Mpc \citep{Leroy2019}, the physical optical radius of NGC~2442  is $R_{25}$ =
17.18 kpc. The stellar mass is $M_{\star}$ = 6.92 $\times$ 10$^{10}$ $M_{\sun}$, or log($M_{\star}/M_{\sun}$) = 10.84 \citep{Leroy2019}. The black hole  mass  is log($M_{BH}/M_{\sun}$) = 7.28$\pm$0.33
\citep{Davis2014}. The radial distribution of the oxygen abundances estimated using the H\,{\sc ii} regions from \citet{Ryder1995} was approximated by the relation
\begin{equation}
12+\log(\rm O/H) = 8.606(\pm 0.050) - 0.050(\pm 0.081) \times R_{g}   
\label{equation:n2442-oh-compil}
\end{equation}
with a scatter of  0.035 dex for eight data points. 

\subsection{NGC~2805}
%=====================

Galaxy NGC~2805 is an SABc galaxy (morphological type code $T$ = 6.9$\pm$0.3). The inclination angle of   NGC~2805 is $i$ = 36$\degr$, and the position angle of the major axis is PA = 123$\degr$
\citep{Erroz2015}. The optical radius is 3.16 arcmin \citep{RC3}. At a distance    of $d$ = 28.7 Mpc \citep{Erroz2015}, the physical optical radius of NGC~2805 is $R_{25}$ = 26.34 kpc.
The stellar mass (rescaled to the adopted distance) is $M_{\star}$ = 9.44 $\times$ 10$^{9}$ $M_{\sun}$, or log($M_{\star}/M_{\sun}$) = 9.98 \citep{Leroy2019}.  The rotation velocity of  
NGC~2805 is 106 km\,s$^{-1}$ \citep{Erroz2015}. The radial abundance distribution based on the H\,{\sc ii} regions from \citet{vanZee1998} was approximated by the relation
\begin{equation}
12+\log(\rm O/H) = 8.477(\pm 0.026) - 0.371(\pm 0.040) \times R_{g}   
\label{equation:n2805-oh-compil}
\end{equation}
with a scatter of 0.033 dex for 17 data points. 

\subsection{NGC~2835}
%=====================

Galaxy NGC~2835 is an Sc galaxy  (morphological type code $T$ = 5.0$\pm$0.4). The inclination angle of   NGC~2835 is $i$ = 41.3$\degr$, and the position angle of the major axis is PA = 1$\degr$
\citep{Lang2020}. The optical radius is 3.30 arcmin \citep{RC3}. At a distance   of $d$ = 12.22 Mpc \citep{Anand2021}, the physical optical radius of NGC~2835 is $R_{25}$ = 11.74 kpc.
The stellar mass is $M_{\star}$ = 1.00 $\times$ 10$^{10}$ $M_{\sun}$, or log($M_{\star}/M_{\sun}$) = 10.00 \citep{Leroy2021}. The mass of the black hole in the NGC~2835 is log($M_{BH}/M_{\sun}$)
= 6.72$\pm 0.30$ \citep{Davis2014}. The abundance gradient for the H\,{\sc ii} regions measured by \citet{Ryder1995} is 
\begin{equation}
12+\log(\rm O/H) = 8.498(\pm 0.045) - 0.305(\pm 0.070) \times R_{g},   
\label{equation:n2835-oh-compil}
\end{equation}
with a scatter of 0.066 dex for 17 data points. 

\subsection{NGC~2997}
%=====================

Galaxy NGC 2997 is a grand design spiral galaxy of the type SABc  (morphological type code $T$ = 5.1$\pm$0.5).  Its inclination angle is $i$ = 33$\degr$, and the position angle of the major
axis is PA = 108$\degr$ \citep{Lang2020}. The optical radius of   NGC~2997 is $R_{25}$ = 4.46 arcmin \citep{RC3}. The distance to   NGC~2997 is 11.3 Mpc \citep{Lang2020}.
The physical optical radius at the adopted distance is $R_{25}$ = 14.66 kpc. The stellar mass of   NGC~2997 is $M_{\star}$ = 3.09 $\times$ 10$^{10}$ $M_{\sun}$, or log($M_{\star}/M_{\sun}$)
= 10.49, rescaled to the adopted distance \citep{Jarrett2019}. The mass of the black hole in   NGC~2997 is log($M_{BH}/M_{\sun}$) = 5.84$\pm$0.75 \citep{Davis2014}. The rotation velocity
of   NGC~2997 is 185 km\,s$^{-1}$ \citep{Peterson1978}. The radial distribution of the oxygen abundances for the H\,{\sc ii} regions from the compilation in \citet{Pilyugin2014} was
approximated by the relation
\begin{equation}
12+\log(\rm O/H) = 8.725(\pm 0.044) - 0.363(\pm 0.088) \times R_{g}   
\label{equation:n2997-oh-compil}
\end{equation}
with a scatter of 0.045 dex for 19 data points. 

\subsection{NGC~3031 (M~81)}
%=====================

Galaxy NGC~3031 (M~81) is an Sab galaxy  (morphological type code $T$ = 2.4$\pm$0.6). The inclination angle of   NGC~3031 is $i$ = 59$\degr$, the position angle of the major axis is PA = 330$\degr$,
and the optical radius is 10.69 arcmin \citep{deBlok2008}. At a distance    of $d$ = 3.63 Mpc \citep{Saha2006}, the physical optical radius of NGC~3031 is $R_{25}$ = 11.29 kpc.
The rotation velocity on the flat part is 215 km\,s$^{-1}$ \citep{deBlok2008,Ponomareva2017}. The mean value of stellar mass out of five estimations is $M_{\star}$ = 4.93 $\times$ 10$^{10}$ $M_{\sun}$,
or log($M_{\star}/M_{\sun}$) = 10.69 \citep{Jarrett2019,Ponomareva2018}. The mass of the black hole in   NGC~3031 is log($M_{BH}/M_{\sun}$) = 7.81$\pm 0.13$ in solar mass \citep{vandenbosch2016}.
The abundance gradient traced by the H\,{\sc ii} regions from the compilation in \citet{Pilyugin2014} and supplemented by the measurements from  \citet{Arellano2016} is\begin{equation}
12+\log(\rm O/H) = 8.638(\pm 0.023) - 0.150(\pm 0.037) \times R_{g}   
\label{equation:n3031-oh-compil}
\end{equation}
with a scatter of 0.041 dex for 77 data points. 

\subsection{NGC~3184}
%=====================

Galaxy NGC~3184 is an SABc spiral galaxy (morphological type code $T$ = 5.9$\pm$0.4).   It is a face-on galaxy. Its inclination angle is $i$ = 16$\degr$, and the position angle of
the major axis is PA = 179$\degr$ \citep{Tamburro2008}. The optical radius of   NGC~3184 is $R_{25}$ = 3.71 arcmin \citep{RC3}.  The estimations of the distance to   NGC~3184
using Type II plateau supernovae as the “standard candle” resulted in the values of  $d$ = 12.5 -- 12.7 Mpc  \citep{Olivares2010}, $d$ = 11.62$\pm$0.29 Mpc \citep{Bose2014},  and
$d$ = 9.68 Mpc \citep{Pejcha2015}. In this work we adopted the distance to   NGC~3184 of $d$ = 11.62  Mpc  obtained by \citet{Bose2014}. The optical radius of   NGC~3184 is $R_{25}$ = 12.53 kpc
at the adopted distance. The value of the stellar mass is $M_{\star}$ = 2.57 $\times$ 10$^{10}$ $M_{\sun},$ or log($M_{\star}/M_{\sun}$) = 10.41 \citep{Leroy2008,Das2020}. The rotation velocity of
  NGC~3184 is $V_{rot}$ = 210 km\,s$^{-1}$ \citep{Leroy2008}.
The radial distribution of the oxygen abundances estimated from the H\,{\sc ii} regions from  \citet{McCall1985,vanZee1998,Berg2020} was approximated by the relation
\begin{equation}
12+\log(\rm O/H) = 8.721(\pm 0.011) - 0.326(\pm 0.020) \times R_{g}   
\label{equation:n3184-oh-compil}
\end{equation}
with a scatter of 0.028 dex for 72 data points. 

\subsection{NGC~3351 (M~95)}
%=====================

Galaxy NGC~3351 (M~95) is an Sb galaxy  (morphological type code $T$ = 3.1$\pm$0.4). The inclination angle of   NGC~3351 is $i$ = 41$\degr$, and the position angle of the major axis is PA = 192$\degr$
\citep{Tamburro2008}. The optical radius is 3.71 arcmin \citep{RC3}. At a distance    of $d$ = 9.96 Mpc \citep{Anand2021}, the physical optical radius of NGC~3351 is $R_{25}$ = 10.74 kpc.
The stellar mass is $M_{\star}$ = 2.34 $\times$ 10$^{10}$ $M_{\sun}$, or log($M_{\star}/M_{\sun}$) = 10.37 \citep{Leroy2021}. The mass of the black hole in   NGC~3351 is log($M_{BH}/M_{\sun}$) =
6.52$^{+0.26}_{-6.52}$ \citep{vandenbosch2016}. The maximum value of the rotation velocity in  NGC~3351 is 210 km\,s$^{-1}$ \citep{Tamburro2008}, and the rotation velocity on the flat part of
the rotation curve is 196 km\,s$^{-1}$ \citep{Leroy2008}. We adopted $V_{rot}$ = 196 km\,s$^{-1}$.
The abundance distribution based on the $R$ calibration-based abundances in H\,{\sc ii} regions from the compilation in \citet{Pilyugin2014} was approximated by the relation
\begin{equation}
12+\log(\rm O/H) = 8.696(\pm 0.005) - 0.076(\pm 0.013) \times R_{g}   
\label{equation:n3351-oh-compil}
\end{equation}
with a scatter of 0.016 dex for 20 data points. 

\subsection{NGC~3359}
%=====================

Galaxy NGC~3359 is an Sc galaxy (morphological type code $T$ = 5.2$\pm$0.5). The inclination angle of   NGC~3359 is $i$ = 53$\degr$, and the position angle of the major axis is PA = 350$\degr$
\citep{Rozas2000}. The optical radius is 3.62 arcmin \citep{RC3}. At a distance    of $d$ = 22.6 Mpc \citep{Leroy2019}, the physical optical radius of NGC~3359 is $R_{25}$ = 23.81 kpc.
The stellar mass is $M_{\star}$ = 1.51 $\times$ 10$^{10}$ $M_{\sun}$, or log($M_{\star}/M_{\sun}$) = 10.18 \citep{Leroy2019}.  The rotation velocity of   NGC~3359 is 145 km\,s$^{-1}$ \citep{Rozas2000}.   
The radial distribution of the oxygen abundances estimated through the $R$ calibration in H\,{\sc ii} regions from \citet{Zahid2011} was approximated by the relation
\begin{equation}
12+\log(\rm O/H) = 8.401(\pm 0.066) - 0.436(\pm 0.138) \times R_{g}   
\label{equation:n3359-oh-compil}
\end{equation}
with a scatter of 0.067 dex for 11 data points. We note that \citet{Zahid2011} suggested the existence of a break in the radial abundance distribution in this galaxy. However,
the number of points is small and the scatter is large, thus preventing us from making a solid conclusion about the break.

\subsection{NGC~3621}
%=====================

 Galaxy
NGC~3621 is an SBcd galaxy  (morphological type code $T$ = 6.9$\pm$0.5). The inclination angle of   NGC~3621 is $i$ = 65$\degr$, and the position angle of the major axis is PA = 345$\degr$
\citep{deBlok2008}. The optical radius is 4.88 arcmin \citep{Ho2011}. At a distance    of $d$ = 7.06 Mpc \citep{Anand2021}, the physical optical radius of NGC~3621 is $R_{25}$ = 10.03 kpc.
The stellar mass is $M_{\star}$ = 1.15 $\times$ 10$^{10}$ $M_{\sun}$, or log($M_{\star}/M_{\sun}$) = 10.06 \citep{Leroy2019}. The black hole  mass  is log($M_{BH}/M_{\sun}$) = 6.00$^{+0.48}_{-6.00}$
\citep{vandenbosch2016}. The rotation velocity of   NGC~3621 is 140~km\,s$^{-1}$ \citep{deBlok2008}. The abundance gradient traced by the H\,{\sc ii} regions from the   compilation in
\citet{Pilyugin2014} was approximated by the relation
\begin{equation}
12+\log(\rm O/H) = 8.704(\pm 0.022) - 0.579(\pm 0.040) \times R_{g}   
\label{equation:n3621-oh-compil}
\end{equation}
with a scatter of 0.057 dex for 49 data points.

\subsection{NGC~4254 (M~99)}
%=====================

Galaxy NGC~4254 (M~99) is a bright Sc galaxy  (morphological type code $T$ = 5.2$\pm$0.7) in the Virgo Cluster. The inclination angle of  NGC~4254 is $i$ = 34$\degr$, and the position angle of the
major axis is PA = 68$\degr$ \citep{Lang2020}. The optical radius is 2.68 arcmin \citep{RC3}. At a distance    of $d$ = 13.1 Mpc \citep{Anand2021}, the physical optical radius of NGC~4254
is $R_{25}$ = 10.23 kpc. The stellar mass is $M_{\star}$ = 2.63 $\times$ 10$^{10}$ $M_{\sun}$, or log($M_{\star}/M_{\sun}$) = 10.42 \citep{Leroy2021}.  The rotation velocity of  NGC~4254 is 183~km\,s$^{-1}$
\citep{Lang2020}. The radial distribution of the oxygen abundances for the H\,{\sc ii} regions from the compilation in \citet{Pilyugin2014} was approximated by the relation
\begin{equation}
12+\log(\rm O/H) = 8.662(\pm 0.023) - 0.254(\pm 0.040) \times R_{g}   
\label{equation:n4254-oh-compil}
\end{equation}
with a scatter of 0.036 dex for 17 data points. 

\subsection{NGC~4258 (M~106)}
%=====================

Galaxy NGC~4258 (M~106) is an Sbc galaxy  (morphological type code $T$ = 4.0$\pm$0.2). The inclination angle of   NGC~4258 is $i$ = 72$\degr$, and the position angle of the major axis is PA = 331$\degr$
\citep{Ponomareva2017}. The optical radius is 9.31 arcmin \citep{RC3}. At   distance   of $d$ = 7.58 Mpc \citep{Anand2021}, the physical optical radius of NGC~4258  is $R_{25}$ = 20.53 kpc.
The stellar mass is $M_{\star}$ = 5.15 $\times$ 10$^{10}$ $M_{\sun}$ or log($M_{\star}/M_{\sun}$) = 10.71 \citep{Jarrett2019}. The black hole  mass  is log($M_{BH}/M_{\sun}$) = 7.58$\pm$0.03
\citep{vandenbosch2016}.  The rotation velocity of  NGC~4258 is 200 km\,s$^{-1}$ \citep{Ponomareva2017}. The abundance gradient traced by the H\,{\sc ii} regions from \citet{Bresolin1999} is 
\begin{equation}
12+\log(\rm O/H) = 8.600(\pm 0.023) - 0.157(\pm 0.042) \times R_{g},   
\label{equation:n4258-oh-compil}
\end{equation}
with a scatter of 0.013 dex for four data points.

\subsection{NGC~4303 (M~61)}
%==========================

Galaxy NGC~4303 (M~61) is an Sbc galaxy  (morphological type code $T$ = 4.0$\pm$0.1) in the Virgo Cluster. The inclination angle of  NGC~4303 is $i$ = 27$\degr$, and the position angle of the major
axis is PA = 318$\degr$ \citep{Guhathakurta1988}. The optical radius is 3.23 arcmin \citep{RC3}. At a distance  of $d$ = 16.99 Mpc \citep{Anand2021}, the physical optical
radius is $R_{25}$ = 15.95 kpc. The stellar mass is $M_{\star}$ = 3.24 $\times$ 10$^{10}$ $M_{\sun}$, or log($M_{\star}/M_{\sun}$) = 10.51 \citep{Leroy2021}. The black hole  mass  is
log($M_{BH}/M_{\sun}$) = 6.58$^{+0.07}_{-0.26}$ \citep{Davis2019}.  The rotation velocity of  NGC~4303 is 150 km\,s$^{-1}$ \citep{Guhathakurta1988},   although  $V_{rot}$ = 178 km\,s$^{-1}$ has been
offered more recently \citep[see][]{Lang2020}.   
The radial distribution of the oxygen abundances estimated through the $R$ calibration in H\,{\sc ii} regions from a compilation in \citet{Pilyugin2014}, is approximated by the relation 
\begin{equation}
12+\log(\rm O/H) = 8.668(\pm 0.041) - 0.406(\pm 0.100) \times R_{g}   
\label{equation:n4303-oh-compil}
\end{equation}
with a scatter of 0.078~dex for the 20 data points.

\subsection{NGC~4321 (M~100)}
%=====================

Galaxy NGC~4321 (M~100) is an SABb galaxy  (morphological type code $T$ = 4.0$\pm$0.3) in the Virgo Cluster. The inclination angle of  NGC~4321 is $i$ = 27$\degr$, and the position angle of the major axis is
PA = 153$\degr$ \citep{Guhathakurta1988}. The optical radius is 3.71 arcmin \citep{RC3}. At a distance    of $d$ = 15.21 Mpc \citep{Anand2021}, the physical optical radius of NGC~4321
is $R_{25}$ = 16.40 kpc. The stellar mass is $M_{\star}$ = 5.62 $\times$ 10$^{10}$ $M_{\sun}$, or log($M_{\star}/M_{\sun}$) = 10.75 \citep{Leroy2021}. The black hole  mass  is log($M_{BH}/M_{\sun}$) =
6.67$^{+0.17}_{-6.67}$ \citep{vandenbosch2016}.  The rotation velocity of   NGC~4321 is 270 km\,s$^{-1}$ \citep{Guhathakurta1988} for  $i$ = 27$\degr$, while \citet{Lang2020} found  $V_{rot}$ = 181 km\,s$^{-1}$
for  $i$ = 38.5$\degr$. The radial distribution of the oxygen abundances in the H\,{\sc ii} regions from the compilation in \citet{Pilyugin2014} was approximated by the relation
\begin{equation}
12+\log(\rm O/H) = 8.616(\pm 0.021) - 0.079(\pm 0.038) \times R_{g}   
\label{equation:n4321-oh-compil}
\end{equation}
with a scatter of 0.022 dex for nine data points. 

\subsection{NGC~4395}
%=====================

Galaxy NGC~4395 is an Sm galaxy (morphological type code $T$ = 8.8$\pm$0.5) in front of the Virgo Cluster. The inclination angle of   NGC~4395 is $i$ = 46$\degr$, and the position angle of
the major axis is PA = 324$\degr$ \citep{Swaters1999}. The optical radius is 6.59 arcmin \citep{RC3}. At a distance (mean value for north and south fields) 
 of $d$ = 4.51 Mpc \citep{Sabbi2018}, the physical optical radius of NGC~4395 is $R_{25}$ = 8.65 kpc. The stellar mass is $M_{\star}$ = 2.63 $\times$ 10$^{9}$ $M_{\sun}$, or log($M_{\star}/M_{\sun}$)
= 9.42 \citep{Jarrett2019}. The black hole  mass  is log($M_{BH}/M_{\sun}$) = 5.64$^{+0.22}_{-0.12}$ \citep{vandenbosch2016}. The rotation velocity of   NGC~4395 is 80 km\,s$^{-1}$ 
\citep{Swaters1999}.  The abundance gradient traced by the H\,{\sc ii} regions from the compilation in \citet{Pilyugin2014} is
\begin{equation}
12+\log(\rm O/H) = 8.063(\pm 0.016) + 0.016(\pm 0.066) \times R_{g},   
\label{equation:n4395-oh-compil}
\end{equation}
with a scatter of 0.054 dex for 14 data points. 

\subsection{NGC~4501 (M~88)}
%=====================

Galaxy NGC~4501 (M~88) is an Sb galaxy  (morphological type code $T$ = 3.3$\pm$0.6). The inclination angle of   NGC~4501 is $i$ = 64$\degr$, and the position angle of the major axis is PA = 141$\degr$
\citep{Wong2004}. The optical radius is 3.46 arcmin \citep{RC3}. At a distance    of $d$ = 16.8 Mpc \citep{Leroy2019}, the physical optical radius of NGC~4501 is $R_{25}$ = 16.90 kpc.
The stellar mass is $M_{\star}$ = 1.0 $\times$ 10$^{11}$ $M_{\sun}$, or log($M_{\star}/M_{\sun}$) = 11.00 \citep{Leroy2019}. The black hole  mass  is log($M_{BH}/M_{\sun}$) = 7.13$\pm$0.08
\citep{Davis2019}.  The rotation velocity of   NGC~4501 is 280 km\,s$^{-1}$ \citep{Wong2004}. The radial distribution of the oxygen abundances in the H\,{\sc ii} regions from
\citet{Skillman1996} was approximated by the relation
\begin{equation}
12+\log(\rm O/H) = 8.774(\pm 0.076) - 0.330(\pm 0.154) \times R_{g}   
\label{equation:n4501-oh-compil}
\end{equation}
with a scatter of 0.034 dex for four data points. 

\subsection{NGC~4625}
%=====================

Galaxy NGC~4625 is an SABm galaxy  (morphological type code $T$ = 8.7$\pm$0.9). The inclination angle of   NGC~4625 is $i$ = 31$\degr$, and the position angle of the major axis is PA = 303$\degr$
\citep{Kaczmarek2012}. The optical radius is 1.09 arcmin \citep{RC3}. At a distance    of $d$ = 11.75 Mpc \citep{McQuinn2017}, the physical optical radius of NGC~4625 is $R_{25}$ = 3.74 kpc.
The stellar mass is $M_{\star}$ = 1.19 $\times$ 10$^{9}$ $M_{\sun}$, or log($M_{\star}/M_{\sun}$) = 9.08 \citep{Leroy2019}.  The rotation velocity of   NGC~4625 is 60 km\,s$^{-1}$ \citep{Kaczmarek2012}. 
The radial distribution of the oxygen abundances from the H\,{\sc ii} regions from \citet{Goddard2011} was approximated by the relation
\begin{equation}
12+\log(\rm O/H) = 8.624(\pm 0.017) - 0.136(\pm 0.056) \times R_{g}   
\label{equation:n4625-oh-compil}
\end{equation}
with a scatter of 0.025 dex for 12 data points. 

\subsection{NGC~5055  (M~63)}
%=====================

Galaxy NGC~5055 (M~63) is an Sbc galaxy  (morphological type code $T$ = 4.0$\pm$0.2). The inclination angle of   NGC~5055 is $i$ = 59$\degr$, the position angle of the major axis is PA = 102$\degr$,
and the optical radius is 5.87 arcmin \citep{deBlok2008,Walter2008}. At a distance    of $d$ = 8.87 Mpc \citep{McQuinn2017}, the physical optical radius of NGC~5055 is $R_{25}$ = 15.16 kpc.
The value of the rotation velocity on the flat part is 192 km\,s$^{-1}$ \citep{deBlok2008,Ponomareva2017}. The mean value of the stellar mass from estimations by \citet{Leroy2008} and
\citet{Jarrett2019} and rescaled to the adopted distance is $M_{\star}$ = 5.40 $\times$ 10$^{10}$ $M_{\sun}$, or log($M_{\star}/M_{\sun}$) = 10.73. The mass of the black hole in   NGC~5055 is
log($M_{BH}/M_{\sun}$) = 8.92$\pm 0.10$ in solar mass \citep{vandenbosch2016}. The abundance gradient traced by the four H\,{\sc ii} regions measured by \citet{McCall1985} is
\begin{equation}
12+\log(\rm O/H) = 8.693(\pm 0.035) - 0.280(\pm 0.065) \times R_{g}   
\label{equation:n5055-oh-compil}
\end{equation}
with a scatter of 0.017 dex for four data points. 

\subsection{NGC~5068}
%=====================

Galaxy NGC~5068 is an Sc galaxy (morphological type code $T$ = 6.0$\pm$0.4). The inclination angle of   NGC~5068 is $i$ = 35$\degr$, and the position angle of the major axis is PA = 342$\degr$
\citep{Lang2020}. The optical radius is 3.62 arcmin \citep{RC3}. At a distance    of $d$ = 5.20 Mpc \citep{Anand2021}, the physical optical radius of NGC~5068 is $R_{25}$ = 5.48 kpc.
The stellar mass is $M_{\star}$ = 2.57 $\times$ 10$^{9}$ $M_{\sun}$, or  log($M_{\star}/M_{\sun}$) = 9.41 \citep{Leroy2021}. The radial distribution of the oxygen abundances in the
H\,{\sc ii} regions from \citet{Ryder1995} was approximated by the relation
\begin{equation}
12+\log(\rm O/H) = 8.474(\pm 0.050) - 0.158(\pm 0.115) \times R_{g}   
\label{equation:n5068-oh-compil}
\end{equation}
with a scatter of 0.058 dex for 16 data points. 

\subsection{NGC~5194 (M~51a)}
%=====================

The nearby galaxy NGC~5194 (M~51a, the Whirlpool Galaxy) is an SABb spiral galaxy (morphological type code $T$ = 4.0$\pm$0.3). The bright disc of NGC~5194 ends abruptly at about 5 arcmin
radius in both the optical images and the  H\,{\sc i}. The velocity structure of the gas in NGC~5194 is extremely complicated and difficult to interpret \citep{Rots1990}. 
There is the misalignment of the major axes of the H\,{\sc i} distribution and the velocity field. Therefore, the geometrical parameters of  NGC~5194 are rather uncertain. 
\citet{Tamburro2008} derived the following geometrical projection parameters of the NGC~5194 galaxy disc: The position angle is PA = 172$\degr,$ and the inclination is $i$ = 42$\degr$.
\citet{Colombo2014} undertook a detailed kinematic study of NGC~5194 and found a position angle PA = (173 $\pm$ 3)$\degr$ and an inclination $i$ = (22 $\pm$ 5)$\degr$. 
The geometrical parameters of   NGC~5194 obtained by \citet{Colombo2014} are used here. We adopt the optical radius of   NGC~5194  $R_{25}$ = 5.61 arcmin \citep{RC3}.
We note that the value of the optical radius of $R_{25}$ = 3.88~arcmin was used for NGC~5194 within THINGS \citep{Walter2008}. 
There are recent distance estimations for   NGC~5194 made through the tip of the red giant branch method based on Hubble Space Telescope measurements. 
\citet{Tikhonov2015} found the distance to NGC~5194 to be in the range from 8.88 to 9.09~Mpc. \citet{McQuinn2017} have measured the distance to NGC~5194 to be
8.58 $\pm$ 0.10 (statistical uncertainty) $\pm$ 0.28 (systematic uncertainty) Mpc. \citet{Sabbi2018} determined the distances for the central pointing (d = 7.6 $\pm$ 0.8 Mpc)
and for two outer fields (d = 7.2 $\pm$ 0.6~Mpc and d = 7.6 $\pm$ 0.6~Mpc). The distance to   NGC~5194 that we adopted in this work is $d$ = 8.58~Mpc, which was obtained by \citet{McQuinn2017}.
The optical radius of   NGC~5194 is $R_{25}$ = 14.00~kpc with the adopted distance. 
The value of the stellar mass rescaled to the adopted distance is $M_{\star}$ = 4.54 $\times$ 10$^{10}$ $M_{\sun}$, or  log($M_{\star}/M_{\sun}$) = 10.66 \citep{Leroy2008,Jarrett2019}. 
The rotation curve of NGC~5194 has been measured in several investigations \citep[e.g.][]{Tilanus1991,Leroy2008,Oikawa2014}. We adopted the rotation velocity of $V_{rot}$ = 219~km\,s$^{-1}$
from \citet{Leroy2008}.  We note that the value of the $V_{rot}$ for NGC~5194 can include a significant error because of the low inclination of the galaxy. 
The radial distribution of the oxygen abundances estimated through the $R$ calibration in the H\,{\sc ii} regions from the compilation in \citet{Pilyugin2014} and supplemented by the
measurements from  \citet{Croxall2015} was approximated by the relation
\begin{equation}
12+\log(\rm O/H) = 8.700(\pm 0.008) - 0.143(\pm 0.020) \times R_{g}   
\label{equation:n5194-oh-compil}
\end{equation}
with a scatter of 0.030~dex for 87 data points. 

\subsection{NGC~5236 (M~83)}
%=====================

Galaxy NGC~5236 (M~83) is an Sc galaxy  (morphological type code $T$ = 5.0$\pm$0.3). The inclination angle of   NGC~5236 is $i$ = 24$\degr$, and the position angle of the major axis is PA = 225$\degr$
\citep{Leroy2021}. The optical radius is 6.44 arcmin \citep{RC3}. At   distance    of $d$ = 4.89 Mpc \citep{Anand2021}, the physical optical radius of NGC~5236 is $R_{25}$ = 9.16 kpc.
The stellar mass is $M_{\star}$ = 3.39 $\times$ 10$^{10}$ $M_{\sun}$, or  log($M_{\star}/M_{\sun}$) = 10.53 \citep{Jarrett2019,Leroy2021}. The rotation velocity of   NGC~5236 is 190 km\,s$^{-1}$
\citep{Lundgren2004}.   The abundance gradient traced by the H\,{\sc ii} regions from the compilation in \citet{Pilyugin2014} was approximated by the relation
\begin{equation}
12+\log(\rm O/H) = 8.688(\pm 0.006) - 0.120(\pm 0.012) \times R_{g}   
\label{equation:n5236-oh-compil}
\end{equation}
with a scatter of 0.022 dex for 51 data points. 

\subsection{NGC~5248}
%=====================

Galaxy NGC~5248 is an SABb galaxy  (morphological type code $T$ = 4.0$\pm$0.3). The inclination angle of   NGC~5248 is $i$ = 47$\degr$, and the position angle of the major axis is PA = 109$\degr$
\citep{Lang2020}. The optical radius is 3.08 arcmin \citep{RC3}. At a distance    of $d$ = 14.87 Mpc \citep{Anand2021}, the physical optical radius of NGC~5248 is $R_{25}$ = 13.34 kpc.
The stellar mass  is $M_{\star}$ = 2.57 $\times$ 10$^{10}$ $M_{\sun}$, or  log($M_{\star}/M_{\sun}$) = 10.41 \citep{Leroy2021}.  The black hole  mass  is log($M_{BH}/M_{\sun}$) = 6.30$\pm$0.38
\citep{vandenbosch2016}.  The rotation velocity of   NGC~5248 is 196 km\,s$^{-1}$ \citep{Lang2020}. The radial distribution of the oxygen abundances from the H\,{\sc ii} regions from the
compilation in \citet{Pilyugin2014} was approximated by the relation
\begin{equation}
12+\log(\rm O/H) = 8.517(\pm 0.017) + 0.033(\pm 0.061) \times R_{g}   
\label{equation:n5248-oh-compil}
\end{equation}
with a scatter of 0.040 dex for 11 data points. 

\subsection{NGC~5457 (M~101)}
%=====================

The giant nearby galaxy NGC~5457 (M~101, the Pinwheel) is a prototype of  Sc spiral galaxy (morphological type code $T$ = 5.9$\pm$0.3). It is a face-on galaxy.
Its inclination angle is $i$ = 18$\degr$, and the position angle of the major axis is PA = 37$\degr$ \citep{Kamphuis1993}. The optical radius of   NGC~5457 is $R_{25}$ = 14.42 arcmin
\citep{RC3}. There are 79 independent distance measurements of NGC~5457 after the year 2000, including those using Cepheids and tip of the read giant branch \citep{Lomeli2022}. 
The obtained distances are within the range of $\sim$6 to $\sim$9 Mpc. In this work, we adopted the distance to NGC~5457 used in our previous study: $d$ = 6.85 \citep{Pilyugin2014}. 
The optical radius of   NGC~5457 is $R_{25}$ = 28.73 kpc with the adopted distance. The stellar mass of   NGC~5457 is $M_{\star}$ = 3.81 $\times$ 10$^{10}$ $M_{\sun}$, or
log($M_{\star}/M_{\sun}$) = 10.58, and the mean value of the estimations from \citet{vanDokkum2014} and \citet{Jarrett2019} were rescaled to the adopted distance. The mass of the black hole in
the NGC~5457 is log($M_{BH}/M_{\sun}$) = 6.41$^{+0.08}_{-6.41}$ \citep{vandenbosch2016}. The abundance gradient traced by the H\,{\sc ii} regions from the compilation in \citet{Pilyugin2014}
and supplemented by the measurements from  \citet{Croxall2016} and \citet{Esteban2020} is 
\begin{equation}
12+\log(\rm O/H) = 8.688(\pm 0.009) - 0.793(\pm 0.021) \times R_{g},   
\label{equation:n5457-oh-compil}
\end{equation}
with a scatter of 0.061 dex for 213 data points. 

\subsection{NGC~6384}
%=====================

Galaxy NGC~6384 is an Sbc galaxy  (morphological type code $T$ = 3.6$\pm$0.6). The inclination angle of   NGC~6384 is $i$ = 55$\degr$, and the position angle of the major axis is PA = 31$\degr$
\citep{Mitchell2018}. The optical radius is 3.08 arcmin \citep{RC3}. At a distance    of $d$ = 25.9 Mpc \citep{Leroy2019}, the physical optical radius of NGC~6384 is $R_{25}$ = 23.23 kpc.
The stellar mass is $M_{\star}$ = 5.74 $\times$ 10$^{10}$ $M_{\sun}$, or log($M_{\star}/M_{\sun}$) = 10.76 \citep{Leroy2019}. The rotation velocity of   NGC~6384 is 230 km\,s$^{-1}$ \citep{Mitchell2018}.   
The radial distribution of the oxygen abundances estimated through the $R$ calibration in H\,{\sc ii} regions from the compilation in \citet{Pilyugin2014} was approximated by the relation
\begin{equation}
12+\log(\rm O/H) = 8.783(\pm 0.055) - 0.373(\pm 0.094) \times R_{g}   
\label{equation:n6384-oh-compil}
\end{equation}
with a scatter of 0.025 dex for eight data points. 

\subsection{NGC~6744}
%=====================

Galaxy NGC~6744 is an Sbc galaxy  (morphological type code $T$ = 4.0$\pm$0.2). The inclination angle of   NGC~6744 is $i$ = 50$\degr$, and the position angle of the major axis is PA = 16$\degr$
\citep{Ryder1999}. The optical radius is 9.98 arcmin \citep{RC3} (although \citet{Ho2011}   found around half that optical radius). At a distance   of $d$ = 9.39 Mpc
\citep{Anand2021}, the physical optical radius of NGC~6744 is R$_{25}$ = 27.25 kpc. The stellar mass (mean value) is M$_{\star}$ = 5.92 $\times$ 10$^{10}$ M$_{\sun}$, or log($M_{\star}/M_{\sun}$) = 10.77
\citep{Jarrett2019,Leroy2021}. The black hole  mass  is log($M_{BH}/M_{\sun}$) = 6.89$\pm$0.34 \citep{Davis2014}. The rotation velocity of   NGC~6744 is 200 km\,s$^{-1}$ \citep{Ryder1999}.   
The abundance gradient traced by the H\,{\sc ii} regions from \citet{Ryder1995} is 
\begin{equation}
12+\log(\rm O/H) = 8.840(\pm 0.025) - 0.640(\pm 0.054) \times R_{g},   
\label{equation:n6744-oh-compil}
\end{equation}
with a scatter of 0.030 dex for 17 data points. 

\subsection{NGC~6946}
%=====================

Galaxy NGC~6946 is an SABc galaxy  (morphological type code $T$ = 5.9$\pm$0.3). The inclination angle of   NGC~6946 is $i$ = 33$\degr$, and the position angle of the major axis is PA = 243$\degr$
\citep{deBlok2008}. The optical radius is 5.74 arcmin \citep{RC3}. At a distance    of $d$ = 7.34 Mpc \citep{Anand2021}, the physical optical radius of NGC~6946 is $R_{25}$ = 12.26 kpc.
The stellar mass (mean value) is $M_{\star}$ = 2.85 $\times$ 10$^{10}$ $M_{\sun}$, or  log($M_{\star}/M_{\sun}$) = 10.45 \citep{Jarrett2019,Leroy2019}.  The rotation velocity of   NGC~6946
is 186 km\,s$^{-1}$ \citep{deBlok2008,Leroy2008}. The radial distribution of the oxygen abundances estimated using the H\,{\sc ii} regions from the compilation in \citet{Pilyugin2014} was
approximated by the relation
\begin{equation}
12+\log(\rm O/H) = 8.648(\pm 0.050) - 0.258(\pm 0.054) \times R_{g}   
\label{equation:n6946-oh-compil}
\end{equation}
with a scatter of 0.047 dex for 12 data points. 

\subsection{NGC~7331}
%=====================

Galaxy NGC~7331 is an Sbc galaxy  (morphological type code $T$ = 3.9$\pm$0.3). The inclination angle of   NGC~7331 is $i$ = 76$\degr$, and the position angle of the major axis is PA = 168$\degr$
\citep{deBlok2008}. The optical radius is 5.24 arcmin \citep{RC3}. At a distance    of $d$ = 14.7 Mpc \citep{Walter2008}, the physical optical radius of NGC~7331  is $R_{25}$ = 22.39 kpc.
The stellar mass (mean value) is $M_{\star}$ = 1.0 $\times$ 10$^{11}$ $M_{\sun}$, or  log($M_{\star}/M_{\sun}$) = 11.00 \citep{Jarrett2019,Leroy2019}. The black hole  mass  is log($M_{BH}/M_{\sun}$)
= 8.02$\pm$0.18 \citep{vandenbosch2016}.  The rotation velocity of   NGC~7331 is 244 km\,s$^{-1}$ \citep{deBlok2008,Leroy2008}. The abundance gradient traced by the H\,{\sc ii} regions
from \citet{Bresolin1999} is 
\begin{equation}
12+\log(\rm O/H) = 8.590(\pm 0.058) - 0.045(\pm 0.117) \times R_{g},   
\label{equation:n7331-oh-compil}
\end{equation}
with a scatter of 0.014 dex for four data points. 

\subsection{NGC~7518}
%=====================

Galaxy NGC~7518 is an SABa galaxy  (morphological type code $T$ = 1.1$\pm$0.6). The inclination angle of   NGC~7518 is $i$ = 47$\degr$, and the position angle of the major axis is PA = 294$\degr$
\citep{Amorin2009}. The optical radius is 0.71 arcmin \citep{RC3}. At a distance    of $d$ = 47.56 Mpc \citep{Amorin2009}, the physical optical radius of NGC~7518 is $R_{25}$ = 9.77 kpc.
The stellar mass is $M_{\star}$ = 1.48 $\times$ 10$^{10}$ $M_{\sun}$, or   log($M_{\star}/M_{\sun}$) = 10.17 \citep{Leroy2019}. The radial distribution of the oxygen abundances estimated for the
H\,{\sc ii} regions from \citet{Robertson2012} was approximated by the relation
\begin{equation}
12+\log(\rm O/H) = 8.675(\pm 0.012) - 0.103(\pm 0.025) \times R_{g}   
\label{equation:n7518-oh-compil}
\end{equation}
with a scatter of 0.019 dex for 12 data points. 

\subsection{NGC~7529}
%=====================

Galaxy NGC~7529 is an Sbc galaxy  (morphological type code $T$ = 4.3$\pm$2.2). The inclination angle of   NGC~7529 is $i$ = 29$\degr$, and the position angle of the major axis is PA = 157$\degr$
\citep{Zurita2021}. The optical radius is 0.43 arcmin \citep{RC3}. At a distance    of $d$ = 63.2 Mpc \citep{Zurita2021}, the physical optical radius of NGC~7529 is $R_{25}$ = 7.82 kpc.
The stellar mass is $M_{\star}$ = 7.34 $\times$ 10$^{9}$ $M_{\sun}$, or   log($M_{\star}/M_{\sun}$) = 9.87 \citep{Leroy2019}. The abundance gradient traced by the H\,{\sc ii} regions from
\citet{Robertson2012} is 
\begin{equation}
12+\log(\rm O/H) = 8.637(\pm 0.043) - 0.263(\pm 0.073) \times R_{g},   
\label{equation:n7529-oh-compil}
\end{equation}
with a scatter of 0.045 dex for 11 data points. 

\subsection{NGC~7591}
%=====================

Galaxy NGC~7591 is an SBbc galaxy  (morphological type code $T$ = 3.6$\pm$0.6). The inclination angle of   NGC~7591 is $i$ = 68$\degr$, and the position angle of the major axis is PA = 148$\degr$
\citep{Rubin1988}. The optical radius is 0.98 arcmin \citep{RC3}. At a distance    of $d$ = 67.3 Mpc \citep{Leroy2019}, the physical optical radius of NGC~7591 is $R_{25}$ = 19.08 kpc.
The stellar mass  is $M_{\star}$ = 3.72 $\times$ 10$^{10}$ $M_{\sun}$, or log($M_{\star}/M_{\sun}$) = 10.57 \citep{Leroy2019}. The rotation velocity of   NGC~7591 is 199 km\,s$^{-1}$ \citep{Rubin1988}.   
The radial distribution of the oxygen abundances estimated through the $R$ calibration in H\,{\sc ii} regions from \citet{Robertson2012} was approximated by the relation
\begin{equation}
12+\log(\rm O/H) = 8.638(\pm 0.016) - 0.085(\pm 0.031) \times R_{g}   
\label{equation:n7591-oh-compil}
\end{equation}
with a scatter of 0.025 dex for 11 data points. 

\subsection{NGC~7793}
%=====================

Galaxy NGC~7793 is  an Scd galaxy  (morphological type code $T$ = 7.4$\pm$0.6). The inclination angle of   NGC~7793 is $i$ = 50$\degr$, the position angle of the major axis is PA = 290$\degr$,
and the optical radius is 5.24 arcmin \citep{deBlok2008}. At a distance    of $d$ = 3.62 Mpc \citep{Anand2021}, the physical optical radius of NGC~7793 is $R_{25}$ = 5.51 kpc.
The maximum value of the rotation velocity of NGC~7793 is 118 km\,s$^{-1}$, and the value of the rotation velocity on the flat part is 95 km\,s$^{-1}$ \citep{Carignan1990,deBlok2008,Ponomareva2017}.
We adopted $V_{rot}$ = 95 km\,s$^{-1}$. The mean value out of six estimations of the stellar mass of NGC~7793 \citep{Ponomareva2018,Jarrett2019,Leroy2021} is $M_{\star}$ = 2.54 $\times$ 10$^{9}$ $M_{\sun}$,
or  log($M_{\star}/M_{\sun}$) = 9.40. The radial distribution of the oxygen abundances for the H\,{\sc ii} regions from compilation in \citet{Pilyugin2014} was approximated by the relation
\begin{equation}
12+\log(\rm O/H) = 8.477(\pm 0.024) - 0.343(\pm 0.055) \times R_{g}   
\label{equation:n7793-oh-compil}
\end{equation}
with a scatter of 0.068 dex for 38 data points.

\subsection{IC~342}
%===========================

Galaxy IC~342 is an SABc galaxy  (morphological type code $T$ = 6.0$\pm$0.3). The inclination angle of  IC~342 is $i$ = 31$\degr$ and, the position angle of the major axis is PA = 37$\degr$
\citep{Crosthwaite2000}. The optical radius is 10.69 arcmin \citep{RC3}. At a distance     $d$ = 3.45 Mpc \citep{Anand2021}, the physical optical radius of IC~342 is $R_{25}$
= 10.73 kpc. The stellar mass (mean value) is $M_{\star}$ = 2.35 $\times$ 10$^{10}$ $M_{\sun},$ or log($M_{\star}/M_{\sun}$) = 10.37 \citep{Jarrett2019,Leroy2019}. (We note the large uncertainties in
the mass estimations: $M_{\star}$ = 1.45 $\times$ 10$^{10}$ $M_{\sun}$, or log($M_{\star}/M_{\sun}$) = 10.16 \citep{Leroy2019}, and  $M_{\star}$ = 3.25 $\times$ 10$^{10}$ $M_{\sun}$,
or log($M_{\star}/M_{\sun}$) = 10.51 \citep{Jarrett2019}). The rotation velocity of  IC~342 is 170~km\,s$^{-1}$ \citep{Crosthwaite2000}, although \citet{Newton1980}  found $V_{rot}$ = 191~km\,s$^{-1}$
for $i$ = 25$\degr$ and PA = 39$\degr$. 
The radial distribution of the oxygen abundances estimated through the $R$ calibration in H\,{\sc ii} regions from \citet{McCall1985} was approximated by the relation
\begin{equation}
12+\log(\rm O/H) = 8.713(\pm 0.030) - 0.426(\pm 0.058) \times R_{g}   
\label{equation:i0342-oh-compil}
\end{equation}
with a  scatter of 0.030~dex for four data points.

\subsection{IC~5201}
%=====================

Galaxy IC~5201 is an Sc galaxy  (morphological type code $T$ = 6.1$\pm$0.6). The inclination angle of   IC~5201 is $i$ = 67$\degr$, and the position angle of the major axis is PA = 206$\degr$
\citep{Kleiner2019}. The optical radius is 4.26 arcmin \citep{RC3}. At a distance    of $d$ = 9.20 Mpc \citep{Leroy2019}, the physical optical radius of IC~5201 is $R_{25}$ = 11.39 kpc.
The stellar mass is $M_{\star}$ = 7.59 $\times$ 10$^{9}$ $M_{\sun}$, or  log($M_{\star}/M_{\sun}$) = 9.88 \citep{Leroy2019}.  The rotation velocity of   IC~5201 is 98 km\,s$^{-1}$ \citep{Kleiner2019}. 
The abundance gradient traced by the H\,{\sc ii} regions from \citet{Ryder1995} is 
\begin{equation}
12+\log(\rm O/H) = 8.349(\pm 0.082) - 0.530(\pm 0.169) \times R_{g},   
\label{equation:i5201-oh-compil}
\end{equation}
with a scatter of 0.085 dex for six data points. 

\subsection{IC~5309}
%=====================

Galaxy IC~5309 is an Sb galaxy  (morphological type code $T$ = 3.1$\pm$0.4). The inclination angle of   IC~5309 is $i$ = 63$\degr$, and the position angle of the major axis is PA = 20$\degr$
\citep{Amram1992}. The optical radius is 0.67 arcmin \citep{RC3}. At a distance    of $d$ = 55.7 Mpc \citep{Leroy2019}, the physical optical radius of IC~5309 is $R_{25}$ = 10.93 kpc.
The stellar mass is $M_{\star}$ = 1.62 $\times$ 10$^{10}$ $M_{\sun}$, or  log($M_{\star}/M_{\sun}$) = 10.21 \citep{Leroy2019}.  The rotation velocity of   IC~5309 is 152 km\,s$^{-1}$ \citep{Amram1992}. 
The radial distribution of the oxygen abundances estimated through the $R$ calibration in H\,{\sc ii} regions from \citet{Robertson2012} was approximated by the relation
\begin{equation}
12+\log(\rm O/H) = 8.614(\pm 0.031) - 0.049(\pm 0.065) \times R_{g}   
\label{equation:i5309-oh-compil}
\end{equation}
with a scatter of 0.058 dex for 14 data points. 

\end{document}